\documentclass[final,3p,times]{elsarticle}       

\usepackage{latexsym}
\usepackage{pstricks}   
\usepackage{bm}
\usepackage{amsmath}
\usepackage{amssymb}
\numberwithin{equation}{section}
\begin{document}
\title{Massive Spin-2: Field-equations, Propagators, \\   
           Massless-limit, and Perihelion Precessions 
	     \footnote{For earlier version see: Report THEF-NIJM 24.01,
	     https://nn-online.org/eprints(2024)}
	     }                                 
\author[1]{Th.A.\ Rijken and J.W.\ Wagenaar}
\address[1]{IMAPP, Radboud University, Nijmegen, The Netherlands}

\date{version of: \today}

\begin{abstract}
\begin{description}
 \item[\bf Background:]
 This paper presents the quantization of massive and massless 
 spin-2 particles using the auxiliary-field method.
The issue, the so-called vDVZ-discontinuity, whether the perihelion precessions 
		for a massive graviton are in agreement
 with the data, is studied in the context of this spin-2 theory in tree-approximation.
 \item[\bf Purpose:]     
	 The aim is to study the massless limit and to investigate the perihelion
	 of the planets as a function of the graviton mass, and to calculate
	 the effects to first order in the graviton mass.
 \item[\bf Method:]      
The field theory for the 
spin-2 particles is constructed using the Dirac quantization method.
 In order to impose sufficient constraints on the spin-2 tensor 
 $h^{\mu\nu}(x)$-field, an auxiliary vector-ghost field $\eta^\mu(x)$ 
and a imaginary scalar-ghost field $\epsilon(x)$ are introduced.
The $h^{\mu\nu}(x)$-field is coupled to a conserved energy-momentum tensor, which 
results in a dependence of the $h^{\mu\nu}$-propagator on the $\epsilon(x)$-field for
 the massive case.


 \noindent  The gravitational interaction between the 
 the sun and the planets (treated as scalar particles) is introduced as in the 
weak-field approximation in general relativity, {\it i.e.} by a coupling 
of the $h^{\mu\nu}(x)$-field to the energy-momentum tensors.


 \item[\bf Results:]    
A general Lagrangian, containing parameters A and $b_2$, 
for the massive spin-2 (tensor) formalism using the auxiliary
spin-1 (vector) and spin-0 (scalar) fields is reviewed.                    
It is found that only A=-1 leads to a physical theory.
Furthermore, it appeared that for the proper transition
to a massless spin-2 theory the limit $b_2 \sim b \rightarrow \pm\infty$ is required.
By making a suitable field-transformation, a theory is obtained for 
massive and massless spin-2 fields with an imaginary-scalar-ghost $\epsilon(x)$ and 
a vector-ghost $\rho^\mu(x)$ field, both satisfying free Klein-Gordon equations.
Furthermore, it is shown that the $\rho$-field is eliminated from this model for $b=\pm\infty$. 
The quantization of the $\epsilon$ ghost-field is analyzed.
Using a standard Gupta subsidiary-condition for physical states, 
taking care of the $\epsilon(x)$ ghost-states in the standard manner, 
a massive spin-2 quantum field theory with a spin-0 scalar-ghost is reached. 

 \noindent Coupling the $h^{\mu\nu}(x)$-field to the energy-momentum
 tensor for the sun and planets, the non-newtonian correction
 to the perihelion precession is in accordance with the
 Einstein result, also in the case of a non-zero graviton mass.
 \item[\bf Conclusions:]
In the context of this setting, massive gravitation with a imaginary scalar-ghost,
it is found that, in tree-approximation, a (small) spin-2 graviton mass is compatible 
with the perihelion-precession of Mercury etc., and there is no vDVZ-discontinuity.
\end{description}
\end{abstract}
\maketitle
 
\section{Introduction}                                          
\label{sec:-1}   

\noindent It is the aim of this investigation to find a massive 
quantized spin-2 theory,  which, in tree-approximation, allows 
(i) a smooth massless limit, and (ii) a perturbative 
expansion in the small ("graviton") mass M
\footnote{In sections I-VIII the graviton mass is notated by M$_2$ or M 
and after that by $\mu_G$.}, avoiding the vDVZ-discontinuity \cite{VDV70,Zak70}.
We exploit the auxiliary field (AF) method with a vector and a scalar field
to construct a linearized relativistic gravitation theory, henceforth
referred to as RGT-AF. For the latter to be meaningful, it is 
necessary that the theory satisfies the following requirements: 
(i) unitarity, and (ii) a correct massless limit.
This would open the possibility of giving a small mass to the graviton without
destroying e.g. the correct prediction for the perihelion of Mercury. 
In the literature this issue has been discussed both in Minkowski 
\cite{VDV70,Zak70,Bou72} and in de Sitter space \cite{Por01,Kog01,Duff01}.

\noindent It appears that this is impossible without making use of complex ghost
fields. The possibility to exploit complex ghost-fields is analyzed in detail 
in this paper using methods discussed in the literature, see \cite{Nak72}.
By making a simple field-transformation the $\eta_\mu(x)$ and $\epsilon(x)$
fields can be decoupled and a model is obtained for massive spin-2 field 
with free auxiliary fields $\rho_\mu(x)$ and $\epsilon(x)$. 
Using Gupta subsidiary-conditions \cite{Gupta50} for the latter a quantum field theory
with an imaginary scalar-ghost field is reached. 
This massive spin-2 formalism is similar to the so-called
"B-field formalism" for QED, QCD, and for massive vector-mesons, 
as developed and described in Refs.~\cite{Lau67,NO90}.

\noindent In this RTG-AF formalism the massless limit can be studied
in detail, and 
it is shown that in this formalism a small spin-2 (graviton)
mass is compatible with the Einstein non-newtonian correction of the 
perihelion-precession of Mercury etc..            
From recent observations \cite{Damour91,Finn02,Choud04} and studies
\cite{Gold10} the upper limit for the graviton mass seems to be
 $\mu_G \leq 2\ 10^{-38}\ m_e= 18.22\times 10^{-66}\ g$, and is estimated in these notes to give 
a really tiny correction to the perihelion precession of Mercury.

The contents of this paper consist of three parts. 
In the first part, the attention is focused on the field equations,   
the quantization with the Dirac method, the commutation relations, and the Feynman propagator.
In the second part, the final field theory model RGT-AF is formulated, which is designed for the
computation of the perihelion precession, having a smooth massless limit. The precession
of Mercury is calculated with a finite mass-correction for the "graviton".
The third  part contains (i) the (causal) quantization of the imaginary auxiliary ghost-field, and
(ii) several appendices containing supporting material.

\noindent {\bf First part}:
In section~\ref{sec:20} the general Lagrangian for the spin-2 $h^{\mu\nu}(x)$, 
the auxiliary spin-1 $\eta^\mu(x)$, and
 spin-0 $\epsilon(x)$ fields is given, which contains the graviton-mass $M_2$-, 
 the scale-mass ${\cal M}$-, the A and b-parameter.
Here also, the field equations are derived. Furthermore, the decoupling of the vector and
the scalar auxiliary field is achieved via a field transformation.
Here also, the coupled Klein-Gordon equations 
for the spin-2, the spin-1, and spin-0 fields are given.            
In section~\ref{sec:30}, the  Dirac's Hamiltonian method, appropriate for the quantization of constrained
systems, is reviewed for the spin-2 tensor-field exploiting the auxiliary field method with
a vector and scalar field.
The canonical momenta are defined, the Hamiltonian is given, the canonical momenta and the 
equal-time commutation (ETC) relations are listed.
In section~\ref{sec:40} an integral representation for solutions of the free Klein-Gordon
equations for the tensor, vector, and scalar fields is used to obtain the non-equal-time 
commutation (NETC) relations. This path also leads to the Feynman propagator for the tensor-field.
Section~\ref{sec:4} is devoted to the question whether a representation for the 
spin-2 propagator etc. can be found in the "{\it b-parameter space}" 
that (i) allows a smooth massless limit, 
and (ii) such that for $M \neq 0$ the theory contains besides the spin-2 propagator also a 
physically acceptable spin-0 propagator. This leads to $b \rightarrow \pm\infty$.
We summarize our preliminary conclusions with respect to the impossibility of
a massive spin-2 theory with a no-ghost scalar field and a correct 
prediction for the perihelia.\\
Section~\ref{sec:50} (i) It is shown that the $\rho$-field can be eliminated from the model 
for $b \rightarrow \pm\infty$, 
leaving only the auxiliary (ghost) $\epsilon$-field.
In section~\ref{sec:50b} the asymptotic fields are introduced and the physical contents
of the $h_{\mu\nu}(x)$-fields is analyzed for the massive and massless case.
Spin-2 $U_{\mu\nu}(x)$-fields are introduced, and it is shown explicitly that there is a
smooth transition from the massive to the massless spin-2 field formalism.
This is an advantage of the auxiliary-field method compared to the Proca method.   

\noindent {\bf Second part}:
In section~\ref{sec:10} the 
massive spin-2 model of this paper is applied to a computation of the
non-Newtonian correction of the perihelion precession of Mercury.
The gravitational interaction between the sun and planets is introduced
via the coupling of the $h^{\mu\nu}(x)$ spin-2 field to the energy-momentum tensors.
In section~\ref{app:S}, the contribution to the perihelion precession
is computed for the scalar and scalar-ghost part of the tensor-field propagator. 
The results on the perihelion precession are summarized and compared 
with solar-system data.
The finite-mass corrections are computed  and shown to be negligible.       
In section~\ref{sec:11}, we discuss the results and compare this RTG-AF
with other models in the literature.\\


\noindent {\bf Appendices}: 
In \ref{app:B} the Lie algebra of the little 
group L(p) corresponding to $\bar{p}^\mu=(p^0,0,0,p)$ is constructed.
We analyze the (smooth) massless-limit transition from the little group SO(3) 
for $\bar{p}^\mu=(p^0=M,0,0,0)$ to the little group E(2) pertaining to
$\bar{p}^\mu= (p,0,0,p)$. 
To avoid the vDVZ-discontinuity, a smooth transition is required between
the little groups SO(3) and E(2), when the graviton mass goes to zero.

\noindent In \ref{sec:9} the (causal) quantization of the imaginary $\epsilon(x)$-field
is treated, and the propagator is derived in detail.
In \ref{app:helicity-coupling}, the smooth massless limit of the model is
analyzed considering the decoupling of the "false helicities" $\lambda \neq \pm 2$.
In \ref{app:BS}, starting with the Bethe-Salpeter equation for 
two-scalar particles, {\it e.g.} the sun and planet,
the local and non-local potentials for the 
Schr\"{o}dinger equation are derived for the planetary motion.
In \ref{app:SIG}, the matrix element of the scalar particle interaction 
for imaginary-ghost exchange is worked out in detail.
In \ref{app:PC3}, the gravitational cross-term field energy for the 
planet-sun system, giving a -1/6 correction 
to the perihelion precession, is explicitly evaluated.
In \ref{app:COSMO}, 
the cosmological term is incorporated in the spin-2 formalism in the weak field
approximation.  The possible relation between the cosmological
constant and graviton mass is discussed, and consequences for cosmological
parameters are estimated.
Finally, \ref{app:Miscel} contains the derivation of $\sqrt{-g}$ up to
second order in the $h^{\mu\nu}$-field.\\

\section{Massive Gravitation Field, Euler-Lagrange Equations}                          
\label{sec:20}   
\noindent In our work on the quantization of the spin-2 fields \cite{WR09}, we used
 the symmetric $h_{\mu\nu}$-tensor field, and 
the two auxiliary (ghost-)fields $\eta^\mu(x)$ and $\epsilon(x)$. The Lagrangian
consists of three parts 
${\cal L}_{2,\eta\epsilon} = {\cal L}_2 + {\cal L}_{GF}+{\cal L}_{int}$ which are 
specified below.
In \cite{Nath65} the most general ${\cal L}_2$ is parametrized in terms of the
parameters A,B, and C, with $B=(3A^2+3A+1)/2$ and $C=3A^2+3A+1$ 
\footnote{These relations between A,B,C are connected to the constraints
$h^\mu_\mu=0$ and $\partial_\mu h^{\mu\nu}=0$.}
where, see \cite{WR09}, 
\begin{eqnarray}
 {\cal L}_2 &=&                                        
 \frac{1}{4}\partial^\alpha h^{\mu\nu} \partial_\alpha h_{\mu\nu} 
 -\frac{1}{2}\partial_\mu h^{\mu\nu} \partial^\alpha h_{\alpha\nu} 
 -\frac{1}{4}B\partial_\nu h^\beta_\beta \partial^\nu h^\alpha_\alpha 
 -\frac{1}{2}A\partial_\alpha h^{\alpha\beta} \partial_\beta h^\nu_\nu \nonumber\\
 && -\frac{1}{4} M_2^2 h^{\mu\nu} h_{\mu\nu} 
+ \frac{1}{4} C M_2^2 h^\mu_\mu h^\nu_\nu, 
\label{eq:20.1}\end{eqnarray}
and the application of the variation principle via the Euler-Lagrange (${\cal E.L.}$)
equations gives the equation of motion (EoM) for the spin-2 field as well as the constraints,
for $A \ne -1/2$, {\it i.e.} analogous as the Proca formalism for spin-1.
However, in the Proca formalism, $M_2 \ne 0$ is essential and hence prevents a useful 
massless limit \cite{NO90}. Since for our purpose the massless limit is essential, we use 
auxiliary fields, henceforth referred to 
as the B-field method. 
\noindent For the symmetric $h_{\mu\nu}$-tensor field and the auxiliary
$\eta^\mu(x)$ and $\epsilon(x)$ fields, the Lagrangian is
${\cal L}_{\eta\epsilon}={\cal L}_2+{\cal L}_{GF}+{\cal L}_{int}$ where
\footnote{In the following, we often denote the mass by $M_2 \equiv M$.}
\begin{subequations}\label{eq:20.2}
\begin{eqnarray}
 {\cal L}_2 &=&                                        
 \frac{1}{4}\partial^\alpha h^{\mu\nu} \partial_\alpha h_{\mu\nu} 
 -\frac{1}{2}\partial^\mu h^{\mu\nu} \partial^\alpha h_{\alpha\nu} 
 -\frac{1}{4}B\ \partial_\nu h^\beta_\beta \partial^\nu h^\alpha_\alpha 
 -\frac{1}{2}A\ \partial_\alpha h^{\alpha\beta} \partial_\beta h^\nu_\nu \nonumber\\
 && -\frac{1}{4} M_2^2 h^{\mu\nu} h_{\mu\nu} 
+ \frac{1}{16} M_2^2 h^\mu_\mu h^\nu_\nu, \label{20.2a}\\
 {\cal L}_{GF} &=& {\cal M}\partial_\mu h^{\mu\nu} \eta_\nu
+ {\cal M}^2 h^\mu_\mu \epsilon + \frac{1}{2} b_2 M_2^2 \eta^\mu \eta_\mu, \label{eq:20.2b}\\
	{\cal L}_{int} &=& \kappa\ h^{\mu\nu} \bigl(t_{M,\mu\nu}+t_{g,\mu\nu}\bigr). \label{eq:20.2c} 
\end{eqnarray} \end{subequations}
Here, we have introduced the scaling mass ${\cal M}$. Since we will 
investigate the massless limit $M_2 \rightarrow 0$ it will be convenient to
distinguish this from the 'dynamical" spin-2 mass $M_2$. In the mass term for
the $\eta^\mu$-field we kept $M_2$, but the $\eta^\mu$-mass is distinct from
the spin-2 mass, namely $M_\eta^2= -b_2M_2^2$. 
\footnote{In studying the limits 
$M_2 \rightarrow 0, b_2 \rightarrow \infty$ we keep $b_2M_2^2$ fixed.
\noindent {\bf Later on in this appendix, it will appear that for a proper
description of the tensor-field commutators, we need to put ${\cal M}=M_2$}.
This implies that we use here the same Lagrangian as in \cite{Wag09a,WR09}.
}
The purpose is to derive the spin-2 propagator via the quantization
	of the $h^{\mu\nu}(x)$-field. For that, we need the constraints which follow 
	from the "free field equations". So, in this section, we take $\kappa=0$.\\

\noindent First, we work out the terms in the Euler-Lagrange equation (\ref{eq:20.1})
\begin{equation}
 \partial_\alpha \frac{\partial{\cal L}}{\partial_\alpha(h_{\mu\nu}(x))}
- \frac{\partial{\cal L}}{\partial h_{\mu\nu}(x)} = 0, 
\label{eq:20.3}\end{equation} 
and similarly for the $\eta^\mu(x)$ and $\epsilon(x)$ fields.
This leads to the equations \cite{WR09}
\begin{subequations}
\label{eq:20.4} 
\begin{eqnarray}
 && \left( g^{\mu\alpha}g^{\nu\beta} \Box -2 \partial^\mu\partial^\alpha
	g^{\nu\beta} - B\ g^{\mu\nu}g^{\alpha\beta}\Box 
	-A\ g^{\mu\nu} \partial^\alpha \partial^\beta 
	+ g^{\mu\alpha}g^{\nu\beta}M_2^2 \right.\nonumber\\ 
	&& \left. -C\ g^{\mu\nu}g^{\alpha\beta}M_2^2 \right) h_{\alpha\beta}(x) 
 + {\cal M}\left(\partial^\mu\eta^\nu+\partial^\nu\eta^\mu\right)(x)
	-2{\cal M}^2 g^{\mu\nu}\ \epsilon(x) = 0, \\
&& h^\mu_\mu=0\ \ ,\ \ b_2 M_2^2\ \eta^\nu(x) + 
 {\cal M} \partial_\mu h^{\mu\nu}(x)=0.
\end{eqnarray} 
\end{subequations}
Using $h^\alpha_\alpha=0$ gives
\begin{subequations}
\label{eq:20.5} 
\begin{eqnarray}
 \left(\Box + M_2^2\right)\ h^{\mu\nu}(x) &=& 
 2\partial^\mu(\partial_\alpha h^{\alpha\nu}(x)) 
	+A\ g^{\mu\nu} \partial_\alpha\partial_\beta h^{\alpha\beta}(x) 
 \nonumber\\ && 
 - {\cal M}\left(\partial^\mu\eta^\nu+\partial^\nu\eta^\mu\right)(x)
 +2{\cal M}^2 g^{\mu\nu}\ \epsilon(x), \\
 h^\mu_\mu &=& 0\ \ ,\ \ 
 \partial_\mu h^{\mu\nu}(x)= 
 -\left(\frac{b_2 M_2^2}{\cal M}\right)\ \eta^\nu(x).  
\end{eqnarray} 
\end{subequations}
With the last equation for $\partial_\alpha h^{\alpha\nu}(x)$
gives for $h^{\mu\nu}(x)$:
\begin{eqnarray}
	\left(\Box + M_2^2\right)\ h^{\mu\nu}(x) &=& 
 -{\cal M}\left(1+b_2\frac{M_2^2}{{\cal M}^2}\right)
 \left(\partial^\mu\eta^\nu+\partial^\nu\eta^\mu\right)(x)
 \nonumber\\ && 
	-g^{\mu\nu}\left(A\ b_2\frac{M_2^2}{\cal M}(\partial\cdot\eta(x)) 
 -2{\cal M}^2\epsilon(x)\right).
\label{eq:20.6}\end{eqnarray} 
Using $h^\mu_\mu=0$ in the last equation above again, we obtain
\begin{eqnarray*}
 && 0 = -2{\cal M}\left(1+b_2\frac{M_2^2}{{\cal M}^2}\right) \partial\cdot\eta(x)
-4A b_2\frac{M_2^2}{\cal M} \partial\cdot\eta(x) + 8 {\cal M}^2\ \epsilon(x), 
\end{eqnarray*}
which gives the relation
\begin{equation}
	\partial\cdot\eta(x) = 4{\cal M}
	\left[1+(2A+1)b_2\frac{M_2^2}{{\cal M}^2}\right]^{-1}\ \epsilon(x).
\label{eq:20.7}\end{equation} 
Finally, we now substitute this relation into the field-equation for $h^{\mu\nu}$
and obtain
\begin{eqnarray}
	&& \left(\Box + M_2^2\right)\ h^{\mu\nu}(x) = 
 -{\cal M}\left(1+b_2\frac{M_2^2}{{\cal M}^2}\right)
 \left(\partial^\mu\eta^\nu+\partial^\nu\eta^\mu\right)(x)
 \nonumber\\ && 
	+2{\cal M}^2\left(1+\frac{b_2 M_2^2}{{\cal M}^2}\right)\ 
	\left[1+(2A+1)\frac{b_2M_2^2}{{\cal M}^2}\right]^{-1}\ g^{\mu\nu}\ \epsilon(x).
\label{eq:20.8}\end{eqnarray} 
Next we derive the field-equations for $\eta^\mu(x)$ and $\epsilon(x)$. 
For that purpose, we introduce the abbreviation 
\footnote{In the following we will assume that ${\cal M}=M_2$. 
In the formulas, we occasionally still use ${\cal M}$ and $M_2$.}
\begin{equation}
 b \equiv b_2\ (M_2^2/{\cal M}^2), 
\label{eq:20.9}\end{equation} 
and the equations have the form
\begin{subequations}
\label{eq:20.10}
\begin{eqnarray}
	\left(\Box + M_2^2\right)\ h^{\mu\nu}(x) &=& 
 -{\cal M}(1+b)\left(\partial^\mu\eta^\nu+\partial^\nu\eta^\mu\right)(x)
 \label{eq:20.10a} \nonumber\\ && 
	+2{\cal M}^2(1+b)\bigl[1+(2A+1)\ b\bigr]^{-1}\ g^{\mu\nu}\ \epsilon(x), \\
  h^\mu_\mu(x) &=& 0, \ \ \partial_\mu h^{\mu\nu}(x) = -b {\cal M} \eta^\nu,\ \ 
	\partial\cdot\eta = 4{\cal M}\bigl[1+(2A+1)\ b\bigr]^{-1}\ \epsilon(x).
	\label{eq:20.10b}
\end{eqnarray} 
\end{subequations}
For the derivation of $\eta^\nu$ we start with 
$\partial_\mu h^{\mu\nu}= -b{\cal M} \eta^\nu$ which gives the relation
\begin{equation}
 \partial_\mu (\Box + M_2^2) h^{\mu\nu}= -b{\cal M}(\Box+M_2^2)\eta^\nu
\label{eq:20.11}\end{equation} 
Using the field-equation for $h^{\mu\nu}$ on the l.h.s. we get
\begin{eqnarray}
 \partial_\mu (\Box + M_2^2) h^{\mu\nu}&=& -{\cal M}(1+b)\left(\Box \eta^\nu +
	\partial^\nu \partial\cdot\eta\right)+2{\cal M}^2\ (1+b)
	\bigl[1+(2A+1)b\bigr]^{-1}\partial^\nu\epsilon(x)
 \nonumber\\ &=& 
 -{\cal M}(1+b)\Box \eta^\nu 
	-2{\cal M}^2(1+b)\left[1+(2A+1)b\right]^{-1}
	\partial^\nu\epsilon(x)
\label{eq:20.12}\end{eqnarray} 
The combination of (\ref{eq:20.11}) and (\ref{eq:20.12}) gives the $\eta^\nu$-equation:
\begin{equation}
\left(\Box + M_\eta^2\right) \eta^\nu(x) = 
	-2{\cal M}\frac{1+b}{1+(2A+1)b} \partial^\nu\epsilon(x)\ 
 \ {\rm with}\ M_\eta^2 = - bM_2^2.
\label{eq:20.13}\end{equation} 
The equation for $\epsilon(x)$ is obtained by differentiation of (\ref{eq:20.13})
and using (\ref{eq:20.7}) which leads to
\begin{equation}
	\left(\Box + M_\epsilon^2\right)\epsilon(x) = 0\ \ {\rm with}\ \ 
 M_\epsilon^2 = -\frac{2b}{3+b} M_2^2.
\label{eq:20.16}\end{equation} 
\noindent \underline{Decoupling vector and scalar field:} Making the transformation 
$\eta^\mu \rightarrow \rho^\mu+\partial^\mu\Lambda$ with
\begin{eqnarray}
	\Lambda(x) &=& -2\frac{{\cal M}}{M_2^2}
	\frac{(3+b)}{b\left[1+(2A+1)b\right]} \epsilon(x),
\label{eq:20.17}\end{eqnarray}
we arrive at the equations
\begin{subequations} \label{eq:20.18}  
\begin{eqnarray}
	&&	\left(\Box -bM_2^2\right) \rho(x)=0,\ \partial\cdot\rho(x)=0, \\
	&&\bigl(\Box + M_2^2\bigr)\ h^{\mu\nu}(x) = 
 -{\cal M}(1+b)\left(\partial^\mu\rho^\nu+\partial^\nu\rho^\mu\right)(x)
 \nonumber\\ &&
 +2{\cal M}^2(1+b)[1+(2A+1)b]^{-1}\ \left[g^{\mu\nu} -2\frac{3+b}{b}\frac{\partial^\mu\partial^\nu}{M_2^2}\right]
	\epsilon(x).   
\end{eqnarray}\end{subequations}

\begin{center}
\fbox{\rule[-5mm]{0cm}{1.0cm} \hspace{5mm}
\begin{minipage}[]{12.9cm}
Summary of the field equations:
\begin{subequations} \label{eq:20.19} 
\begin{eqnarray}
&&\bigl(\Box + M_2^2\bigr)\ h^{\mu\nu}(x) = 
 -{\cal M}(1+b)\left(\partial^\mu\rho^\nu+\partial^\nu\rho^\mu\right)(x)
 \nonumber\\ &&
 +2{\cal M}^2(1+b)[1+(2A+1)b]^{-1}\ \left[g^{\mu\nu} 
	-2\frac{3+b}{b}\frac{\partial^\mu\partial^\nu}{M_2^2}\right]
\epsilon(x), \\
&&	\left(\Box -bM_2^2\right) \rho(x)=0, \\
&&  \left(\Box - \frac{2b}{3+b} M_2^2\right) \epsilon(x) = 0,                 
\end{eqnarray}\end{subequations}
and the constraints:
\begin{subequations} \label{eq:20.20} 
\begin{eqnarray}
  h^\mu_\mu(x) &=& 0, \ \ \partial\cdot\rho(x) = 0, \\
  \partial_\mu h^{\mu\nu}(x) &=& -b {\cal M}\ \left(\rho^\nu
	+\frac{2{\cal M}}{M_2^2}\frac{3+b}{b(1-b)}\partial^\nu\epsilon\right)(x).    
\end{eqnarray}\end{subequations}
	{\it	The vector and scalar auxiliary fields $\rho^\mu(x)$ and $\epsilon(x)$
	are free (ghost) fields with an imaginary mass.}
\end{minipage} \hspace{5mm} }\\
\end{center}

\noindent From equations (\ref{eq:20.17}) and (\ref{eq:20.18}) one easily derives 
for the "free" $h^{\mu\nu}$-field, that
\begin{subequations} \label{eq:20.21} 
\begin{eqnarray}
&& (\Box+M_\rho^2) \rho^\mu = 0, \label{eq:20.19a}\\
&& (\Box+M_\epsilon^2)(\Box+M_\rho^2)(\Box+M_2^2) h^{\mu\nu} = 0. \label{eq:20.19b}
\end{eqnarray} \end{subequations}

\section{Quantization                                                 
	}
\label{sec:30}   
Dirac's Hamiltonian method is appropriate for the quantization of constrained
systems \cite{Dirac50,WR09}. The canonical momenta are defined as
\begin{equation}
	\pi^{\mu\nu} = \frac{\partial{\cal L}}{\partial\dot{h}_{\mu\nu}}\ ,\
	\pi^{\mu} = \frac{\partial{\cal L}}{\partial\dot{\eta}_\mu}\ ,\
	\pi^{\epsilon} = \frac{\partial{\cal L}}{\partial\dot{\epsilon}}.  
\label{eq:30.21} \end{equation}

\noindent For the general Lagrangian ${\cal L}_{2,\eta\epsilon}$ one obtains
\begin{eqnarray}
	\pi_{\mu\nu}(x) &=& \frac{1}{2}\partial_0 h_{\mu\nu}(x)
	-\delta_{\mu 0}\partial^\alpha h_{\alpha\nu}(x)
	-\frac{1}{2}B\ \partial_0 h(x)\ \eta_{\mu\nu}
	\hspace{7mm} \pi_{\eta, \nu} = 0, \hspace{7mm} \nonumber\\
	\nonumber\\ && 
	-\frac{1}{2}A \left(\delta_{\mu 0} \partial_\nu h(x)
	+ \partial_\alpha h^{\alpha 0}(x)\ \eta_{\mu\nu}\right)
	+{\cal M} \delta_{\mu 0} \eta_\nu, 
	\hspace{4mm} \pi_{\epsilon} = 0.                          
\label{eq:30.20} \end{eqnarray}
Explicitly, these momenta are
\footnote{\it Note that for A=-1, B=C=1, $\pi_2^{00}$ etc. agree with \cite{WR09}. 
	The $\pi_2^{0m}$ leads to a constraint because of $\partial_\mu h^{\mu\nu}|f\rangle =0$.}
\begin{subequations}\label{eq:30.23}
\begin{eqnarray}
	\pi^{00}_2(x) &=& -\frac{1}{2}(1+2A+B)\ \dot{h}^{00}(x)
	-(1+\frac{1}{2}A) \partial_n h^{n0}(x) -\frac{1}{2}(A+B) \dot{h}^n_n(x), 
	\label{eq:20.23a} \\ 
	\pi^{0m}_2(x) &=& +\frac{1}{2} \dot{h}^{0m}(x) -\partial_nh^{nm}(x)
	-\frac{1}{2}A\ \partial^m \left(h^{00}(x)+h^n_n(x)\right), \label{eq:20.23b} \\
	\pi^{nm}_2(x) &=& \frac{1}{2}\dot{h}^{nm}(x)
	-\frac{1}{2}B \left(\dot{h}^{00}(x)+\dot{h}^k_k(x)\right)\ \eta^{nm}
	-\frac{1}{2} A\left(\partial_kh^{k 0}(x) +\dot{h}^{00}(x)\right) \eta^{mn}, 
	\label{eq:20.23c} \\
	\pi^{\ k}_{2k}(x) &=& +\frac{1}{2}(1-3B)\ \dot{h}^k_k(x)
	-\frac{3}{2}\left(A+B\right) \dot{h}^{00}(x)
		-\frac{3}{2} A\ \partial_kh^{k0}(x). \label{eq:20.23d}   
\end{eqnarray}\end{subequations}
	From the constraint $\partial_\mu h^{\mu\nu}=0 ({\cal M}=0)$ it follows that
	$\dot{h}^{0m} = -\partial_n h^{0n}$ and this gives 
	\begin{eqnarray*}
		\pi^{0m}_2(x) &=& -\frac{1}{2} \partial_k h^{0k}(x) -\partial_nh^{nm}(x)
	-\frac{1}{2}A\ \partial^m \left(h^{00}(x)+h^n_n(x)\right), \\
	\end{eqnarray*}
	{\it i.e.} a constraint and not an independent canonical-momentum variable. 
	\footnote{
	{\bf Remark:} In the Dirac method of quantization, the 
	${\cal E.L.}$ equations are always valid and can be used, for example for
	the time dependences etc, for example, leading to new constraints \cite{Dirac50}.}

\noindent With $B=(3A^2+2A+1)/2$ the momenta in (\ref{eq:30.23}) become
\begin{subequations}\label{eq:32.23}
\begin{eqnarray}
	\pi^{00}_2(x) &=& -\frac{3}{4}(A+1)^2\ \dot{h}^{00}(x)
	-\frac{1}{2}(A+2) \partial_n h^{n0}(x) -\frac{1}{4}(3A+1)(A+1) \dot{h}^n_n(x), 
	\label{eq:32.23a} \\ 
	\pi^{0m}_2(x) &=& +\frac{1}{2} \dot{h}^{0m}(x) -\partial_nh^{nm}(x)
	-\frac{1}{2}A\ \partial^m \left(h^{00}(x)+h^n_n(x)\right), \label{eq:32.23b} \\
	\pi^{nm}_2(x) &=& \frac{1}{2}\dot{h}^{nm}(x)
	-\frac{1}{4}(3A+1)(A+1)\ \dot{h}^{00}(x)\eta^{nm} 
	-\frac{1}{4}(3A^2+2A+1)\ \dot{h}^k_k(x)\ \eta^{nm} \nonumber\\ && 
	-\frac{1}{2} A\ \partial_kh^{k 0}(x)\  \eta^{mn}, 
	\label{eq:32.23c} \\
	\pi^{\ k}_{2k}(x) &=& -\frac{1}{4}(3A+1)^2\ \dot{h}^k_k(x)
	-\frac{3}{4} (3A+1)(A+1)\ \dot{h}^{00}(x)
		-\frac{3}{2} A\ \partial_kh^{k0}(x). \label{eq:32.23d}   
\end{eqnarray}\end{subequations}
Trying to solve for $\dot{h}^{00}$ and $\dot{h}^n_n$ leads to the equation
\begin{eqnarray*}
\left(\begin{array}{c} \pi_2^{00}+\frac{1}{2}(A+2)\partial_nh^{n0} \\
\pi^{\ k}_{2 k}+\frac{3}{2}A\partial_k h^{k0} \end{array}\right) =
\left(\begin{array}{cc} 
-\frac{3}{4}(A+1)^2 & -\frac{1}{4}(3A+1)(A+1) \\
-\frac{3}{4}(3A+1)(A+1) & -\frac{1}{4}(3A+1)^2 \end{array}\right)\
\left(\begin{array}{c} \dot{h}^{00} \\ \dot{h}^k_k \end{array}\right)
\end{eqnarray*}
Obviously, the determinant is zero. There are two solutions: A=-1 and A=-1/3.
	The case A=-1 leads to the velocity
	$\dot{h}^k_k = -\pi^{\ k}_{2 k}+3 \partial_k h^{k0}/2$ and the constraint
	$\theta^{00} = \pi^{00}+\partial_nh^{n0}/2=0$.\\
	The case A=-1/3 leads to the velocity
	$\dot{h}^{00}= -3\pi_2^{00}-5\partial_nh^{n0}/2$ and the constraint
	$\theta^{\ k}_{2 k} = \pi^{\ k}_{2 k}-\partial_k h^{k0}=0$. 
	For gravity, this solution is unphysical. Therefore, henceforth we will only consider the case A=-1,
	which is treated in Ref.~\cite{WR09}.

\noindent The velocities are
\begin{subequations}\label{eq:30.23b}
\begin{eqnarray}
	\dot{h}^{nm} &=& 2\pi_2^{nm} -\eta^{nm} \pi^{\ k}_{2 k} 
	+\frac{1}{2}\eta^{nm} \partial_kh^{k0}, \\
 \dot{h}^{\ k}_{2 k} &=& -\pi^{\ k}_{2 k} +\frac{3}{2}\partial_k h^{k0}, 
\end{eqnarray}\end{subequations}
	and the {\it primary} constraints
\begin{eqnarray}
	\theta_2^{00} &=& \pi_2^{00} +\frac{1}{2}\partial_nh^{n0}-M_2 \eta^0\ , 
	\hspace{0.5cm} \theta^0_\eta = \pi_\eta^0\ , \nonumber\\
	\theta_2^{0m} &=& \pi_2^{0m} +\partial_n h^{nm} -\frac{1}{2}\partial^m h^{00}\ ,  
	\hspace{2mm} \theta_\eta^m= \pi_\eta^m\ , \nonumber\\ &&
	-\frac{1}{2} \partial^m h^n_n -M_2 \eta^m\ , 
	\hspace{1.2cm} \theta_\epsilon = \pi_\epsilon\ . 
\label{eq:30.27}\end{eqnarray}
which vanish in the {\it weak} sense \cite{Dirac50,Sudarshan74}.
\subsection{Dirac-theory: The Hamiltonian and Constraints
\protect\footnote{The material in this section is taken from Ref.'s~\cite{WR09,Wag09a}
and is included here for completeness.}
	}                         
\label{sec:1a}   
The Hamiltonian is $H_{2,\eta\epsilon} = \int d^3x\ {\cal H}_{2,\eta\epsilon}$ with \cite{WR09}
\begin{eqnarray}
	{\cal H}_{2,\eta\epsilon} &=& 
	\pi_2^{nm} \pi_{2,nm}-\frac{1}{2}\pi^{\ n}_{2n} \pi^{\ m}_{2m} 
        +\frac{1}{2}\pi^{\ n}_{2n}\partial^m h_{m0}
	-\frac{1}{2}\partial^k h^{n0}\partial_k h_{n0}
	\nonumber\\ &&
	-\frac{1}{4}\partial^k h^{nm}\partial_k h_{nm}
	+\frac{1}{8}\partial_n h^{n0}\partial^m h_{m0}
	+\frac{1}{2}\partial_n h^{nm}\partial^k h_{km}
	\nonumber\\ &&
	+\frac{1}{2}\partial_m h^{00}\partial^m h^n_n 
	+\frac{1}{4}\partial_m h^n_n \partial^m h^k_k 
	-\frac{1}{2}\partial_n h^{nm}\partial_m h_{00}
	-\frac{1}{2}\partial_n h^{nm}\partial_m h^k_k 
	\nonumber\\ &&
	+\frac{1}{2} M_2^2 h^{n0}h_{n0}+\frac{1}{4}M_2^2 h^{nm}h_{nm}
	-\frac{1}{2}M_2^2 h^0_0 h^m_m -\frac{1}{4} M_2^2 h^n_n h^m_m
	\nonumber\\ &&
	-\frac{1}{2}cM_2^2\eta^\mu\eta_\mu-M_2\partial_nh^{n0}-M_2\partial_nh^{nm}\eta_m
	-M_2^2 h^0_0\epsilon-M_2^2h^k_k\epsilon \nonumber\\ &&
	+\lambda_{2,00} \theta_2^{00} + \lambda_{2,0m} \theta_2^{0m}  
	+\lambda_{0,\eta} \theta^0_\eta 
	+\lambda_{m,\eta}\theta^m_\eta+\lambda_\epsilon \theta_\epsilon\ .
\label{eq:30.28}\end{eqnarray}

\noindent In Ref.~\cite{WR09,Wag09a}, the primary, secondary, and tertiary constraints are
worked out, giving the Lagrange parameters $\lambda$ in (\ref{eq:30.28}). Furthermore, here
also the Dirac-brackets (DB's) are derived, enabling the quantization.

\noindent The ETC relations are obtained by considering the fields as operators, 
replacing the Db's by commutators, and adding a factor i. The result is
\begin{eqnarray}
	\left[h^{00}(x), h^{0l}(y)\right]_0 &=& \frac{4i}{3M_2^4}
\partial^j\partial_j\partial^l \delta^3(x-y), \nonumber\\
\left[h^{0m}(x), h^{kl}(y)\right]_0 &=& \frac{-i}{M_2^2} \left[
	\frac{4}{3M_2^2}\partial^m\partial^k\partial^l -\frac{2}{3}\partial^m g^{kl}
	+\partial^k g^{ml}+\partial^l g^{mk}\right]\ \delta^3(x-y), 
\nonumber\\ 
	\left[\dot{h}^{00}(x), h^{00}(y)\right]_0 &=& -\frac{4i}{3M_2^4}
\partial^i\partial_i\partial^j\partial_j \delta^3(x-y), \nonumber\\
	\left[\dot{h}^{0m}(x), h^{0l}(y)\right]_0 &=& \frac{i}{M_2^2}
	\left[\frac{4}{3M_2^2}\partial^m\partial^l\partial^j\partial_j
	+\frac{1}{3}\partial^m\partial^l+\partial^j\partial_j g^{ml}\right]\
	\delta^3(x-y), \nonumber\\ 
	\left[\dot{h}^{00}(x), h^{kl}(y)\right]_0 &=& \frac{i}{M_2^2}
	\left[\frac{4}{3M_2^2}\partial^k\partial^l\partial^j\partial_j
	+2\partial^k\partial^l-\frac{2}{3}\partial^j\partial_j g^{kl}\right]\
	\delta^3(x-y), \nonumber\\ 
 \left[\dot{h}^{nm}(x), h^{kl}(y)\right]_0 &=& i \left[
 		-g^{nk}g^{ml}-g^{nl}g^{mk}+\frac{2}{3}g^{nm}g^{kl} \right.\nonumber\\
 		&& \left. -\frac{1}{M_2^2}\left( \partial^n\partial^k g^{ml}+
 		\partial^m\partial^k g^{nl}+ \partial^n\partial^l g^{mk}+
 		\partial^m\partial^l g^{mk}\right) \right.\nonumber\\ && \left.
 		+\frac{2}{3M_2^2}\left(
 		\partial^n\partial^m g^{kl}+ g^{nm} \partial^k\partial^l\right)
 		-\frac{4}{3M_2^4} \partial^n \partial^m \partial^k \partial^l \right]
 		\ \delta^3(x-y).              
\label{eq:30.40}\end{eqnarray}
The ETC containing the auxiliary fields $\eta^\mu(x)$ and $\epsilon(x)$ are
\begin{eqnarray}
\left[h^{00}(x), \eta^0(y)\right]_0 &=& \frac{3i}{M_2(3+b)}\ \delta^3(x-y), \nonumber\\
\left[h^{0n}(x), \eta^m(y)\right]_0 &=& \frac{i}{M_2}\ g^{nm}\ \delta^3(x-y), \nonumber\\
\left[h^{0n}(x), \epsilon(y)\right]_0 &=& -\frac{i}{M_2^2}\ 
\left(\frac{1-b}{3+b}\right) \partial^n \delta^3(x-y), \nonumber\\
\left[h^{mn}(x), \eta^0(y)\right]_0 &=& -\frac{i}{M_2(3+b)}\ g^{nm}\ \delta^3(x-y), \nonumber\\
\left[\eta^0(x), \eta^m(y)\right]_0 &=& \frac{i}{M_2(3+b)}\ \partial^m \delta^3(x-y), \nonumber\\
\left[\eta^0(x), \epsilon(y)\right]_0 &=& \frac{3i}{2M_2}\ \frac{(1-b)}{(3+b)^2}\delta^3(x-y).            
\label{eq:30.41}\end{eqnarray}
Here, as mentioned in \cite{WR09}, not shown are the ETC's among time derivatives of 
the fields in (\ref{eq:30.41}), which are of importance for the calculation of the 
non-equal times commutation (NETC) relations below. 
These follow from the constraints $\partial_\mu h^{\mu\nu}(x) = -bM_2 \eta^\nu(x),\ 
\partial\cdot\eta(x)= 4M_2(1-b)^{-1}\epsilon(x)$ in (\ref{eq:20.10}) for A=-1,
which solves the time derivatives 
\begin{eqnarray}
	\dot{h}^{00} &=& \partial_n h^{0n}-bM_2\eta^0\ ,\ 
	\dot{h}^{0n} = \partial_k h^{kn}-bM_2\eta^0, \nonumber\\ 
	\dot{\eta}^0 &=& \partial_k\eta^k +4M_2(1-b)^{-1}\ \epsilon. 
\label{eq:30.42}\end{eqnarray}
This  enables the derivation of the ETC's (and NETC's) for these time derivatives in terms of those
in (\ref{eq:30.40}) and (\ref{eq:30.41}) immediately.

In \ref{app:SIG}, the quantization of imaginary-ghost $\epsilon(x)$-field is worked out in detail.                   
 
\section{Field-Commutators and Spin-2 Field Propagator}                       
\label{sec:40}   
	The solutions of the homogeneous equations in (\ref{eq:20.19}) and (\ref{eq:20.20})
	satify the identities, see \cite{WR09,NO90},
\begin{eqnarray}
	h^{\mu\nu}(x) &=& \int d^3z\left[\partial^z_0 \Delta(x-z; M_2^2)\cdot h^{\mu\nu}(z)
	-\Delta(x-z; M_2^2)\ \partial^z_0 h^{\mu\nu}(z)\right] 
	+\frac{1}{M_2^2-M_\rho^2}
	\cdot\nonumber\\ && \times \int d^3z\biggl[\partial_0^z
	\biggl(\Delta(x-z;M_\rho^2)-\Delta(x-z; M_2^2)\biggr) 
	-\biggl(\Delta(x-z;M_\rho^2)-\Delta(x-z; M_2^2)\biggr)\partial_0^z\biggr] 
 	\cdot\nonumber\\ && \times
	(\Box+M_2^2)\ h^{\mu\nu}(z)  
	+\frac{1}{(M_\rho^2-M_\epsilon^2)(M_2^2-M_\rho^2)(M_2^2-M_\epsilon^2)}
	\cdot\nonumber\\ && \times \int d^3z\biggl[\partial_0^z 
	\biggl((M_2^2-M_\rho^2)\Delta(x-z;M_\epsilon^2)
		-(M_2^2-M_\epsilon^2)\Delta(x-z;M_\rho^2)
		+(M_\rho^2-M_\epsilon^2)\Delta(x-z;M_2^2)\biggr) \nonumber\\ &&
-\biggl( (M_2^2-M_\rho^2)\Delta(x-z;M_\epsilon^2)
		-(M_2^2-M_\epsilon^2)\Delta(x-z;M_\rho^2)
		+(M_\rho^2-M_\epsilon^2))\Delta(x-z;M_2^2)\biggr)\partial_0^z \biggr]
		\cdot\nonumber\\ && \times 
		(\Box+M_\rho^2)(\Box+M_2^2)\ h^{\mu\nu}(z).
\label{eq:40.1}\end{eqnarray}
and 
\begin{eqnarray}
	\rho^\mu(x) &=& \int d^3z\left[\partial^z_0 \Delta(x-z; M_\rho^2)\ \rho^{\mu}(z)
	-\Delta(x-z; M_\rho^2)\ \partial^z_0 \rho^{\mu}(z)\right] \\ 
	\epsilon(x) &=& \int d^3z\left[\partial^z_0 \Delta(x-z; M_\epsilon^2)\ \epsilon(z)
	-\Delta(x-z; M_\epsilon^2)\ \partial^z_0 \epsilon(z)\right] 
\label{eq:40.2}\end{eqnarray}
Using this identity the commutation relation can be calculated from the ETC in
Eqn's~(\ref{eq:30.40}-(\ref{eq:30.42}) with the
result \cite{WR09}
\begin{eqnarray}
\left[h^{\mu\nu}(x),h^{\alpha\beta}(y)\right] &=& \biggl\{
\left(\eta^{\mu\alpha}\eta^{\nu\beta}
 +\eta^{\mu\beta}\eta^{\nu\alpha}
-\frac{2}{3}\eta^{\mu\nu}\eta^{\alpha\beta}\right) \nonumber\\ && \hspace{0cm}
+\frac{1}{M_2^2}\left(
\partial^\mu\partial^\alpha \eta^{\nu\beta}
 +\partial^\nu\partial^\alpha \eta^{\mu\beta}
 +\partial^\mu\partial^\beta  \eta^{\nu\alpha}
 +\partial^\nu\partial^\beta  \eta^{\mu\alpha}\right) \nonumber\\ && \hspace{0cm}
 -\frac{2}{3M_2^2}\left(
 \partial^\mu\partial^\nu\eta^{\alpha\beta}+\eta^{\mu\nu}\partial^\alpha\partial^\beta\right)
 + \frac{4}{3M_2^2}\partial^\mu\partial^\nu\partial^\alpha\partial^\beta\biggr\}\ 
   i\Delta(x-y;M_2^2) 
\nonumber\\ && 
 -\frac{1}{M_2^2}\biggl\{ \partial^\mu\partial^\alpha\eta^{\nu\beta}
+\partial^\nu\partial^\alpha\eta^{\mu\beta} +\partial^\mu\partial^\beta \eta^{\nu\alpha}
+\partial^\nu\partial^\beta \eta^{\mu\alpha}
\nonumber\\ && 
 + \frac{4}{M_\eta^2}\partial^\mu\partial^\nu\partial^\alpha\partial^\beta\biggr\}\ 
   i\Delta(x-y;M_\rho^2) 
\nonumber\\ && 
	-\biggl\{\frac{1}{3}\frac{b}{3+b} \eta^{\mu\nu}\eta^{\alpha\beta}-\frac{2}{3M_2^2}
	\left(\partial^\mu\partial^\nu\eta^{\alpha\beta}
	+\eta^{\mu\nu}\partial^{\alpha}\partial^\beta\right)
\nonumber\\ && 
 +\frac{4(3+b)}{3bM_2^4}\partial^\mu\partial^\nu\partial^\alpha\partial^\beta\biggr\}
 i\Delta(x-y;M_\epsilon^2) \nonumber\\ &&
	= 2P_2^{\mu\nu\alpha\beta}(\partial)\ i\Delta(x-y;M_2^2) + 
	  P_\rho^{\mu\nu\alpha\beta}(\partial)\ i\Delta(x-y;M_\rho^2) 
	  \nonumber\\ &&
	 +P_\epsilon^{\mu\nu\alpha\beta}(\partial)\ i\Delta(x-y;M_\epsilon^2). 
\label{eq:40.3}\end{eqnarray}
where $P_2(\partial)$ etc are the (on mass-shell) spin projection operators, see
{\it e.g.} \cite{Car71}. 
Defining $\bar{\eta}^{\mu\nu}= \eta^{\mu\nu}+\partial^\mu\partial6\nu/M_2^2,
\hat{\eta}^{\mu\nu}= \eta^{\mu\nu}+\partial^\mu\partial^\nu/M_\rho^2$, and
$\widetilde{\eta}^{\mu\nu}= \eta^{\mu\nu}+\partial^\mu\partial^\nu/M_\epsilon^2$ the  
commutator takes on the compact form
\begin{eqnarray}
\left[h^{\mu\nu}(x),h^{\alpha\beta}(y)\right] &=& \biggl\{
\left(\bar{\eta}^{\mu\alpha}\bar{\eta}^{\nu\beta}
+\bar{\eta}^{\mu\beta}\bar{\eta}^{\nu\alpha}
-\frac{2}{3}\bar{\eta}^{\mu\nu}\eta^{\alpha\beta}\right) \nonumber\\ && \hspace{0cm}
+\frac{b}{M_\rho^2}\biggl[
\partial^\mu\partial^\alpha\hat{\eta}^{\nu\beta}+
\partial^\nu\partial^\alpha\hat{\eta}^{\mu\beta}+
\partial^\mu\partial^\beta \hat{\eta}^{\nu\alpha}+
\partial^\nu\partial^\beta \hat{\eta}^{\mu\alpha}\biggr] i\Delta(x-y;M_\rho^2)
\nonumber\\ && -\frac{b}{3+b}\biggl\{
\frac{1}{3}\widetilde{\eta}^{\mu\nu}\widetilde{\eta}^{\alpha\beta}
+\frac{1}{M_\epsilon^2}\left(
\partial^\mu\partial^\nu \widetilde{\eta}^{\alpha\beta}+\widetilde{\eta}^{\mu\nu}
\partial^\alpha\partial^\beta\right)
+\frac{3}{M_\epsilon^4}\partial^\mu\partial^\nu\partial^\alpha\partial^\beta\biggr\} 
i\Delta(x-y;M_\epsilon^2)
	\nonumber\\ &=& 
	 2P_2^{\mu\nu\alpha\beta}(\partial)\ i\Delta(x-y;M_2^2) + 
	  P_\rho^{\mu\nu\alpha\beta}(\partial)\ i\Delta(x-y;M_\rho^2) 
	  \nonumber\\ &&
	 +P_\epsilon^{\mu\nu\alpha\beta}(\partial)\ i\Delta(x-y;M_\epsilon^2). 
\label{eq:40.3c}\end{eqnarray}
and the Feynman spin-2 propagator is
\begin{eqnarray}
	D_F^{\mu\nu\alpha\beta}(x-y) &=& 
	 2P_2^{\mu\nu\alpha\beta}(\partial)\ \Delta_F(x-y;M_2^2) + 
	  P_\rho^{\mu\nu\alpha\beta}(\partial)\ \Delta_F(x-y;M_\rho^2) 
	  \nonumber\\ &&
	 +P_\epsilon^{\mu\nu\alpha\beta}(\partial)\ \Delta_F(x-y;M_\epsilon^2). 
\label{eq:40.3cc}\end{eqnarray}
The masses $M_\rho=M_\eta$ and $M_\epsilon$ are given in (\ref{eq:20.13}) and (\ref{eq:20.16}).
In taking the limit $b \rightarrow \infty$ we use the form (\ref{eq:40.3c}), which is
important when studying the spectral representation later on. 
The spin-2 Feynman propagator, in the absence of the auxiliary fields, becomes \cite{WR09}
	\begin{eqnarray}
		D_F^{\mu\nu\alpha\beta}(x-y) &=& -i\langle 0|T\left[h^{\mu\nu}(x)
		h^{\alpha\beta}(y)\right]|0 \rangle \nonumber\\ &=&
		-i\theta(x^0-y^0)\ 2P_2^{\mu\nu\alpha\beta}(\partial)\Delta^{(+)}(x-y;M_2^2)
		\nonumber\\ &&
		-i\theta(y^0-x^0)\ 2P_2^{\mu\nu\alpha\beta}(\partial)\Delta^{(-)}(x-y;M_2^2)
		+ \ldots
		\nonumber\\ &=& 2P_2^{\mu\nu\alpha\beta}(\partial)\ \Delta_F(x-y;M_2^2)
		+ (\rm{ non-covariant\ local\ terms}) +\ldots
\label{eq:40.4}\end{eqnarray}

In the following, we often denote the mass by $M_2 \equiv M$.
\noindent For the normalization of our solutions, the commutation relations of the
field operators are important. Using the Dirac quantization method, and using
a vector and a scalar auxiliary field, the obtained field commutators read \cite{Wag09a,WR09}
\footnote{In contrast to \cite{Wag09a} we use here for the Minkowski-metric
the notation $\eta^{\mu\nu}$. There should be no confusion with the
 auxiliary vector-field $\eta^\mu(x)$.}
\begin{subequations}
\label{eq:40.6} 
\begin{eqnarray}
 \left[\epsilon(x),\epsilon(y)\right] &=& 
 -\frac{3}{4}\frac{b(1-b)^2}{(3+b)^3}\left(\frac{M^2}{{\cal M}^2}\right)^2\
 i\Delta(x-y; M_\epsilon^2), \label{40.6a}\\
 \left[\eta^\mu(x),\epsilon(y)\right] &=& 
 -\frac{3}{2}\frac{(1-b)}{(3+b)^2}\frac{M^2}{{\cal M}^2} 
 \frac{\partial^\mu}{\cal M}\ i\Delta(x-y; M_\epsilon^2), \label{40.6b}\\
 \left[\eta^\mu(x),\eta^\nu(y)\right] &=& 
 \left[\eta^{\mu\nu}-\frac{\partial^\mu\partial^\nu}{bM^2}\right]
 i\Delta(x-y; M_\eta^2)\ \nonumber\\ && +
 \frac{3}{b(3+b)}\frac{\partial^\mu\partial^\nu}{{\cal M}^2}\
 i\Delta(x-y; M_\epsilon^2), \label{40.6c}\\
 \left[\epsilon(x), h^{\mu\nu}(y)\right] &=& \frac{(1-b)}{(3+b)}
 \left[\frac{\partial^\mu\partial^\nu}{M^2}-\frac{1}{2}\frac{b}{(3+b)}
 \eta^{\mu\nu}\right]\ i\Delta(x-y;M_\epsilon^2), \label{40.6d}\\
\left[\eta^\alpha(x),h^{\mu\nu}(y)\right] &=& 
\frac{1}{M}\left[\partial^\mu\eta^{\alpha\nu}+\partial^\nu\eta^{\alpha\mu}
+\frac{2}{M_\eta^2} \partial^\alpha\partial^\mu\partial^\nu\right]\ i\Delta(x-y;M_\eta^2)
\nonumber\\ &&  
-\frac{1}{M}\left[\frac{1}{(3+b)}\partial^\alpha\eta^{\mu\nu}
+\frac{2}{M_\eta^2}\partial^\alpha\partial^\mu\partial^\nu\right]\ i\Delta(x-y;M_\epsilon^2),
\label{40.6e}\\
 \left[h^{\mu\nu}(x),h^{\alpha\beta}(y)\right] &=& 
 \left[\left(\eta^{\mu\alpha}\eta^{\nu\beta}+\eta^{\mu\beta}\eta^{\nu\alpha}\right)
 -\frac{2}{3}\eta^{\mu\nu}\eta^{\alpha\beta} + \ldots\right]
 i\Delta(x-y; M^2) \nonumber\\ && 
 -\left[\frac{1}{3}\frac{b}{3+b} \eta^{\mu\nu}\eta^{\alpha\beta} + \ldots \right]
 i\Delta(x-y; M_\epsilon^2). \label{40.6f} 
\end{eqnarray} 
\end{subequations}
The ellipses in the square brackets above denote terms with $\partial^\mu, ...., \partial^\beta$. 
These are taken care of in section~\ref{sec:50b} and 
are unimportant due to $\partial_\mu t_M^{\mu\nu}(x)=0$ and $(t_M)^\mu_\nu(x)=0$. 
The masses $M_\eta$ and $M_\epsilon$ are given in \cite{WR09} in terms of
$M_2$ and the $b$-parameter                              
\begin{equation}
 M_\eta^2 = -b M^2\ \ ,\ \ M_\epsilon^2 = -\frac{2b}{3+b} M^2\ .
\label{eq:40.7}\end{equation} 

\noindent 
The field commutators for the $h^{\mu\nu}(x), \rho^\mu(x)$, and $\epsilon(x)$ fields are
readily derived. Starting from Eqs.~(\ref{eq:40.1}) one obtains 
\begin{subequations}
\label{eq:40.8} 
\begin{eqnarray}
 \left[\epsilon(x),\epsilon(y)\right] &=& -\frac{3}{4}\frac{b(1-b)^2}{(3+b)^3}
 \left(\frac{M^2}{{\cal M}^2}\right)^2\ i\Delta(x-y; M_\epsilon^2), \label{40.8a}\\
 \left[\rho^\mu(x),\epsilon(y)\right] &=& 
 0, \label{40.8b}\\
 \left[\rho^\mu(x),\rho^\nu(y)\right] &=& 
 \left[\eta^{\mu\nu}-\frac{\partial^\mu\partial^\nu}{bM^2}\right]\
 i\Delta(x-y; M_\rho^2), \label{40.8c}\\                   
 \left[\rho^\alpha(x), h^{\mu\nu}(y)\right] &=& \frac{1}{M}          
	\left[\left(\partial^\mu\eta^{\alpha\nu}+\partial^\nu\eta^{\alpha\mu}\right)
 -\frac{2}{bM^2}\partial^\alpha\partial^\mu\partial^\nu\right]\
	i\Delta(x-y;M_\rho^2), \label{40.8d}\\
 \left[\epsilon(x), h^{\mu\nu}(y)\right] &=& \frac{(1-b)}{(3+b)}
 \left[\frac{\partial^\mu\partial^\nu}{M^2}-\frac{1}{2}\frac{b}{(3+b)}
 \eta^{\mu\nu}\right]\ i\Delta(x-y;M_\epsilon^2), \label{40.8e}\\
 \left[h^{\mu\nu}(x),h^{\alpha\beta}(y)\right] &=& 
 \left[\left(\eta^{\mu\alpha}\eta^{\nu\beta}+\eta^{\mu\beta}\eta^{\nu\alpha}\right)
 -\frac{2}{3}\eta^{\mu\nu}\eta^{\alpha\beta} + \ldots\right]
 i\Delta(x-y; M^2), \nonumber\\ && 
 -\left[\frac{1}{3}\frac{b}{3+b} \eta^{\mu\nu}\eta^{\alpha\beta} + \ldots \right]
 i\Delta(x-y; M_\epsilon^2), \label{40.8f} 
\end{eqnarray} 
\end{subequations}
where $M_\rho = M_\eta$, and in the commutator of $\rho^\alpha$ and $h^{\mu\nu}$ 
we have set ${\cal M}=M$, anticipating with what will be done later.
Again, the ellipses in the
square brackets above denote terms with $\partial^\mu, ...., \partial^\beta$, 
do not contribute and are dealt with in section~\ref{sec:50b}.                                                  
\footnote{Lateron we will consider the double limit $b \rightarrow \infty$ and 
$M \rightarrow 0$. In doing so, we keep $M_\rho^2=M_\eta^2= -b M^2$ finite.
}

\noindent It is important to note that upon quantization, the sign in $[\rho^\mu(x),\rho^\nu(y)]$ 
on the r.h.s. means negative norm. Likewise for $b/(3+b)>0$ we have negative norm states for 
$\epsilon(x)$. This would require setting up for physical states $|f\rangle $ 
the {\it subsidiary conditions} for the positive-frequency parts
\begin{equation}
 \rho^{\mu (+)}(x) |f\rangle =0\ \ ,\ \ \epsilon^{(+)}(x)|f\rangle =0.
\label{eq:40.9}\end{equation} 

\noindent It is one of the aims of this investigation to find a theory 
which allows (i) a smooth and correct massless limit, and (ii)  a perturbation
expansion in the small mass M.\\
For the latter to be meaningful, it is 
necessary that the theory satisfies the following requirements: (i) no-ghosts,
(ii) unitarity, and (iii) a correct massless limit.
This would open the possibility of giving a small mass to the graviton without
destroying {\it e.g.}, the correct prediction for the perihelion of Mercury. 

\vspace{3mm}
\begin{center}
\fbox{\rule[-5mm]{0cm}{1.0cm} \hspace{5mm}
\begin{minipage}[]{12.9cm}
	{\blue \underline{Corollary}: As follows from the commutator (\ref{40.8f}) 
	the correct massless limit 
 requires the double limit $ b \rightarrow \pm \infty, M \rightarrow 0$.}
\end{minipage} \hspace{5mm} }\\
	\end{center}


\section{Massless Limit: Scalar-tensor or Imaginary-ghost Theory}   
\label{sec:4}   
In the previous section, we concluded that the double limit: 
$M \rightarrow 0; b \rightarrow \infty$ leads to the proper massless graviton propagator.
It is the purpose of this section to 
analyze the distinctive physical contents of the propagator for the 
$h^{\mu\nu}$-field for the different regions of the parameter 
$-\infty < b < +\infty$. In particular in the massless limit
we want to investigate the possibility of a smooth decoupling of the
"false helicities".

\begin{figure}[hhhhtb]
 \resizebox*{10cm}{8cm} 
 {\includegraphics[  0,225][400,575]{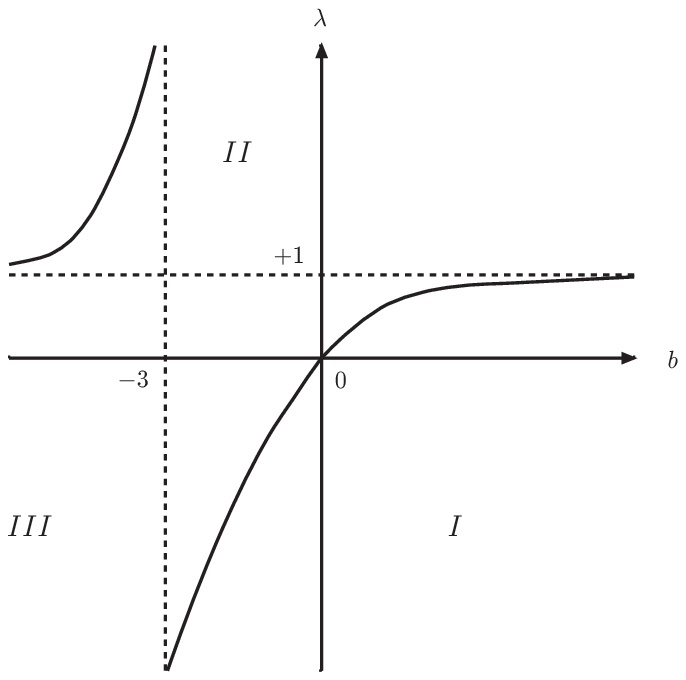}}
  \caption{\sl Three regions in $(b,\lambda)$-space. Region I, III:
 spin-2 and spin-0 imaginary ghost. Region II: (massive) spin-2 and 
spin-0.  }                         
  \label{fig:regions}
 \end{figure}

\noindent From the commutators in (\ref{eq:40.8}), we obtain the propagator by the
replacement
\begin{equation}
 \Delta(x-y; M^2) \rightarrow \Delta_F(x-y;M^2)\ .
\label{eq:4.1}\end{equation} 
Then, we have apart from irrelevant terms for the $h^{\mu\nu}$-field the
Feynman propagator
\footnote{ In the limit $M \rightarrow 0, b \rightarrow \infty$ 
this propagator becomes the standard one, see {\it e.g.} \cite{VDam74}.
}
\begin{eqnarray}
 D_F^{\mu\nu,\alpha\beta}(x-y) &=& \left[\frac{1}{2}\left(
\eta^{\mu\alpha}\eta^{\nu\beta}+\eta^{\mu\beta}\eta^{\nu\alpha}\right)-\frac{1}{3}
\eta^{\mu\nu}\eta^{\alpha\beta}\right]\ \Delta_F(x-y;M^2) \nonumber\\
&& -\frac{1}{6}\frac{b}{3+b}
\eta^{\mu\nu}\eta^{\alpha\beta}\ \Delta_F(x-y;-\frac{2b}{3+b}M^2)\ .               
\label{eq:4.2a}\end{eqnarray} 
Introducing the parameter $\lambda$, defined as
\begin{equation}
 \lambda =: b/(3+b), 
\label{eq:4.2b}\end{equation} 
we can distinghuish three regions, see Fig.~\ref{fig:regions},
 for the b-parameter
\begin{eqnarray*}
 I\ \ &:& \hspace*{3mm} 0\ \leq b < +\infty\ 
  (\hspace{2mm} 0 \leq \lambda \leq 1), \\
 II\ &:& -3\ \leq b \leq 0\ \hspace*{5mm} ( -\infty < \lambda \leq 0), \\
 III &:& -\infty < b \leq -3\ \ ( 1 < \lambda < +\infty).    
\end{eqnarray*}
One sees from (\ref{eq:4.2a}) that for $M \neq 0$ for region II 
the contents is a massive spin-2, and a massive spin-0
particle. For regions I and III the contents is besides a massive
spin-2, a spin-0 ghost particle with an imaginary mass.

\noindent The main goal of this investigation is to find a theory 
which allows (i) a smooth massless limit, and (ii)  a perturbation
expansion in the small mass M. In a \underline{ no-ghost scenario}, it is 
necessary that the theory satisfies the following requirements:\\
\noindent 1.\ No-ghost: $M_\epsilon^2 > 0 \rightarrow b/(3+b) < 0$, \\
\noindent 2.\ Unitarity: $b/(3+b) < 0$, \\
\noindent 3.\ Correct massless limit: 
$ b/(3+b) \rightarrow +1-\Delta,\ \ \Delta >0 $\ , \\
where $\Delta \equiv 3/(3+b)$.



\noindent {\it Clearly, requirement 3) conflicts with 1) and 2) if $\Delta =0$, i.e.
for a pure spin-2 theory in the massless limit. So, with exclusively 
physical fields, at best we can end up with a scalar-tensor type of theory!}\\

\noindent 


\noindent In order to analyze the massless limit in more detail we write 
the $h^{\mu\nu}$-propagator as follows
\begin{eqnarray}
 D_F^{\mu\nu,\alpha\beta}(x-y) &=& \left[\frac{1}{2}\left(
\eta^{\mu\alpha}\eta^{\nu\beta}+\eta^{\mu\beta}\eta^{\nu\alpha}\right)-\frac{1}{2}
\eta^{\mu\nu}\eta^{\alpha\beta}\right]\ \Delta_F(x-y;M^2) + \nonumber\\
&& \frac{1}{6} \eta^{\mu\nu}\eta^{\alpha\beta}\left[ \Delta_F(x-y;M^2) -  
\lambda\ \Delta_F(x-y;-2\lambda\ M^2)\right]\ \nonumber\\
&\equiv& \bar{D}_F^{\mu\nu,\alpha\beta}(x-y,M^2)
 +\Delta D_F^{\mu\nu,\alpha\beta}(x-y;\lambda M^2)
\label{eq:4.4}\end{eqnarray} 
 In the limit $M \rightarrow 0$ for $\lambda=1\ (b \rightarrow \pm\infty)$
the extra piece $\Delta D_F^{\mu\nu;\alpha\beta} \rightarrow 0$, and we
get the proper massless spin-2 propagator.
Then, with $\lambda=1$ we have for $M \neq 0$
a theory with (i) a massive spin-2,  and (ii) an "imaginary" spin-0 ghost 
particle. Below, we will show that the latter will satisfy a free field
equation, which can be quantized \cite{Nak72} and taken care off using a 
{\it Gupta type subsidiary-condition}, see below in section~\ref{sec:9}.\\

\noindent In momentum space we have
\begin{eqnarray}
\Delta\widetilde{F}_F^{\mu\nu;\alpha\beta}(p) &=& 
\frac{i}{6} \eta^{\mu\nu}\eta^{\alpha\beta}\left[ \frac{1}{p^2-M^2+i\delta}
-\frac{\lambda}{p^2 + 2\lambda M^2 + i\delta}\right] \nonumber\\
&=& 
\frac{1}{6}(1-\lambda) \eta^{\mu\nu}\eta^{\alpha\beta}
\frac{p^2+3\lambda M^2/(1-\lambda)}{p^2 + 2\lambda M^2 + i\delta}\cdot
\frac{1}{p^2-M^2+i\delta}\ .
\label{eq:4.5}\end{eqnarray} 


	\begin{center}
\fbox{\rule[-5mm]{0cm}{1.0cm} \hspace{5mm}
\begin{minipage}[]{12.9cm}
	\vspace*{2mm}
\underline{Preliminary summary and prospect}: 
We found in this section that by choosing the constants suitably, and performing a couple of 
gauge transformations, we can eliminate the unwanted helicity components in the
massless limit. Thereby, we arrive at a satisfactorily massless spin-2 theory. 
This is in accordance with the Dirac quantization method for spin-2 fields
using auxiliary vector and scalar (ghost) fields.\\
\noindent There are models of the kind: (i) $-\infty < \lambda \leq 0$:
 massive spin-2 and spin-0
 particles within the massless limit, a kind of scalar-tensor" model,
(ii) $ \lambda=1$: a massive spin-2 and an imaginary-ghost spin-0 
particles with 
a proper massless limit giving a relativistic theory of gravitation
in Minkowski space (RGT-AF). 
The latter we will investigate further in this paper to find out 
whether it is possible to give  a small mass to the graviton, without
destroying the correct prediction for the perihelion of Mercury. 
This in contrast to the formalism 
considered by Van Dam and Veltman \cite{VDV70,Bou72}.

	\vspace*{2mm}
\end{minipage} \hspace{5mm} }\\
	\end{center}

\section{Elimination $\rho(x)$-field, New field Commutators }                   
\label{sec:50}   

\vspace{3mm}
\fbox{\rule[-5mm]{0cm}{1.0cm} \hspace{5mm}
\begin{minipage}[]{14.9cm}
\noindent {\it The conditions for the massless limit
leads to the limit $b \rightarrow \pm\infty$ as the only solution, see Eqn.~(\ref{40.8f}). 
 For the proper behavior of the propagators the $\rho$-field is not necessary, and 
 therefore we effectively eliminate the $\rho$-field from the equations by restricting the
	Hilbert-space (I). Also, this can be done by out-integrating the $\rho$-field (II).
This leads us finally to the spin-2 model of this
paper, besides the $h^{\mu\nu}$-fields only the scalar-ghost field $\epsilon(x)$ 
	with the Gupta subsidiary-condition $\epsilon^{(+)}(x)|\Psi\rangle =0$
	for the states $|\Psi\rangle \in {\cal V}(h,\epsilon)$.}
\end{minipage} \hspace{5mm} }\\

\vspace{3mm}
\subsection{Elimination $\rho(x)$-field}                       
\label{sec:50a}   
\underline{ Elimination $\rho(x)$-field I}:             
The total Hilbert space ${\cal V}_T={\cal V}(h,\rho,\epsilon)$ is an indefinite-metric space. 
From now on we restrict ourselves to ${\cal V}(h,\epsilon) \subset {\cal V}_T$, {\it i.e.} 
states $|\Psi\rangle \in {\cal V}(h,\epsilon)$ which do not contain $\rho$-field quanta. Then, 
for all (field) equations, we consider states such that 
$\langle \Phi| f(\rho)|\Psi\rangle=0,\ \forall f\ {\rm with}\ f(0)=0$.
Therefore, in the rest of this work $\rho(x) \equiv 0$.

\vspace{5mm}
	\underline{ Elimination $\rho(x)$-field II}:             
Notice that the $\rho(x)$-field occurs
only in ${\cal L}_{GF}$ in a quadratic form without derivatives.
Consider the generating functional 
\begin{eqnarray*}
Z &=& \int {\cal D} h_{\mu\nu} {\cal D} \rho_\lambda {\cal D} \epsilon\
	\exp\left[ (i/\hbar) \int {\cal L}\ d^4x\right], 
\end{eqnarray*}
where ${\cal L} = {\cal L}_2+ {\cal L}_{GF}$. 
Ignoring 1/b-terms the gauge-fixing Lagrangian is
\begin{eqnarray*}
{\cal L}_{GF} = 
\frac{1}{2} b\ {\cal M}^2 \rho^\mu(x)\rho_\mu(x) 
+{\cal M}\left[\partial_\mu h^{\mu\nu}(x)+
\widetilde{\gamma} {\cal M}\ \partial^\nu \epsilon(x)\right]\cdot \rho_\nu(x) 
+{\cal M}^2 h(x) \epsilon(x).
\end{eqnarray*}
Introducing 
$\chi^\nu(x) = \partial_\mu h^{\mu\nu}(x)+\widetilde{\gamma} {\cal M}\ \partial^\nu \epsilon(x)$,
the $\rho(x)$-field can be out-integrated and yields
\begin{eqnarray*}
	&& \int {\cal D} \rho_\lambda\ \exp\left\{(i/\hbar)\int d^4x\left[
\frac{1}{2} b{\cal M}^2\rho^\mu(x)\rho_\mu(x)+{\cal M}\ \chi^\nu(x) \rho_\nu(x)\right]
\right\}
\nonumber\\ &&
	= {\cal N}\exp\left\{(i/\hbar) \int d^4x\ \left[-\frac{b{\cal M}^2}{2b^2{\cal M}^2}
\chi^\mu(x)\chi_\mu(x)\right]\right\}
\end{eqnarray*}
This means that ${\cal L}_{GF} \rightarrow {\cal L}_{GF}'$ where
\begin{eqnarray*}
{\cal L}_{GF}' &=& -\frac{b {\cal M}^2}{2b^2{\cal M}^2}
 \left(\partial_\mu h^{\mu\nu}+\widetilde{\gamma}{\cal M}\partial^\nu\epsilon\right)
 \left(\partial^\alpha h_{\alpha\nu}+\widetilde{\gamma}{\cal M}\partial_\nu\epsilon\right)
+{\cal M}^2 h(x)\ \epsilon(x).
\end{eqnarray*}
So, for $b \rightarrow \infty$ the first (contact)
interaction vanishes, only the second one survives.
This defines the model for $b \rightarrow \infty$ without the $\rho$-field. (QED)\\

	\noindent {\bf Remark}: ${\cal N} \sim 1/\sqrt{b} \rightarrow 0$. So the functional 
	integral vanishes for $b=\infty$. This is consistent with the Riemann-Lebesque lemma. Therefore,
	in the limit $\rho(x)$ has to be set to zero.

\subsection{Commutators and Propagator spin-2 for $b \rightarrow \infty$} 
\label{sec:50c}   
Since we are interested in the correct massless limit, we take 
$b \rightarrow \pm \infty$, i.e. $\lambda=1$. Leaving out the $\rho$-field terms
the commutators in Eqn.~(\ref{eq:40.3}) become
\begin{subequations}
\label{HMUNU.3} 
\begin{eqnarray}
 \left[\epsilon(x),\epsilon(y)\right] &=& -\frac{3}{4}
 \left(\frac{M^2}{{\cal M}^2}\right)^2\ i\Delta(x-y; M_\epsilon^2), 
 \label{HMUNU.3a}\\
 \left[\epsilon(x),h^{\mu\nu}(y)\right] &=& -\left[
 \frac{\partial^\mu\partial^\nu}{M^2}
 -\frac{1}{2} \eta^{\mu\nu}\right]\
 i\Delta(x-y; M_\epsilon^2), \label{HMUNU.3b} \\
\left[h^{\mu\nu}(x),h^{\alpha\beta}(y)\right] &=& \biggl\{
\left(\eta^{\mu\alpha}\eta^{\nu\beta}
 +\eta^{\mu\beta}\eta^{\nu\alpha}
-\frac{2}{3}\eta^{\mu\nu}\eta^{\alpha\beta}\right) \nonumber\\ && \hspace{0cm}
+\frac{1}{M_2^2}\left(
\partial^\mu\partial^\alpha \eta^{\nu\beta}
 +\partial^\nu\partial^\alpha \eta^{\mu\beta}
 +\partial^\mu\partial^\beta  \eta^{\nu\alpha}
 +\partial^\nu\partial^\beta  \eta^{\mu\alpha}\right) \nonumber\\ && \hspace{0cm}
 -\frac{2}{3M_2^2}\left(
 \partial^\mu\partial^\nu\eta^{\alpha\beta}+\eta^{\mu\nu}\partial^\alpha\partial^\beta\right)
 + \frac{4}{3M_2^4}\partial^\mu\partial^\nu\partial^\alpha\partial^\beta\biggr\}\ 
   i\Delta(x-y;M_2^2) 
\nonumber\\ && 
	-\biggl\{\frac{1}{3} \eta^{\mu\nu}\eta^{\alpha\beta}-\frac{2}{3M_2^2}
	\left(\partial^\mu\partial^\nu\eta^{\alpha\beta}
	+\eta^{\mu\nu}\partial^{\alpha}\partial^\beta\right)
 +\frac{4}{3M_2^4}\partial^\mu\partial^\nu\partial^\alpha\partial^\beta\biggr\}
 i\Delta(x-y;M_\epsilon^2). \label{HMUNU.3c}
\end{eqnarray} 
\end{subequations}
The compact form of (\ref{HMUNU.3c}) reads
\begin{eqnarray}
\left[h^{\mu\nu}(x),h^{\alpha\beta}(y)\right] &=& \biggl\{
	\bar{\eta}^{\mu\alpha}\bar{\eta}^{\nu\beta}
	+\bar{\eta}^{\mu\beta}\bar{\eta}^{\nu\alpha}
	-\frac{2}{3}\bar{\eta}^{\mu\nu}\bar{\eta}^{\alpha\beta}\biggr\} i\Delta(x-y;M_2^2) 
\nonumber\\ && 
	-\biggl\{\frac{1}{3} \widetilde{\eta}^{\mu\nu}\widetilde{\eta}^{\alpha\beta}+\frac{1}{M_\epsilon^2}
	\left(\partial^\mu\partial^\nu\widetilde{\eta}^{\alpha\beta}
	+\widetilde{\eta}^{\mu\nu}\partial^{\alpha}\partial^\beta\right)
 +\frac{3}{M_\epsilon^4}\partial^\mu\partial^\nu\partial^\alpha\partial^\beta\biggr\}
 i\Delta(x-y;M_\epsilon^2), 
\label{HMUNU.3d} 
\end{eqnarray} 
with the notation
$\bar{\eta}^{\mu\nu} \equiv \eta^{\mu\nu}+\partial^\mu\partial^\nu/M^2$   
 and $\widetilde{\eta}^{\mu\nu} \equiv \eta^{\mu\nu}+\partial^\mu\partial^\nu/M_\epsilon^2$.
For the spin-2 Feynman propagator from (\ref{eq:40.3cc}) one obtains 
\begin{eqnarray}
D_F^{\mu\nu\alpha\beta}(x-y) &=& -i\langle 0|T\left[h^{\mu\nu}(x)
h^{\alpha\beta}(y)\right]|0 \rangle =
2P_2^{\mu\nu\alpha\beta}(\partial)\ \Delta_F(x-y;M_2^2)
	\nonumber\\ &&
+P_\epsilon^{\mu\nu\alpha\beta}(\partial)\ \Delta_F(x-y;M_\epsilon^2)
\label{eq:HMUNU.3e}\end{eqnarray}
because $\lim_{ b \rightarrow \infty} \Delta_F(x-y;M_\rho^2)=0$.\\

\noindent The derivation of the sum rules for the spectral function $\rho(s)$, 
see below, uses for example the ETC relations
\begin{subequations}
\label{HMUNU.31} 
\begin{eqnarray}
\left[h^{00}(x),h^{0l}(y)\right]_0 &=& +\frac{2i}{M_2^2} \partial^l 
	\delta^3({\bf x}-{\bf y}), \\
\left[h^{0m}(x),h^{kl}(y)\right]_0 &=& -\frac{i}{M_2^2}\left( 
	\eta^{ml}\partial^k+\eta^{mk}\partial^l\right)
	\delta^3({\bf x}-{\bf y})
\end{eqnarray} 
\end{subequations}
Here, the difference with \cite{WR09} Eqn.~(21) is due to the presence of the
$\epsilon(x)$-field.

\section{Asymptotic fields. Physical Contents $h_{\mu\nu}$-field} 
\label{sec:50b}   

\subsection{Massive $M_2 \ne 0$ case}
\label{sec:50b1}   
Coupling the $h_{\mu\nu}$-field to the conserved matter and traceless energy-momentum tensor 
	gives the field
	equations, setting ${\cal M}=M$ and $\alpha_2 \equiv -2\lambda \rightarrow -2$, 
\begin{subequations}\label{HMUNU.4}
\begin{eqnarray}
\left(\Box + M^2\right)\ h_{\mu\nu}(x) +2M^2\left(\eta_{\mu\nu}
+\alpha_2\frac{\partial_\mu\partial_\nu}{M^2}\right)\epsilon(x) =
\kappa\ t_{M,\mu\nu}(x), \\
\left(\Box + \alpha_2 M^2\right)\ \epsilon(x)=0,\ 
	\partial^\mu h_{\mu\nu}(x)+\alpha_2\partial_\nu \epsilon(x) =0.    
\end{eqnarray}\end{subequations}
	Requiring $(\Box + M^2)\ h^\mu_\mu(x)=0$ implies that $(t_M)^\mu_\mu=0$. 
With the spectral representation 

\begin{eqnarray}
&&\kappa^2\left(0|\left[t_{M,\mu\nu}(x),t_{M,\alpha\beta}(y)\right]|0\right) = 
 i\int_{0+}^\infty ds\ s^2\pi(s)\
\bar{\Pi}_{\mu\nu;\alpha\beta}(s)\ \Delta(x-y;s)
\label{HMUNU.5a}\end{eqnarray}
where 
\begin{equation}
\bar{\Pi}_{\mu\nu;\alpha\beta}(s) = 
\biggl\{\left(\bar{\eta}_{\mu\alpha}\bar{\eta}_{\nu\beta}+\bar{\eta}_{\mu\beta}\bar{\eta}_{\nu\alpha}\right)
-\frac{2}{3}\bar{\eta}_{\mu\nu}\bar{\eta}_{\alpha\beta}\biggr\}\ 
\label{HMUNU.5b}\end{equation}
with
\begin{equation}
	\bar{\eta}_{\mu\nu}(\partial,s) = \eta_{\mu\nu}+s^{-1}\partial_\mu \partial_\nu,\ 
\partial^\mu\bar{\eta}_{\mu\nu}=0,\ \eta^{\mu\nu}\bar{\eta}_{\mu\nu}(\partial)=3.
\label{HUMU.5c}\end{equation}
\noindent Using $\left[t_{M, \mu\nu}(x), \epsilon(y)\right]=0$ we get
\begin{eqnarray}
	&& \left(0|\left[h_{\mu\nu}(x),h_{\alpha\beta}(y)\right]|0\right) =
-\frac{i}{3}\left(\eta_{\mu\nu}+\alpha_2\frac{\partial_\mu\partial_\nu}{M^2}\right)
\left(\eta_{\alpha\beta}+\alpha_2\frac{\partial_\alpha\partial_\beta}{M^2}\right)
\Delta(x-y;\alpha_2 M^2) \nonumber\\ && 
	+i\int_0^\infty ds\ \rho(s) \bar{\Pi}_{\mu\nu;\alpha\beta}(s)\ \Delta(x-y; s),\  
 \rho(s) = s(s-M^2)^{-2} \pi(s) \geq 0.
\label{HMUNU.6}\end{eqnarray}
Differentiation (\ref{HMUNU.6}) by $x^0$ and setting $x^0=y^0$ one obtains
\begin{eqnarray}
&& \left(0|\left[\dot{h}_{\mu\nu}(x),h_{\alpha\beta}(y)\right]_0|0\right) =
-i\biggl\{ \bar{\Pi}_{\mu\nu;\alpha\beta}(M^2)
	-\frac{1}{3} \widetilde{\eta}_{\mu\nu} \widetilde{\eta}_{\alpha\beta}
	\nonumber\\ && 
	-\frac{1}{M_\epsilon^2}\left(\partial_\mu\partial_\nu \widetilde{\eta}_{\alpha\beta}
	+\widetilde{\eta}_{\mu\nu} \partial_\alpha\partial_\beta\right)
	-\frac{3}{M_\epsilon^4}\partial_\mu\partial_\nu\partial_\alpha\partial_\beta\biggr\}\
	\delta^3({\bf x}-{\bf y})
 = \nonumber\\ && 
	\biggl[ \frac{i}{3}\left(\eta_{\mu\nu}+\alpha_2\frac{\partial_\mu\partial_\nu}{M^2}\right)
\left(\eta_{\alpha\beta}+\alpha_2\frac{\partial_\alpha\partial_\beta}{M^2}\right)
\Delta(x-y;\alpha_2 M^2) \nonumber\\ && 
-i\int_0^\infty ds\ \rho(s) \bar{\Pi}_{\mu\nu;\alpha\beta}(s)\biggr]\ \delta^3({\bf x}-{\bf y}),
\label{HMUNU.6a}\end{eqnarray}
Taking $\mu,\nu,\alpha,\beta=m,n,k,l$ and comparing the $\eta_{mn}\eta_{kl}$ terms gives the sum
rule
	\begin{equation}
		1 = \int_0^\infty ds\ \rho(s).
\label{HMUNU.6b}\end{equation}
The ETC relations derived from (\ref{HMUNU.3d}) and (\ref{HMUNU.6}) give the equations
\begin{eqnarray*}
	&&\left(0|\left[h_{00}(x),h_{0l}(y)\right]_0|0\right) = -\frac{i\alpha_2}{M^2} 
	\partial_l \delta^3({\bf x}-{\bf y}) = \nonumber\\ &&
	\left[+\frac{i\alpha_2}{3M^2}\left((1-\alpha_2^2)+\alpha_2\bm{\nabla}^2/M^2\right)
	 -\frac{4i}{3}\int_0^\infty ds\ (\rho(s)/s^2) \bm{\nabla}^2\right]\  
	 \partial_l\delta^3({\bf x}-{\bf y}), 
	 \nonumber\\ 
&&\left(0|\left[h_{0m}(x),h_{kl}(y)\right]_0|0\right) = 
-\frac{i}{M^2}\left(\eta_{ml}\partial_k+\eta_{mk}
	\partial_l\right) \delta^3({\bf x}-{\bf y}) =
\nonumber\\
&& \left[+\frac{i\alpha_2}{3M^2}\left(\eta_{kl}+\frac{\alpha_2}{M^2}\partial_k\partial_l\right)\partial_m
-i\int_0^\infty ds\ (\rho(s)/s)\biggl\{
\left(\eta_{ml}\partial_k+\eta_{mk}\partial_l-\frac{2}{3}\eta_{kl}\partial_m\right)
+\frac{4}{3}s^{-1}\partial_k\partial_l\partial_m\biggr\} \right]\ \delta^3({\bf x}-{\bf y}). 
\end{eqnarray*}
With $\alpha_2=-2$, these relations lead to the sum rules
\begin{eqnarray}
		(1/M^2)&=& \int_0^\infty ds\ \rho(s)/s\ ,\ 
		(1/M^4) =  \int_0^\infty ds\ \rho(s)/s^2.   
\label{HMUNU.6c}\end{eqnarray}
	Let $M_r$ be the physical ("renormalized") mass of $h_{\mu\nu}$, then
	\begin{equation}
		\rho(s) = Z\delta(s-M_r^2) + \theta(s-s_0) \sigma(s)
\label{HMUNU.6d}\end{equation}
with $s_0 > 0$. The sum rules (\ref{HMUNU.6b}) and (\ref{HMUNU.6c}) give                
\begin{subequations}\label{HMUNU.6e}
\begin{eqnarray}
	&& 0 < Z=1-\int_{s_0}^\infty ds\ \sigma(s) \leq 1, \\
	&& \frac{1}{M^{2n}} = \frac{Z}{M_r^{2n}}+\int_{s_0}^\infty ds\ \sigma(s)/s^n\ (n=1,2)     
\end{eqnarray}\end{subequations}
which shows that if $M \rightarrow 0$ then $M_r \rightarrow 0$. Thus, in this limit the 
	$h_{\mu\nu}$-field can not acquire a non-zero physical mass. The condition for the
	validity is that $t_{M,\mu\nu}$ has no massless discrete spectrum when $M \ne 0$.
	This is the analog of the same theorem as for the massive vector-field in the 
	B-field formalism \cite{NO90}.\\

\noindent The commutation relations in (\ref{HMUNU.3}) and before in this paper were for
free fields, which we denote further on in this section as 'asymptotic" fields
$h^{as}_{\mu\nu}(x), \epsilon^{as}(x)$,
which satisfy (\ref{HMUNU.3}) with ${\cal M}=M$, and (\ref{HMUNU.4}).

\noindent The asymptotic field $h^{as}_{\mu\nu}$ and $\epsilon^as$ are defined 
\begin{equation}
Z^{-1/2} h_{\mu\nu} \rightarrow h^{as}_{\mu\nu}\ ,\
k_2\ Z^{1/2} \epsilon = \epsilon^{as},
\label{HMUNU.6f}\end{equation}
where  
\begin{subequations}\label{HMUNU.6g}
\begin{eqnarray}
k_2 &\equiv& Z^{-1} M_r^2/M^2 =1+ Z^{-1}M_r^2\int_{s_0}^\infty ds\ \sigma(s)/s \geq 1,\\
k_4 &\equiv& Z^{-1} M_r^4/M^4 =1+ Z^{-1}M_r^4\int_{s_0}^\infty ds\ \sigma(s)/s^2 \geq 1,   
\end{eqnarray}\end{subequations}
where we also introduced $k_4$.
Then, with $\rho(s)= Z \delta(s-M_r^2)$, 
	\begin{eqnarray}
\left[h^{as}_{\mu\nu}(x),h^{as}_{\alpha\beta}(y)\right] &=& \frac{i}{3} Z^{-1}
\left(\eta_{\mu\nu}+\alpha_2\frac{\partial_\mu\partial_\nu}{M^2}\right)
\left(\eta_{\alpha\beta}+\alpha_2\frac{\partial_\alpha\partial_\beta}{M^2}\right)
\Delta(x-y;\alpha_2M^2) \nonumber\\ &&
+i\bar{\Pi}_{\mu\nu;\alpha\beta}(M_r^2)\ \Delta(x-y;M_r^2)
\nonumber\\ &=& \frac{i}{3} Z^{-1}
		\left(\eta_{\mu\nu}+\bar{\alpha}_{2r}\frac{\partial_\mu\partial_\nu}{M_r^2}\right)
		\left(\eta_{\alpha\beta}+\bar{\alpha}_{2r}\frac{\partial_\alpha\partial_\beta}{M_r^2}\right)
		\Delta(x-y;\alpha_{2r} M_r^2) \nonumber\\ &&
+i\bar{\Pi}_{\mu\nu;\alpha\beta}(M_r^2)\ \Delta(x-y;M_r^2),
\label{HMUNU.6h}\end{eqnarray}
	where  $M^2\alpha_2=M_r^2\alpha_{2r}$ and 
	$\bar{\alpha}_{2r} \equiv (M_r/M)^4 \alpha_{2r} = k_4Z\ \alpha_{2r}$.
	Notice that the commutator is completely expressed as a function of "renormalized" quantities.
	For the other commutators, we have
	\begin{eqnarray}
		\left[\epsilon^{as}(x),h^{as}_{\alpha\beta}(y),\epsilon^{as}(y)\right] &=& +\frac{i}{2} k_2\left(
\eta_{\alpha\beta}+\bar{\alpha}_{2r}\frac{\partial_\alpha\partial_\beta}{M_r^2}\right)
 \Delta(x-y;\alpha_rM_r^2), \\
 \left[\epsilon^{as}(x),\epsilon^{as}(y)\right] &=&  -\frac{3i}{4}\ k_2\left(\frac{M_r}{M}\right)^2
		 \Delta(x-y;\alpha_rM_r^2).    
\label{HMUNU.6i}\end{eqnarray}
To determine the "renormalized" $h^{as}_{\mu\nu}$-equation we get from (\ref{HMUNU.6h}) the relation
\begin{eqnarray*}
(\Box+M_r^2) \left[h^{as}_{\mu\nu}(x),h^{as}_{\alpha\beta}(y)\right] &=& 
\frac{i}{3} Z^{-1} (1-\alpha_{2r})M_r^2
\left(\eta_{\mu\nu}+\bar{\alpha}_{2r}\frac{\partial_\mu\partial_\nu}{M_r^2}\right)
\left(\eta_{\alpha\beta}+\bar{\alpha}_{2r}\frac{\partial_\alpha\partial_\beta}{M_r^2}\right)
\Delta(x-y;\alpha_{2r} M_r^2) 
\nonumber\\ &=&
-\frac{2}{3} (k_2Z)^{-1} (1-\alpha_{2r})M_r^2
\left(\eta_{\mu\nu}+\bar{\alpha}_{2r}\frac{\partial_\mu\partial_\nu}{M_r^2}\right)
	\left[\epsilon^{as}(x),h^{as}_{\alpha\beta}(y)\right], 
\end{eqnarray*}
which leads to the "renormalized" equation
	\begin{eqnarray}
		(\Box+M_r^2)\ h^{as}_{\mu\nu}(x) +\frac{2}{3}\left(1-\alpha_{2r}\right) (k_2Z)^{-1}
		M_r^2\left(\eta_{\mu\nu}+\bar{\alpha}_{2r}\frac{\partial_\mu\partial_\nu}{M_r^2}\right)\
		\epsilon^{as}(x)=0.
\label{HMUNU.6j}\end{eqnarray}
\noindent Next, we introduce the asymptotic physical spin-2 field by
	\begin{eqnarray}
		U^{as}_{\mu\nu}(x) &=& 
		h^{as}_{\mu\nu}(x) +\frac{2}{3}\left(1-\alpha_{2r}\right) (k_2Z)^{-1}
		M_r^2\left(\eta_{\mu\nu}+\bar{\alpha}_{2r}\frac{\partial_\mu\partial_\nu}{M_r^2}\right)\
		(\Box+M_r^2)^{-1}\ \epsilon^{as}(x) \nonumber\\ &=&
		h^{as}_{\mu\nu}(x) +\frac{2}{3} (k_2Z)^{-1}
		\left(\eta_{\mu\nu}+\bar{\alpha}_{2r}\frac{\partial_\mu\partial_\nu}{M_r^2}\right)\
		 \epsilon^{as}(x).                 
\label{HMUNU.7}\end{eqnarray}
which satisfies 
\begin{subequations}\label{HMUNU.8}
\begin{eqnarray}
&& \left[U^{as}_{\mu\nu}(x), U^{as}_{\alpha\beta}(y)\right] =
\biggl\{\left(\bar{\eta}_{\mu\alpha}\bar{\eta}_{\nu\beta}+\bar{\eta}_{\mu\beta}\bar{\eta}_{\nu\alpha}\right)
-\frac{2}{3}\bar{\eta}_{\mu\nu}\bar{\eta}_{\alpha\beta}\biggr\}\ 
i\Delta(x-y;M_r^2), \\
&&\left[U^{as}_{\mu\nu}(x), \epsilon^{as}(y)\right]=0, \\
&& \left(\Box+M_r^2\right)\ U^{as}_{\mu\nu}(x)=0,\ \partial^\mu U^{as}_{\mu\nu}(x)=0, 
\end{eqnarray}\end{subequations}

With the interaction, the field equation reads 
$(\Box +M_r^2) U_{\mu\nu}(x)= \kappa\ t_{M,\mu\nu}(x)$
which can be integrated to a kind of Yang-Feldman equation \cite{Yan50}
\begin{subequations}\label{HMUNU.9}
\begin{eqnarray}
U_{\mu\nu}(x) &=& U^{as}_{\mu\nu}(x) + \int d^4y\
\Delta^{as}_{\mu\nu;\alpha\beta}(x-y; M_r^2)\ t_M^{\alpha\beta}(y), \\
\Delta^{as}_{\mu\nu;\alpha\beta}(x-y;M_r^2) &=& 
\biggl\{\left(\bar{\eta}_{\mu\alpha}\bar{\eta}_{\nu\beta}+\bar{\eta}_{\mu\beta}\bar{\eta}_{\nu\alpha}\right)
-\frac{2}{3}\bar{\eta}_{\mu\nu}\bar{\eta}_{\alpha\beta}\biggr\}\ 
i\Delta^{as}(x-y;M_r^2) \\
 &\rightarrow& 
\biggl\{\left(\eta_{\mu\alpha}\eta_{\nu\beta}+\eta_{\mu\beta}\eta_{\nu\alpha}\right)
-\frac{2}{3}\eta_{\mu\nu}\eta_{\alpha\beta}\biggr\}\ 
i\Delta^{as}(x-y;M_r^2), 
\end{eqnarray}\end{subequations}
where $\bar{\eta}_{\mu\nu} \equiv \bar{\eta}(\partial,M_r^2)$ etc., and
$U_{\mu\nu}^{as}= U_{\mu\nu}^{in}$ or $U_{\mu\nu}^{out}$ and correspondingly
$\Delta^{as}$ is the retarded $\Delta_R$ or the advanced $\Delta_A$.
	In (\ref{HMUNU.9}) the replacement $\bar{\eta}_{\mu\nu} \rightarrow \eta_{\mu\nu}$ etc. follows from 
	$\partial^\mu t_{M,\mu\nu}= (t_M)^\mu_\mu=0$ and partial integration.
	\\

\noindent {\it The physical Hilbert space ${\cal V}_{phys}$ consists of the totality of states $|f\rangle$
which satisfy the Gupta subsidiary-condition $\epsilon^{(+)}(x)|f\rangle=0$. Representatives of equivalent 
classes are obtained by applying products of creation operators associated with the $U^{as}_{\mu\nu}$ field.
}
\subsection{Massless $M=M_2=0$ case}
\label{sec:50b2}   
In the limit $M, M_r \rightarrow 0$ the sum rules (\ref{HMUNU.6g}) reveal that
	\begin{eqnarray}
		&& \lim_{M,M_r \rightarrow 0} (M_r^2/M^2)^n =Z\ (n=1,2) \Rightarrow Z \rightarrow 1,
	\label{HMUNU.10}\end{eqnarray}
	which means the zero-mass limit $M=M_r \rightarrow 0$. Analyzing this limit for 
	Eqn.~(\ref{HMUNU.6j}) we note that with
	$(k_2Z)^{-1}M_r^2=M^2$ one obtains for the second term 
	\begin{eqnarray*}
		(k_2Z)^{-1}M_r^2\left(\eta_{\mu\nu}+\bar{\alpha}_{2r}\frac{\partial_\mu\partial_\nu}{M_r^2}\right)
		= M^2 \eta_{\mu\nu}+\left(\frac{M_r}{M}\right)^2 \alpha_2\ \partial_\mu\partial_\nu \rightarrow
		\alpha_2 \partial_\mu\partial_\nu.
	\end{eqnarray*}
Therefore, we obtain for the massless limit the equations
\begin{subequations}\label{HMUNU.11}
\begin{eqnarray}
&& \Box h^{as}_{\mu\nu}(x)+ 
2\alpha_2 \partial_\mu\partial_\nu \epsilon^{as}(x)=0, \\
&& \Box \epsilon^{as}(x)=0,\ \partial^\mu h_{\mu\nu}^{as}(x)+\alpha_2 \partial_\nu\epsilon^{as}(x)=0, 
\end{eqnarray}\end{subequations}
	where we added the last equation from (\ref{HMUNU.4}).
Again, the derivatives in $\Delta^{as}_{\mu\nu;\alpha\beta}$ can be omitted. 
	Next we take the limit $M \rightarrow 0$ which 
changes (\ref{HMUNU.9}) to the massless form
\begin{equation}
		\Delta^{as}_{\mu\nu;\alpha\beta}(x-y) \rightarrow 
	\left(\eta_{\mu\alpha}\eta_{\nu\beta}+\eta_{\mu\beta}\eta_{\nu\alpha}
	-\eta_{\mu\nu}\eta_{\alpha\beta}\right)\ i D(x-y), 
\label{HMUNU.11b}\end{equation}
and changes (\ref{HMUNU.6h}) and (\ref{HMUNU.6i}) into the commutators
\begin{subequations}\label{HMUNU.12}
\begin{eqnarray}
&& \left[h^{as}_{\mu\nu}(x), h^{as}_{\alpha\beta}(y)\right] =
\biggl\{\left(\eta_{\mu\alpha}\eta_{\nu\beta}+\eta_{\mu\beta}\eta_{\nu\alpha}\right)
-\eta_{\mu\nu}\eta_{\alpha\beta}\biggr\}\ iD(x-y), \\
&&\left[h^{as}_{\mu\nu}(x), \epsilon^{as}(y)\right]=
\frac{i}{2}\left(\eta_{\mu\nu}D(x-y)+\alpha_2 \partial^\mu\partial^\nu \Box^{-2}E(x-y)\right), \\
	&& \biggl[\epsilon^{as}(x),\epsilon^{as}(y)\bigr] = -\frac{3i}{4} D(x-y),   
\end{eqnarray}\end{subequations}
	where $D(x) = \Delta(x;M^2)|_{M=0}$ and $E(x)= \Box D(x)$.   
	Furthermore, in the massless limit the $U^{as}_{\mu\nu}(x)$ field now satisfies
\begin{subequations}\label{HMUNU.13}
\begin{eqnarray}
&& \left[U^{as}_{\mu\nu}(x), U^{as}_{\alpha\beta}(y)\right] =
\left(\eta_{\mu\alpha}\eta_{\nu\beta}+\eta_{\mu\beta}\eta_{\nu\alpha}
-\eta_{\mu\nu}\eta_{\alpha\beta}\right)\ i D(x-y), \\
&&\left[U^{as}_{\mu\nu}(x), \epsilon^{as}(y)\right]=0, \\
&& \Box U^{as}_{\mu\nu}(x)=0,\ \partial^\mu U^{as}_{\mu\nu}(x)=0, 
\end{eqnarray}\end{subequations}
describing a massless graviton, demonstrating the smooth massless limit $M_2 \rightarrow 0$.\\

\noindent In Appendix~\ref{sec:9} the quantization of $\epsilon(x)$-field is described
and the propagator is derived in detail. 
Because of the (-)-sign in the last commutator in (\ref{HMUNU.9}) it has a negative
norm and so a Gupta subsidiary-condition has to be imposed, similarly to that in the case
of the Lorentz-gauge in QED. Then, the matrix elements between physical states
of the equations (\ref{HMUNU.10}) and of the comutators in (\ref{HMUNU.11}) satisfy the
classical equations, {\it i.e.} $\big\langle \Box h^{as}_{\mu\nu}(x)\big\rangle =0,
\big\langle \partial^\mu h_{\mu\nu}(x) \big\rangle =0$.\\

\noindent {\it Note that this auxiliary-field treatment for the massive and massless spin-2 meson is analogous to that for the 
vector-meson in Ref.~\cite{NO90} section 2.4, and distinct from the Proca-formalism used in \cite{Nath65}.
The advantage is a smooth transition between the massive and massless cases.}

\begin{flushleft}
\rule{16cm}{0.5mm}
\end{flushleft}

\subsection{Momentum-space asymptotic states}
\label{sec:50b3}   
\noindent In the expansion
\begin{eqnarray}
U^{as}_{\mu\nu}(x) &=& \int\frac{d^3k}{\sqrt{(2\pi)^3 2\omega_2(k)}}
	\left[u_{\mu\nu}(k)e^{i({\bf k}\cdot{\bf x}-\omega_2t)} + h.c.\right]
	\equiv \int d\widetilde{k}\sum_{\lambda=1}^5\left[
		u^{(\lambda)}(k) \varepsilon^{(\lambda)}_{\mu\nu}(k) 
		e^{-ik\cdot x} + h.c. \right], 
\label{HMUNU.61}\end{eqnarray}
where the helicity annihilation and creation operators satisfy
\begin {eqnarray}
	\left[u^{(\lambda)}(k), u^{(\lambda') \dagger}(k')\right] &=&
	-\eta^{\lambda \lambda'} \delta^3({\bf k}-{\bf k}'),
\label{HMUNU.62}\end{eqnarray}
which implies for the operators $u_{\mu\nu}(k)$ the commutators
	\begin{subequations}\label{HMUNU.63}
\begin{eqnarray}
\left[u_{\mu\nu}(k), u^\dagger_{\alpha\beta}(k')\right] &=&
	\bar{\Pi}_{\mu\nu,\alpha\beta}(k)\ \delta^3({\bf k}-{\bf k}'),\ {\rm with}\
	\bar{\Pi}_{\mu\nu,\alpha\beta}(k) = -\sum_{\lambda \lambda'} \eta^{\lambda \lambda'}
	\varepsilon^{(\lambda)}_{\mu\nu}(k)\varepsilon^{(\lambda')}_{\alpha\beta}(k))
\end{eqnarray}\end{subequations}
		Explicitly \cite{Schw70},
\begin{eqnarray}
	&& \bar{\Pi}_{\mu\nu,\alpha\beta}(k) = 
	\frac{1}{2}\left( \bar{\eta}_{\mu\alpha} \bar{\eta}_{\nu\beta} 
	+\bar{\eta}_{\mu\beta}  \bar{\eta}_{\nu\alpha}\right)
	-\frac{1}{3}\bar{\eta}_{\mu\nu}\bar{\eta}_{\alpha\beta},\ \
	\bar{\eta}_{\mu\nu} = \eta_{\mu\nu}-k^\mu k^\nu/M^2.
\label{HMUNU.64}\end{eqnarray}
The expansion of $h_{\mu\nu}^{as}(x)$ and $\epsilon^{as}(x)$
in annihilation and creation operators is
\begin{subequations}\label{HMUNU.66}
\begin{eqnarray}
h^{as}_{\mu\nu}(x) &=& \int d\widetilde{k}
\left[a_{\mu\nu}^{(+)}(k)e^{i({\bf k}\cdot{\bf x}-\omega_2t)} + h.c.\right], \\
\epsilon^{as}_{\mu\nu}(x) &=& \int d\widetilde{k}
\left[\epsilon(k)^{(+)}(k)e^{i({\bf k}\cdot{\bf x}-\omega_\epsilon t)} + h.c.\right],\ 
\end{eqnarray}\end{subequations}
with the relation
\begin{eqnarray}
a^{(+)}_{\mu\nu}(k) &=& u^{(+)}_{\mu\nu}(k)+\frac{2}{3}\left(
\eta_{\mu\nu}-\alpha_2 \frac{k_\mu k_\nu}{M^2}\right) \epsilon^{(+)}(k)
\label{HMUNU.67}\end{eqnarray}
From $\partial^\mu h_{\mu\nu}+\alpha_2\partial_\nu \epsilon=0$ one has, 
since $k^\mu u^{(+)}_{\mu\nu}(k)=0$,
\begin{eqnarray}
&& k^\mu a^{(+)}_{\mu\nu}(k) +\alpha_2 k_\nu\epsilon^{(+)}(k)=0,\  
k^\mu k^\nu a^{(+)}_{\mu\nu}(k) +\alpha_2 k^2\epsilon^{(+)}(k)=0.    
\label{HMUNU.68}\end{eqnarray}
For the $\epsilon(x)$-field the following commutator applies, see section \ref{sec:9}, 
\begin{equation}
	\left[\epsilon^{(+)}(k), \epsilon^{(+) \dagger}(k')\right]=
-\frac{3}{4} \delta^3({\bf k}-{\bf k}'), 
\label{HMUNU.69}\end{equation}
which leads to
\begin{eqnarray}
	&& \left[a^{(+)}_{\mu\nu}(k),a^{(+)\dagger}_{\alpha\beta}(k')\right] =
	2\left\{\bar{\Pi}_{\mu\nu,\alpha\beta}(k) 
	-\frac{1}{6}\left(\eta_{\mu\nu}-\alpha_2\frac{k_\mu k_\nu}{M^2}\right)
	\left(\eta_{\alpha\beta}-\alpha_2\frac{k_\alpha k_\beta}{M^2}\right)\right\}\
	\delta^3({\bf k}-{\bf k}')   
\label{HMUNU.70}\end{eqnarray}


\begin{flushleft}
\rule{16cm}{0.5mm}
\end{flushleft}


\section{Graviton-mass and the Perihelion-precession of Planets} 
\label{sec:10}   
We want to compute the finite-mass corrections to the massless spin-2 
perihelion-precession of the planets. In this section, and henceforth,
we denote the graviton mass
by $\mu_G$, the mass of the Sun by M, and the mass of Mercury by m.\\

\begin{center}
\fbox{\rule[-5mm]{0cm}{1.5cm} \hspace{5mm}
\begin{minipage}[]{12.9cm}
\noindent The propagator for the (massive) $h^{\mu\nu}$-field, for
 $\lambda=1$, see Eqn.~(\ref{eq:4.4}), reads
\begin{subequations}
\label{eq:10.1a} 
\begin{eqnarray}
 D^{\mu\nu,\alpha\beta}(x;\mu_G^2)_F &=& 
 \frac{1}{2}\left(\eta^{\mu\alpha}\eta^{\nu\beta}+\eta^{\mu\beta}\eta^{\nu\alpha}
 -\eta^{\mu\nu}\eta^{\alpha\beta}\right)\ \Delta_F(x;\mu_G^2) 
\nonumber\\ & & 
 +\frac{1}{6} \eta^{\mu\nu}\eta^{\alpha\beta}
 \left[\Delta_F(x;\mu_G^2) -\Delta_F(x; M^2_\epsilon)\right]  
 \\ &:=& 
 \Delta_{F,0}^{\mu\nu,\alpha\beta}(x;\mu_G^2) + 
 \delta\Delta_F^{\mu\nu,\alpha\beta}(x;\mu_G^2), 
\end{eqnarray} 
\end{subequations}
where $\Delta^{\mu\nu,\alpha\beta}_{F,0}(x)$ in the $\lim_{\mu_G \rightarrow 0}$  is the 
propagator for the massless spin-2 particle, and 
\begin{eqnarray}
 \delta\Delta_{F,\mu\nu,\alpha\beta}(x;\mu_G^2) &\equiv& 
 \Delta^{(S)}_{\mu\nu,\alpha\beta}(x;\mu_G^2) + \Delta^{(SG)}_{\mu\nu,\alpha\beta}(x;\mu_G^2),
\label{eq:10.1b}\end{eqnarray} 
where 
\begin{subequations}
\label{eq:10.1c} 
\begin{eqnarray}
 \Delta^{(S)}_{\mu\nu,\alpha\beta}(x;\mu_G^2) &=& 
+\frac{1}{6} \eta^{\mu\nu}\eta^{\alpha\beta} \Delta_F(x;\mu_G^2), \\
 \Delta^{(SG)}_{\mu\nu,\alpha\beta}(x;\mu_G^2) &=& 
-\frac{1}{6} \eta^{\mu\nu}\eta^{\alpha\beta} \Delta_F(x;-2\mu_G^2)
\end{eqnarray} 
\end{subequations}
\end{minipage} \hspace{5mm} }\\
\end{center}

\noindent {\it As noted before, in the massless $\lim_{\mu_G \rightarrow 0}$ the 
propagator $D_F^{\mu\nu,\alpha\beta}(x;\mu_G^2)$ corresponds to a massless graviton.     
Therefore, it gives the Einstein prediction for the perihelion precession.}

\noindent In \ref{app:helicity-coupling} the smooth massless limit of the model is
considered using the properties of the spin-2 polarizations for $\lim_{\mu_G \downarrow 0}$,
and the conservation of the energy-momentum tensor $t_{M, \mu\nu}$.
\subsection{Interaction Spin-2 Particles with a Scalar-field}
\label{app:C.b}   
In general-relativity \cite{Ein16,Haw-Ellis76} the Lagrangian for a neutral scalar 
field, mass m, invariant under general coordinate transformations in a gravitational
field described by the metrc $g_{\mu\nu}$, is given by
\begin{equation}
 {\cal L}_S = \frac{1}{2} \sqrt{-g}\ \left(
 g^{\mu\nu} D_\mu\phi D_\nu\phi -m^2 \phi^2\right),
\label{app:C.21}\end{equation}
where $g = \det g_{\mu\nu}$, likewise $\eta = \det \eta_{\mu\nu}=-1$.
In the "weak field" approximation we write
\begin{equation}
 g_{\mu\nu} = \eta_{\mu\nu} + \kappa\ h_{\mu\nu},\ \ 
 g^{\mu\nu} = \eta^{\mu\nu} - \kappa\ h_{\mu\nu},\ \ 
\label{app:C.22}\end{equation}
where $\kappa \propto \sqrt{G}$ with $G$ is the Newtonian gravitational constant. 
Note that the signs in (\ref{app:C.22}) are consistent with 
 $g^{\mu\nu} g_{\nu\lambda} = \delta^\mu_{\ \lambda}$.\\

\noindent {\it In the following the lowering and raising of the indices 
 is done with the $\eta_{\mu\nu}=\eta^{\mu\nu}$-tensor}.\\

\noindent Using 
\begin{equation}
  g \approx \eta\ \left( 1 + \kappa\ h^\mu_\mu\right),\ \
 \sqrt{-g} \approx 1+\frac{\kappa}{2}\ h^\mu_\mu, 
\label{app:C.23}\end{equation}
the scalar Lagrangian in the weak-field approximation is
\begin{eqnarray}
 {\cal L}_S &=& \frac{1}{2} \left(
 \eta^{\mu\nu} \partial_\mu\phi \partial_\nu\phi -m^2 \phi^2\right) 
 \nonumber\\ && 
 -\frac{\kappa}{2} h^{\mu\nu}\ \partial_\mu \phi \partial_\nu\phi 
 +\frac{\kappa}{4} h^\lambda_\lambda\ \left(\eta^{\mu\nu} \partial_\mu\partial_\nu
 -m^2 \phi^2\right) \nonumber\\ 
 &\equiv& {\cal L}_S^{(0)} + {\cal L}_S^{int}, 
\label{app:C.24}\end{eqnarray}
with
\begin{eqnarray}
 {\cal L}_S^{int} &=& 
 -\frac{\kappa}{2} h^{\mu\nu}\ \partial_\mu\phi \partial_\nu\phi 
 +\frac{\kappa}{4} h^\lambda_\lambda\ \left(\eta^{\mu\nu} \partial_\mu\partial_\nu
 -m^2 \phi^2\right) \nonumber\\ &=&
 -\frac{\kappa}{2} h^{\mu\nu}\ \left[\partial_\mu\phi \partial_\nu\phi 
 -\frac{1}{2} \eta_{\mu\nu}\ \left(\eta^{\alpha\beta} \partial_\alpha\partial_\beta
 -m^2 \phi^2\right)\right] \nonumber\\ 
 &\equiv& -\frac{\kappa}{2} h^{\mu\nu}\ t^{(S)}_{\mu\nu}, 
\label{app:C.25}\end{eqnarray}
where 
\begin{equation}
 t^{(S)}_{\mu\nu} = 
 \partial_\mu\phi \partial_\nu\phi -\eta_{\mu\nu}\ {\cal L}_S^{(0)}
\label{app:C.26}\end{equation}
the energy-momentum tensor operator for the scalar field.
 

\noindent The momentum expansion for $h_{\mu\nu}(x)$ field reads
\begin{eqnarray}
 h_{\mu\nu}(x) &=& \int\frac{d^3k}{\sqrt{2\omega(k)(2\pi)^3}}
 \sum_{\lambda = 1}^5\ e_{\mu\nu}(k,\lambda)\left[\vphantom{\frac{A}{A}}
 a(k,\lambda) e^{-ik\cdot x} + a^\dagger(k,\lambda) e^{+ik\cdot x}\right], 
\label{app:C.27a}\end{eqnarray}
and for the (neutral) scalar field \cite{BD65}
\begin{eqnarray}
 \phi(x) &=& \int\frac{d^3p}{\sqrt{2\omega(p)(2\pi)^3}}
 \left[\vphantom{\frac{A}{A}}
 a(p) e^{-ip\cdot x} + a^\dagger(p) e^{+ip\cdot x}\right], 
\label{app:C.27b}\end{eqnarray}
with the commutation relation
\begin{equation}
 \left[a({\bf p}), a^\dagger({\bf p}')\right] = \delta^{3}({\bf p}-{\bf p}').
\label{app:C.27c}\end{equation}
For the one-particle scalar states, we use the so-called "non-relativistic
normalization":
\begin{equation}
 | {\bf p}\rangle = a^\dagger({\bf p})|0\rangle,\ \
 \langle {\bf p}| {\bf p}'\rangle = \delta^{3}({\bf p}-{\bf p}').
\label{app:C.27d}\end{equation}

Matrix elements for the energy-momentum operators, using {\bf normal-ordering}, is
\begin{subequations}
\label{app:C.27e}
\begin{eqnarray}
&& \langle p'| :t_{\alpha\beta}(x):| p \rangle =
 \frac{\exp i(p'-p)\cdot x}{(2\pi)^3\sqrt{4 E(p') E(p)}}\
 \left[ \vphantom{\frac{A}{A}}
 \left(p'_\alpha p_\beta + p'_\beta p_\alpha\right)-\eta_{\alpha\beta} 
 \left( p'\cdot p- m^2\right)\right], \\
&& \langle P'|:T_{\mu\nu}:| P \rangle =
 \frac{\exp i(P'-P)\cdot x}{(2\pi)^3\sqrt{4{\cal E}(P') {\cal E}(P)}}\
 \left[ \left(P'_\mu P_\nu + P'_\nu P_\mu\right)-\eta_{\mu\nu} 
 \left( P'\cdot P- M^2\right)\right].    
\end{eqnarray}
\end{subequations}

\noindent The (spin-0)-( spin-2) vertex is given by 
\begin{eqnarray}
 \langle p'|-i\int d^4x\ {\cal H}^{(S)}_{int}(x) |p,k\rangle &\equiv &
 (2\pi)^4 i\delta^4(p'-p-k)\ e^{\mu\nu}(k,\lambda) \Gamma_{\mu\nu}(p',p)  
 \cdot\nonumber\\ && \times
 \left[(2\pi)^9\ 8E(p')E(p)\omega(k)\right]^{-1/2}
\label{app:C.28a}\end{eqnarray}
Using 
\begin{equation}
 \langle 0 | h^{\mu\nu}(x)| k,\lambda\rangle = 
 \left[(2\pi)^3\ 2\omega(k)\right]^{-1/2}\ e^{\mu\nu}(k,\lambda)\
 e^{-i k\cdot x}, 
\label{app:C.28b}\end{equation}
(\ref{app:C.28a}) leads to
\begin{eqnarray}
&& \langle p'|-i\int d^4x\ {\cal H}^{(S)}_{int}(x) |p,k\rangle =
 -i (\kappa/2)\ \left[(2\pi)^9\ 8\omega(p')\omega(p)\omega(k)\right]^{-1/2}
\cdot\nonumber\\ && \times (2\pi)^4\delta^4(p'-p-k)\cdot
 e^{\mu\nu}(k,\lambda)\left\{\vphantom{\frac{A}{A}}
 \left(p_\mu p'_\nu+ p'_\mu p_\nu\right)-\eta_{\mu\nu}(p'\cdot p-m^2)
 \right\}, 
\label{app:C.28c}\end{eqnarray}
which leads to
\begin{equation}
 \Gamma_{\mu\nu}(p',p) = -(\kappa/2)\ \left[\vphantom{\frac{A}{A}}
 \left(p'_\mu p_\nu + p'_\nu p_\mu\right)-\eta_{\mu\nu} 
 \left( p'\cdot p- m^2\right)\right]. 
\label{app:C.29}\end{equation}
We notice that $\Gamma_{\mu\nu}(p',p)=\Gamma_{\nu\mu}(p',p)$ and 
$(p'-p)^\mu \Gamma_{\mu\nu}=0$. This gives for the matrix elements of
the energy-momentum operators $T^{\mu\nu}$ and $t^{\alpha\beta}$
\begin{subequations}
\label{app:C.30}
\begin{eqnarray}
 \widetilde{T}^{\mu\nu}(P',P) &=& 
 \left[\vphantom{\frac{A}{A}}
 \left(P'_\mu P_\nu + P'_\nu P_\mu\right)-\eta_{\mu\nu} 
 \left( P'\cdot P- M^2\right)\right], \\
 \widetilde{t}^{\alpha\beta}(p',p) &=& 
 \left[\vphantom{\frac{A}{A}}
 \left(p'_\alpha p_\beta + p'_\alpha p_\beta\right)-\eta_{\alpha\beta} 
 \left( p'\cdot p- m^2\right)\right].     
\end{eqnarray}
\end{subequations}

 \begin{figure}[hhhhtb]
 \resizebox{10cm}{8cm} 
 {\includegraphics[ 50,475][350,625]{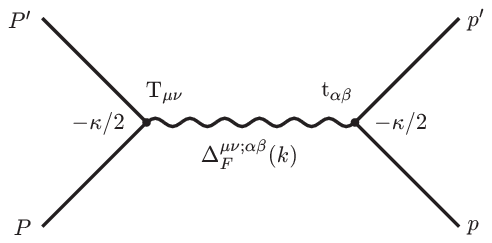}}
 \caption{Graviton-exchange between mass M and mass m.}
 \label{fig.grav1}    
 \end{figure}

Propagators:
\begin{subequations}
\label{app:C.31}
\begin{eqnarray}
 \Delta^{(m)}_{\mu\nu;\alpha\beta}(k) &=& 
 \frac{P^{(m)}_{\mu\nu;\alpha\beta}(k)}{k^2-\mu_G^2+i\delta}, 
 \nonumber\\ 
 P^{(m)}_{\mu\nu;\alpha\beta}(k) &=& \frac{1}{2}\left(  
 \eta_{\mu\alpha} \eta_{\nu\beta}+\eta_{\mu\beta} \eta_{\nu\alpha}\right)
 -\frac{1}{3} \eta_{\mu\nu} \eta_{\alpha\beta}, \\
 \Delta^{(0)}_{\mu\nu;\alpha\beta}(k) &=& 
 \frac{P^{(0)}_{\mu\nu;\alpha\beta}(k)}{k^2-\mu_G^2+i\delta}, 
 \nonumber\\ 
 P^{(0)}_{\mu\nu;\alpha\beta}(k) &=& \frac{1}{2}\left(  
 \eta_{\mu\alpha} \eta_{\nu\beta}+\eta_{\mu\beta} \eta_{\nu\alpha}
 -\eta_{\mu\nu} \eta_{\alpha\beta}\right).
\end{eqnarray}
\end{subequations}
For distinguishing the massless and massive case convenient is the common notation
\begin{equation}
 P^{(a)}_{\mu\nu;\alpha\beta}(k) = \frac{1}{2}\left(  
 \eta_{\mu\alpha} \eta_{\nu\beta}+\eta_{\mu\beta} \eta_{\nu\alpha}\right)
 -a\ \eta_{\mu\nu} \eta_{\alpha\beta}, 
\label{app:C.31c}\end{equation}
where a=1/2 and 1/3 for the massless and massive case, respectively.

\noindent The amplitude corresponding to the Feynman graph Fig.~\ref{fig.grav1},
for the Feynman rules see \cite{Schw62}, chapter 14, is 
\begin{eqnarray}
 S_{fi}^{(2)} &=&
 (+i)^2\ (\kappa^2/4)\ 
 {\cal N}_f(P',p')\ {\cal N}_i(P,p)\ 
 \int\frac{d^4k}{(2\pi)^4} (2\pi)^4\delta^4(P'-P+k)\ 
\cdot\nonumber\\ && \times (2\pi)^4\delta^4(p'-p-k)\cdot 
 \langle P'|T^{\mu\nu}|P\rangle\ i\Delta^{(m)}_{\mu\nu;\alpha\beta}(k)\
 \langle p'|t^{\alpha\beta}|p\rangle \nonumber\\ &=&
-i (\kappa^2/4)\ (2\pi)^4\delta^4(P'+p'-P-p)\ 
 {\cal N}_f(P',p')\ {\cal N}_i(P,p)\ 
 \cdot\nonumber\\ && \times 
 T^{\mu\nu}(P',P)\ \Delta^{(m)}_{\mu\nu;\alpha\beta}(k)\ t^{\alpha\beta}(p',p)
 \nonumber\\ &\equiv & 
 -i(2\pi)^4\delta^4(P'+p'-P-p)\ {\cal N}_f\ M_{fi}^{(2)}\ {\cal N}_i\   
\label{app:C.31b}\end{eqnarray}
with $k=p'-p=P-P'$, and 
\begin{equation}
 {\cal N}_i(P,p) = \left[(2\pi)^6\ {\cal E}(P)\ E(p)\right]^{-1/2},\ \ 
 {\cal N}_f(P',p') = \left[(2\pi)^6\ {\cal E}(P')\ E(p')\right]^{-1/2}.    
\label{app:C.31a}\end{equation}
Here $E(p)=\omega(p), {\cal E}(P)=\omega(P)$ etc.\\
With $a=1/3$ and $a=1/2$ for the massive (m) and massless (0) 
case, respectively, one gets for the invariant amplitude the expression
\begin{eqnarray}
&& M_{fi}^{(2)}(P',p';P,p) =  
 \frac{1}{4}\kappa^2\ 
 \langle P'|T^{\mu\nu}|P\rangle\ \Delta^{(m,0)}_{\mu\nu;\alpha\beta}(k)\
 \langle p'|t^{\alpha\beta}|p\rangle = 
 \frac{1}{4}\kappa^2\cdot
 \nonumber\\ &&
 \times\left[\left(P'_\mu P_\nu+P'_\nu P_\mu\right)-\eta_{\mu\nu}
 (P'\cdot P -M^2)\right]\cdot
 \left\{\frac{1}{2}\left(\eta^{\mu\alpha}\eta^{\nu\beta}
 +\eta^{\mu\beta}\eta^{\nu\alpha}\right)
 -a\ \eta^{\mu\nu}\eta^{\alpha\beta}\right\}\cdot
  \nonumber\\ && \times
 \left[\left(p'_\alpha p_\beta + p'_\beta p_\alpha\right)
 -\eta_{\alpha\beta}(p'\cdot p-m^2)\right]\cdot
 \left[k^2-\mu_G^2+i\delta\right]^{-1} =
 \nonumber\\ && \frac{1}{2}\kappa^2\ 
 \left\{\vphantom{\frac{A}{A}}
 (P'\cdot p')(P\cdot p) +(P'\cdot p) (P\cdot p')
  -2a\ (P'\cdot P)(p'\cdot p)
 \right.\nonumber\\ && \left.
 -2(4a-1)\ M^2 m^2 
 +(4a-1)\left[m^2\ (P'\cdot P) + M^2\ (p'\cdot p)\right]
 \vphantom{\frac{A}{A}}\right\}\cdot
\left[k^2-\mu_G^2+i\delta\right]^{-1}.
\label{app:C.32}\end{eqnarray}
Now, $k=p'-p=P-P'$ which gives with $P^2=P^{\prime 2}=M^2$ and
$p^2=p^{\prime 2}=m^2$, 
\begin{eqnarray*}
&& k^2 = (P'-P)^2= 2M^2-2P'\cdot P = (p'-p)^2= 2m^2-2p'\cdot p, \\
&& P'\cdot P = M^2-\frac{1}{2} k^2,\ \ p'\cdot p = m^2-\frac{1}{2} k^2,   
\end{eqnarray*}
and similar expressions for $P'\cdot p, P\cdot p'$ in the Mandelstam variables
$s=(P+p)^2=(P'+p')^2$ and $u=(P-p')^2=(P'-p)^2$. 
In the C.M.-system, ${\bf P}=-{\bf p}$ and
 ${\bf P}'=-{\bf p}'$, and ${\bf k}= {\bf p}'-{\bf p}$,
 ${\bf q}= ({\bf p}'+{\bf p})/2$.
\begin{equation}
 {\bf p}= {\bf q}-\frac{1}{2} {\bf k},\ \ 
 {\bf p}'= {\bf q}+\frac{1}{2} {\bf k}.     
\label{app:C.33a}\end{equation}
The scalar products in the C.M.-system read, taking the external particles
on the energy shell {\it i.e.}, ${\bf p}^2={\bf p}^{\prime 2}$, 
giving $k^2 = -{\bf k}^2$, ${\bf q}\cdot{\bf k}=0$.     
Evaluating the scalar products in (\ref{app:C.32}) in the C.M.-system, we expand in $1/Mm$.    
Then, we obtain the leading terms,
\begin{eqnarray}
 M_{fi}^{(2)}(P',p';P,p) &\approx& -\frac{1}{4}\kappa^2\ 
 (2M m)^2\left\{ (1-a) 
 + \frac{(M+m)^2}{M^2 m^2}\ ({\bf q}^2+\frac{1}{4}{\bf k}^2)\ 
 \right.\nonumber\\ && \left.
+ \frac{(2a-1)(M^2+m^2)-2M m}{4M^2 m^2}\ {\bf k}^2 \right\}\ 
 \left({\bf k}^2+\mu_G^2-i\delta\right)^{-1}. 
\label{app:C.333}\end{eqnarray}

\begin{center}
\fbox{ \begin{minipage}[b][6.5cm][l]{12.9cm}
In \ref{app:BS}, the connection between the invariant 
amplitude $M^{(2)}_{fi}$ and the potential in the Lippmann-Schwinger (or 
Schr\"{o}dinger) equation is given by (\ref{app:BS.21})              
\begin{eqnarray}
 {\cal V}(p_f,p_i) &\approx& \frac{\pi}{2 M m}\ 
 \left(1-\frac{M^2+m^2}{4M^2 m^2} 
 ({\bf q}^2+{\bf k}^2/4)\right) M^{(2)}(p_f,p_i;W)                     
 \nonumber\\ &\approx& -\frac{\pi}{2} [\kappa^2\ M m]
 \left\{ (1-a)+ \frac{2a(M^2+m^2)-(M+m)^2}{4M^2 m^2}\ {\bf k}^2
 \right.\nonumber\\ && \left.
 + \left[\frac{(M+m)^2}{M^2 m^2} -(1-a)\frac{M^2+m^2}{4M^2 m^2}\right]
 ({\bf q}^2+\frac{1}{4}{\bf k}^2)\right\}\cdot
 \nonumber\\ && \times 
 \left({\bf k}^2+\mu_G^2-i\delta\right)^{-1}. 
\label{app:C.34a}\end{eqnarray}
\end{minipage} }
\end{center}

\vspace{5mm}

\noindent {\bf 1.\ Massless case, a=1/2}:
From (\ref{app:C.34a}) the on energy-shell potential becomes
\begin{eqnarray}
	{\cal V}_{fi}^{(0)}(P',p';P,p) &=&  
 -\frac{\pi}{4}[\kappa^2\ M m]\ 
 \left\{ 1 +\frac{7M^2+16M m+7m^2}{4 M^2 m^2}
 ({\bf q}^2+\frac{1}{4}{\bf k}^2)
 - \frac{{\bf k}^2}{Mm} \right\}\cdot
 \left[{\bf k}^2+i\delta\right]^{-1}.
\label{app:C.41a}\end{eqnarray}
The potential in momentum space from the leading term is 
\begin{eqnarray}
 {\cal V}^{(0)}(r) &=& \int\!\! \frac{d^3k}{(2\pi)^3}\ e^{i{\bf k}\cdot{\bf x}}\
 V^{(0)}({\bf p}',{\bf p}) = 
 -\frac{\kappa^2 M m}{16}\ \frac{1}{r} 
 \Rightarrow -\frac{G M m}{r}, 
\label{app:C.42}\end{eqnarray}
{\it i.e.}i, equating this potential to the Newton potential, we get,
using units $\hbar=c=1$, 
\begin{equation}
 \kappa/4 = \sqrt{G} = 1.616\ \times\ 10^{-33}\ cm\ \
 (\bm{ Planck\ length}).
\label{app:C.43}\end{equation}

\noindent {\bf 2.\ Massive case, a=1/3}:
From (\ref{app:C.34a}) the on energy-shell potential becomes
\begin{eqnarray}
	{\cal V}_{fi}^{(m)}(p_f,p_i)&=&  
 -\frac{2\pi}{6}[\kappa^2\ M m]\ 
 \left\{ 1 +\frac{5M^2+ 12 M m +5m^2}{4M^2 m^2}
 ({\bf q}^2+\frac{1}{4}{\bf k}^2) 
 \right.\nonumber\\ && \left. 
 - \frac{M^2+6M m+m^2}{2M m} \frac{{\bf k}^2}{4Mm} \right\}\cdot
 \left[{\bf k}^2+\mu_G^2\right]^{-1}
\label{app:C.34b}\end{eqnarray}
In the C.M.-system we write   
 $V^{(2)}(P',p';P,p) \equiv (2\pi)^{-6}V^{(m)}({\bf p}',{\bf p})$, and the 
configuration space potential is given by
\begin{eqnarray}
 ({\bf x}'|V|{\bf x}) &=& \int \frac{d^3k}{(2\pi)^3}\ 
 \int \frac{d^3q}{(2\pi)^3}\ 
 e^{i{\bf k}\cdot({\bf x}'+{\bf x})/2}\ e^{i{\bf q}\cdot({\bf x}'-{\bf x})}\
 V({\bf k},{\bf q}).
\label{app:C.36a}\end{eqnarray}
\noindent From (\ref{app:C.34b}) one gets, 
neglecting for the moment the ${\bf k}^2$ and ${\bf q}^2$ terms,
the C.M.-system local potential becomes
\begin{eqnarray}
 {\cal V}^{(m)}(r) &=& \int\!\! \frac{d^3k}{(2\pi)^3}\ e^{i{\bf k}\cdot{\bf x}}\
 V^{(m)}({\bf p}',{\bf p}) = 
 -\frac{\kappa^2 M m}{12}\ \frac{e^{-\mu_G r}}{r}.
\label{app:C.36b}\end{eqnarray}
Note that for very small $\mu_G$ we have 
\begin{equation}
 {\cal V}^{(m)}(r) \approx \frac{4}{3}\ {\cal V}^{(0)}(r),   
\label{app:C.44}\end{equation}
{\it i.e.}, the massless limit of the massive potential is off by a factor 4/3,
a well known fact \cite{VDV70}.

\subsection{Perihelion Precession}                            
\label{app:C.c}   
\noindent {\bf 1.\  Massless case:}
We analyze the ${\bf k}^2$- and ${\bf q}^2$-term in the 
momentum space potential (\ref{app:C.41a}), which we write as
\begin{eqnarray}
 V^{(0)}({\bf p}',{\bf p}) &=& -(4\pi G M m)\ 
 \left\{ 1 - \frac{{\bf k}^2}{Mm}
 +\frac{7(M + m)^2 +2 M m }{4M^2 m^2}
 ({\bf q}^2+\frac{1}{4}{\bf k}^2)\right\}\cdot
  \left[{\bf k}^2+\mu_G^2-i\delta\right]^{-1}.
\label{app:C.45}\end{eqnarray}
{\it Here, we introduced the graviton mass $\mu_G$ because we want to
analyze the massless limit. So, in (\ref{app:C.45}) the massless 
graviton projection operator is used in the propagator}.\\

\noindent {\bf (a)}\ The central term in (\ref{app:C.45}) 
gives the Newtonian potential
\begin{equation}
 {\cal V}^{(0)}(r) = -\left[G M m\right]\ \frac{e^{-\mu_G r}}{r}\ \
 {\rm with}\ \ \mu_G=0.
\label{app:C.45b}\end{equation}

\noindent {\bf (b)}\ The ${\bf k}^2$-terms in (\ref{app:C.34b}) 
and (\ref{app:C.45}) 
give terms with $-4\pi \delta^3({\bf r})$ which in the planetary motion
do not contribute and hence can be dropped.\\

\noindent {\bf (c)}\ The Fourier transformation to configuration space 
of the non-local $({\bf q}^2+{\bf k}^2/4)$-term is 
\footnote{ A potential term 
$V({\bf k},{\bf q})=\widetilde{V}_{n.l.}({\bf k})\left({\bf q}^2+{\bf k}^2/4\right)$ 
gives in coordinate space, see \cite{NRS78} Eqn.~(11), 
\begin{eqnarray*}
	{\cal V}^{(1)}(r) &=& -\frac{1}{2}\left(\bm{\nabla}^2{\cal V}_{n.l.}(r)
	+{\cal V}_{n.l.}(r)\bm{\nabla}^2\right).     
\end{eqnarray*}
}

\begin{equation}
\left( {\bf r}|{\cal V}^{(1)}|\psi\right) = -\frac{7(M+m)^2}{8M^2 m^2}\left(
\vphantom{\frac{A}{A}} 
 \bm{\nabla}^2 {\cal V}^{(0)}(r) + {\cal V}^{(0)}(r) \bm{\nabla^2}
\right)\ \psi({\bf r}).
\label{app:C.46}\end{equation}
\noindent The Schr\"{o}dinger equation reads
\begin{eqnarray*}
	\left(-\frac{\bm{\nabla}^2}{2m_{red}}+ {\cal V}\right)\ \psi = E\ \psi\ \
 \left(m_{red}= \frac{M m}{M+m}\right),
\end{eqnarray*}
and gives the possibility of replacement 
$\bm{\nabla}^2 \rightarrow 2m_{red}({\cal V}_C-E)$, where ${\cal V}_C$ is the total central potential. (The 
spin-spin, tensor and spin-orbit potentials in ${\cal V}$  are of order $1/Mm$ and can be neglected.) 
For (very) small $\mu_G$ we have ${\cal V}_C= -A\ G M m /r = A\ {\cal V}^{(0)}$.
From $\left[\bm{\nabla}^2,{\cal V}_C(r)\right]\psi(r) = 
 \left[ [\bm{\nabla}^2{\cal V}_C(r)]+2\bm{\nabla} {\cal V}_C(r)\cdot\bm{\nabla}\right]\psi(r)$ neglecting 
 $\bm{\nabla}\psi$ \cite{neglect.1} and the $\bm{\nabla}^2$-term, which gives $\approx \delta^3({\bf r})$
 contribution, 
the contribution of the non-local term gives the correction to the
Newton-potential
\begin{eqnarray}
	{\cal V}^{(1)}(r) &\approx&     
 -2m_{red} \frac{7(M+m)^2}{4M^2 m^2}\ {\cal V}^{(0)}(r) \left[
\vphantom{\frac{A}{A}} {\cal V}_C(r) -E \right] 
\nonumber\\ &=&
 -\frac{7}{2m}\left(1+\frac{m}{M}\right) \frac{G M m}{r} \left[
\vphantom{\frac{A}{A}} A\frac{G M m}{r} +E \right] 
 \nonumber\\ 
  & \Rightarrow & -\frac{7A}{2m}{\left[{\cal V}^{(0)}\right]^2}{m} 
 \sim 1/r^2\ \ (M \gg m).
\label{app:C.47}\end{eqnarray}
In the last step, we used that in a planetary orbit E= constant,
and the E-term in (\ref{app:C.47}) gives a correction to the
Newtonian potential, which means a small modification of the orbit.
Henceforth, being interested here only in the $1/r^2$ potential, we
omit the E-term.

\noindent We see that the non-local potential gives a  
$1/r^2$ correction leading to a {\bf perihelion-precession} which is 
$7/6$ Einstein's result! 
This agrees with the treatment of Schwinger \cite{Schw70,VDam74}.  
The remaining $-1/6$ comes from the change in the gravitational field
energy due to the presence of the planet, see for discussion 
\cite{Schw70,VDam74}.\\

\noindent {\bf 2.\  Massive case:} Following the same steps as for the massless case, taking into account
that ${\cal V}^{(m)}_C = (4/3) {\cal V}^{(0)}(r)$ and the non-local term with 5/3 instead of 7/4, we 
obtain 
\begin{eqnarray}
{\cal V}^{(1)}(r) &\Rightarrow& 
 -\frac{40A}{9m} \left[{\cal V}^{(0)}\right]^2.    
\label{app:C.47b}\end{eqnarray}

\noindent {\it Below we show that the contributions from the scalar (section \ref{app:SS}) 
and ghost (section \ref{app:SG}) to
the non-local potential cancel each other in the massless limit. Therefore, the total result for the
 $1/r^2$ correction to the Newtonian potential from the $h^{\mu\nu}$-field is given by 
 (\ref{app:C.47}). Furthermore, corrections to the perihelion precession 
 for a finite mass $\mu_G$ turn out proportional to $\mu_G^2$ and are tiny, 
 see section \ref{app:SPP}.}


\noindent We see that the non-local potential gives a  
$1/r^2$ correction leading to a {\bf perihelion-precession} which is 
$7/6\times$ Einstein's result! 
This agrees with the treatment of Schwinger \cite{Schw70,VDam74}.  
The remaining $-1/6$ comes from the change in the gravitational field
energy due to the presence of the planet, see for discussion 
\cite{Schw70,VDam74}.\\


\begin{flushleft}
\rule{16cm}{0.5mm}
\end{flushleft}
\section{Scalar contributions Perihelion Precession}                   
\label{app:S}   
In this section, we derive the contribution to the perihelion precession
 from the scalar and scalar-ghost terms in the 
$h_{\mu\nu}$-propagator $\Delta_F^{\mu\nu,\alpha\beta}(x;\mu_G^2)$
(\ref{eq:10.1b}). 
\subsection{Scalar-exchange: Perihelion Precession}                            
\label{app:SS}    
The amplitude corresponding to the Feynman graph Fig.~\ref{fig.grav1},
for scalar-exchange is, in analogy with Eqn.~(\ref{app:C.31b}), given by
\begin{eqnarray}
 S_{fi}^{(2)} &=&
 (+i)^2\ (\kappa^2/4)\ 
 {\cal N}_f(P',p')\ {\cal N}_i(P,p)\ 
 \int\frac{d^4k}{(2\pi)^4} (2\pi)^4\delta^4(P'-P+k)\ 
\cdot\nonumber\\ && \times (2\pi)^4\delta^4(p'-p-k)\cdot 
 \langle P'|T^{\mu\nu}|P\rangle\ i\Delta^{(S)}_{\mu\nu;\alpha\beta}(k)\
 \langle p'|t^{\alpha\beta}|p\rangle \nonumber\\ &=&
-i (\kappa^2/4)\ (2\pi)^4\delta^4(P'+p'-P-p)\ 
 T^{\mu\nu}(P',P)\ \Delta^{(S)}_{\mu\nu;\alpha\beta}(k)\ t^{\alpha\beta}(p',p)
 \nonumber\\ &\equiv & 
 -i(2\pi)^4\delta^4(P'+p'-P-p)\ {\cal N}_f\ M_{fi}^{(S)}\ {\cal N}_i\   
\label{app:S.11}\end{eqnarray}
where 
\begin{eqnarray}
 \Delta^{(S)}_{\mu\nu;\alpha\beta}(k) &=& 
 \frac{P^{(S)}_{\mu\nu;\alpha\beta}(k)}{k^2-\mu_G^2+i\delta},\ \ 
 P^{(S)}_{\mu\nu;\alpha\beta}(k) = 
 \frac{1}{6} \eta_{\mu\nu} \eta_{\alpha\beta}.    
\label{app:S.12}\end{eqnarray}
Up to terms of order $1/M^2$ and $1/m^2$, taking only the terms
proportional to the parameter $a$ in the expression (\ref{app:C.34a})
and putting $a=-1/6$, one finds
\begin{eqnarray}
 M_{fi}^{(2)}(P',p';P,p) &\approx& -\frac{1}{24}\kappa^2\ 
 (2M m)^2\left\{ 1  
- \frac{(M^2+m^2)-M m}{2M^2 m^2}\ {\bf k}^2 \right\}\ 
 \left({\bf k}^2+\mu_G^2-i\delta\right)^{-1}. 
\label{app:S.13}\end{eqnarray}
In Appendix~\ref{app:BS}, the connection between the invariant 
amplitude $M^{(2)}_{fi}$ and the potential in the Lippmann-Schwinger (or 
Schr\"{o}dinger) equation is given by (\ref{app:BS.21})              
\begin{eqnarray}
 {\cal V}^{(S)}(p_f,p_i) &\approx& \frac{\pi}{2 M m}\ 
 \left(1-\frac{M^2+m^2}{4M^2 m^2} 
 ({\bf q}^2+{\bf k}^2/4)\right) M^{(S)}(p_f,p_i;W)                     
 \nonumber\\ &\approx& -\frac{\pi}{12}[\kappa^2\ M m]
 \left\{ 1 - \frac{(M^2+m^2)-Mm}{2M^2 m^2}\ {\bf k}^2
 \right.\nonumber\\ && \left. \hspace{1.5cm}
  -\frac{M^2+m^2}{4M^2 m^2}
 ({\bf q}^2+\frac{1}{4}{\bf k}^2)\right\}\cdot
 \left({\bf k}^2+\mu_G^2-i\delta\right)^{-1}. 
\label{app:S.14}\end{eqnarray}
The central potential in the CM-system is, with $\kappa^2=16G$,
\begin{eqnarray}
 {\cal V}^{S}_C(r) &=& -\frac{\pi}{12}\left[\kappa^2\ M m\right]
 \int\frac{d^3k}{(2\pi)^3}\ \frac{e^{(i{\bf k}\cdot{\bf r})}}
 {{\bf k}^2+\mu_G^2} = -\frac{1}{3} \left[G M m\right] \frac{e^{-\mu_G r}}{r}.
\label{app:S.16}\end{eqnarray}

\noindent The contribution of the non-local term gives the correction to
the Newtonian potential 
\begin{eqnarray}
 {\cal V}^{(1)}(r) &\approx& +2m_{red}\frac{M^2+m^2}{4M^2m^2}\
 {\cal V}^{S}_C(r)\left[ {\cal V}_C(r) -E\right]
\nonumber\\ &\Rightarrow& +\frac{A}{2m}\left[{\cal V}^{(0)}\right]^2 =
+\frac{A}{18m} \left[G M m\right]^2\ \frac{e^{-2\mu r}}{r^2}
 \sim 1/r^2\ \ (M \gg m).
\label{app:S.17}\end{eqnarray}

\begin{flushleft}
\rule{16cm}{0.5mm}
\end{flushleft}
 \begin{figure}[hhhhtb]
 \resizebox{10cm}{11cm} 
 {\includegraphics[110,300][410,625]{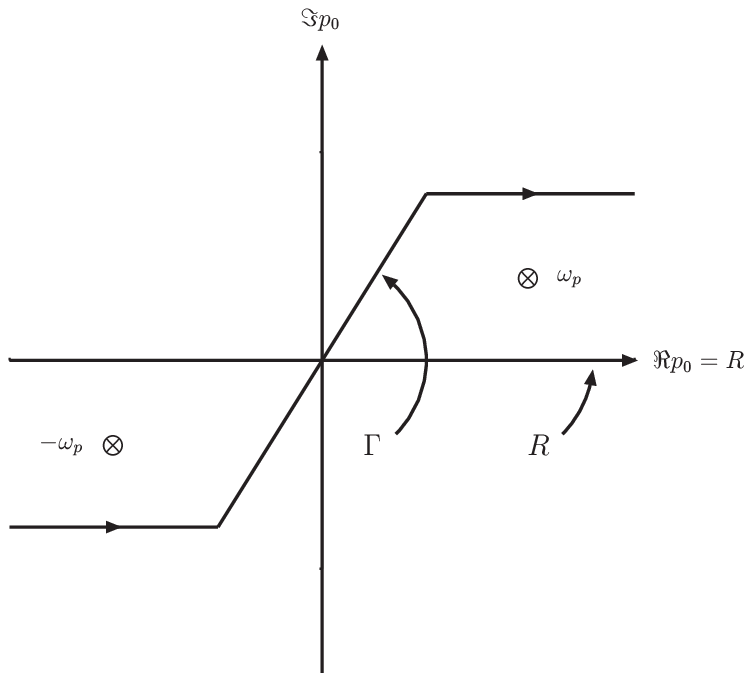}}
  \caption{\sl Complex $p_0$-plane contour 
 $\Gamma = R + \delta(-\omega_p) -\delta(\omega_p)$. Here, $\delta$-operator
 is Leray's coboundary operator.
 }                         
  \label{fig:Gamma.1}
  \end{figure}                     






\subsection{Scalar-ghost-exchange: Perihelion Precession}      
\label{app:SG}   
In \ref{app:SIG}, the details of the scalar interaction with the
imaginary-ghost field is worked out in detail, which results are 
employed in this section.\\
The amplitude corresponding to the Feynman graph Fig.~\ref{fig.grav1},
for scalar-ghost-exchange is, in analogy with Eqn.~(\ref{app:C.31b}), given by
\begin{eqnarray}
 S_{fi}^{(2)} &=&
 (+i)^2\ (\kappa^2/4)\ 
 {\cal N}_f(P',p')\ {\cal N}_i(P,p)\ 
 \int_\Gamma\frac{d^4k}{(2\pi)^4} (2\pi)^4\delta^4(P'-P+k)\ 
\cdot\nonumber\\ && \times (2\pi)^4\delta^4(p'-p-k)\cdot 
 \langle P'|T^{\mu\nu}|P\rangle\ i\Delta^{(SG)}_{\mu\nu;\alpha\beta}(k)\
 \langle p'|t^{\alpha\beta}|p\rangle \nonumber\\ &=&
-i (\kappa^2/4)\ (2\pi)^4\delta^4(P'+p'-P-p)\ 
 T^{\mu\nu}(P',P)\ \Delta^{(SG)}_{\mu\nu;\alpha\beta}(k)\ t^{\alpha\beta}(p',p)
 \nonumber\\ &\equiv & 
 -i(2\pi)^4\delta^4(P'+p'-P-p)\ {\cal N}_f\ M_{fi}^{(SG)}\ {\cal N}_i\   
\label{app:SG.11}\end{eqnarray}
where 
\begin{eqnarray}
 \Delta^{(SG)}_{\mu\nu;\alpha\beta}(k) &=& 
 -\frac{1}{6} \eta_{\mu\nu} \eta_{\alpha\beta}\ 
 \widetilde{\Delta}_F(k;-\mu^2).
\label{app:SG.12}\end{eqnarray}
Here, the imaginary ghost-mass is 
$-\mu^2 \equiv M_\epsilon^2= -2\mu_G^2$. 
The proper integration contour
 $\Gamma$, see Fig.~\ref{fig:Gamma.1}, in the complex $k_0$-plane is given as \cite{Nak72} 
$\Gamma= R-\delta(\omega_k)+\delta(-\omega_k)$, where $\delta(\pm \omega_k)$
denotes a {\it counterclockwise circle} around the poles at $\pm\omega_k$
\cite{Leray}. (Note that for a real mass $M_\epsilon$ the $\Gamma$-contour becomes the 
usual contour for the Feynman propagator $C_F$.)\\
The Feynman propagator function in (\ref{app:SG.12}) is
\begin{subequations}
\label{app:SG.13}
\begin{eqnarray}
 \widetilde{\Delta}(k;M_\epsilon^2) &=& \frac{1}{2}\left[
 \widetilde{\Delta}^{(1)}(k;M_\epsilon^2) + 
 \widetilde{\Delta}^{(2)}(k;M_\epsilon^2)\right], \\
 \widetilde{\Delta}^{(1)}(k;M_\epsilon^2) &=& \frac{1}{k^2-M_\epsilon^2},\ \
 \int\frac{d^4p}{(2\pi)^4} \rightarrow 
 \int\frac{d^3k}{(2\pi)^3}\ \int_\Gamma\frac{dk_0}{(2\pi)}, \\
 \widetilde{\Delta}^{(2)}(k;M_\epsilon^2) &=& 
 \frac{1}{k^2-M_\epsilon^{* 2}},\ \
 \int\frac{d^4p}{(2\pi)^4} \rightarrow 
 \int\frac{d^3k}{(2\pi)^3}\ \int_R\frac{dk_0}{(2\pi)}.    
\end{eqnarray}
\end{subequations}
Up to terms of order $1/M^2$ and $1/m^2$, taking only the terms
proportional to the parameter $a$ in the expression (\ref{app:C.333})
and putting $a=+1/6$, one finds
\begin{eqnarray}
 M_{fi}^{(SG)}(P',p';P,p) &\approx& +\frac{1}{24}\kappa^2\ 
 (2M m)^2\left\{ 1  
- \frac{(M^2+m^2)-M m}{2M^2 m^2}\ {\bf k}^2 \right\}\ 
 {\cal P} \frac{1}{{\bf k}^2-\mu^2}.
\label{app:SG.14}\end{eqnarray}
Here, we used $\widetilde{\Delta}(k,M_\epsilon^2)=
 {\cal P}({\bf k}^2-\mu^2)^{-1}$, 
 which is derived explicitly in Appendix~\ref{app:NAK}.\\
In Appendix~\ref{app:BS}, the connection between the invariant 
amplitude $M^{(2)}_{fi}$ and the potential in the Lippmann-Schwinger (or 
Schr\"{o}dinger) equation is given by (\ref{app:BS.21})              
\begin{eqnarray}
 {\cal V}^{(SG)}(p_f,p_i) &\approx& \frac{\pi}{2 M m}\ 
 \left(1-\frac{M^2+m^2}{4M^2 m^2} 
 ({\bf q}^2+{\bf k}^2/4)\right) M^{(SG)}(p_f,p_i;W)                     
 \nonumber\\ &\approx& +\frac{\pi}{12}[\kappa^2\ M m]
 \left\{ 1 - \frac{(M^2+m^2)-Mm}{2M^2 m^2}\ {\bf k}^2
 \right.\nonumber\\ && \left. \hspace{1.5cm}
  -\frac{M^2+m^2}{4M^2 m^2}
 ({\bf q}^2+\frac{1}{4}{\bf k}^2)\right\}\cdot
 {\cal P} \frac{1}{{\bf k}^2-\mu^2}.
\label{app:SG.15}\end{eqnarray}
The central potential in the CM-system is
\begin{eqnarray}
 {\cal V}^{SG}_C(r) &=& +\frac{\pi}{12}[\kappa^2\ M m]
 \int\frac{d^3k}{(2\pi)^3}\ {\cal P}\ \frac{e^{(i{\bf k}\cdot{\bf r})}}
 {{\bf k}^2-\mu^2} = +\frac{1}{3} \left[G M m\right] \frac{\cos(\mu r)}{r}.
\label{app:SG.16}\end{eqnarray}

\noindent The contribution of the non-local term gives the correction to
the Newtonian potential 
\begin{eqnarray}
 {\cal V}^{(1)}(r) &\approx& +2m_{red}\frac{M^2+m^2}{4M^2m^2}\
 {\cal V}^{(SG)}_C(r)\left[ {\cal V}_C(r) -E\right]
	\nonumber\\ &\Rightarrow& -\frac{A}{2m}\left[{\cal V}^{(SG)}_C {\cal V}^{(0)}\right] =
+\frac{A}{18m} \left[G M m\right]^2\ \frac{\cos(\mu r)}{r^2}
 \sim 1/r^2\ \ (M \gg m).
\label{app:SG.17}\end{eqnarray}

\begin{center}
\fbox{\rule[-5mm]{0cm}{1.0cm} \hspace{5mm}
\begin{minipage}[]{12.9cm}
{\it The total result for the non-local $1/r^2$-potential from massive and scalar-ghost is
\begin{eqnarray}
	{\cal V}^{(1)}(total) &=& {\cal V}^{(1)}_m +{\cal V}^{(1)}_{SG} 
	\rightarrow +\frac{7A}{2m}\left[{\cal V}^{(0)}\right]^2
\label{app:SG.18}\end{eqnarray}
for $\lim \mu_G \rightarrow 0$. Below, we will show that the correction for the
finite graviton mass $\sim \mu_G^2$, which is very small.
}
 \end{minipage} \hspace{5mm} 
 }\\
\end{center}

\begin{flushleft}
\rule{16cm}{0.5mm}
\end{flushleft}
\section{Results Perihelion Precession }                             
\label{app:SPP}   
Taking the definition of the gravitational constant as that which occurs
in the massless case $\mu_G=0$, i.e. $\kappa^2=16G$, 
the gravitational potential in momentum space is of the form
\begin{eqnarray}
{\cal V}(p_f,p_i) &=& -4\pi\ [G M m]\left\{A+B {\bf k}^2/m^2 
	-C \left({\bf q}^2+\frac{1}{4}{\bf k}^2\right)/m^2\right\}\cdot\nonumber\\
 && \times (2\pi)^{-6} \left({\bf k}^2 +\mu_G^2-i\delta\right).
\label{SPP.1}\end{eqnarray}
Here A,B, and C contain the total contributions, {\it i.e.} $A = \sum_i A_i$ etc.,
where $A_i,B_i,C_i$ come from the individual contributions.
In coordinate space, we get for the central and non-local potential
\begin{subequations}\label{SPP.2} 
\begin{eqnarray}
	{\cal V}^{(0)}(r) &\equiv& -[G M m]\ \frac{e^{-\mu_G r}}{r}\ ,\
	{\cal V}_C(r) = A\ {\cal V}^{(0)}(r), \\             
	\left({\bf r}|{\cal V}^{(1)}|\psi\right) &=& +\frac{1}{2}\ C\
	\biggl(\bm{\nabla}^2{\cal V}^{(0)}(r)+{\cal V}^{(0)}(r)\bm{\nabla}^2\biggr)/m^2.
\end{eqnarray}\end{subequations}
Making the approximation $\bm{\nabla}\psi \approx 0$,  and 
	$\bm{\nabla}^2{\cal V}^{(0)} \sim \delta^3({\bf r}) \rightarrow 0$
	for the planetary motion, leads to   
$[\bm{\nabla}^2,{\cal V}^{(0)}(r)] \approx 0$. Then,        
as described above, using the Schr\"{o}dinger equation one makes the replacement
$\bm{\nabla}^2 \rightarrow 2m_{red}({\cal V}_C-E)$ where $m_{red}=m$.
Then, we arrive at the correction to the Newtonian potential, see \cite{NRS78} Eqn.~(33),
\begin{eqnarray}
	{\cal V}^{(1)}(r) &\approx& 2m_{red}\ \left(C/m^2\right)\ 
	{\cal V}^{(0)}(r)\bigl[ {\cal V}_C(r)-E\bigr] \nonumber\\ 
        & \Rightarrow & 2CA\ \frac{\left[{\cal V}^{(0)}\right]^2}{m} 
	= 2\left(AC/m\right)\ [G M m]^2/r^2\ (M \gg m).
\label{SPP.3}\end{eqnarray}
As explained above the contribution of the ${\bf k}^2$-term can be neglected in the 
central potential.\\
\begin{table}[hhhhhbt]
\begin{center}
	\caption{ Coefficients $A,B,C$ for the different exchange types}
	\label{tab:coef-exch}
\begin{tabular}{cc|c|c|c|c} \hline
	\multicolumn{2}{c|}{ Exchange} & Propagator & A & B & C \\ \hline 
	I: & Massless & $\Delta_{\mu\nu,\alpha\beta}^{(0)}$ & 1 & $\frac{m}{M}$ & $-\frac{7}{4}$ \\                
	\hline
	II: & Massive  & $\Delta_{\mu\nu,\alpha\beta}^{(m)}$ & $\frac{4}{3}$ & $\frac{m}{M}-\frac{1}{6}$ & $-\frac{5}{3}$ \\                
	III: & Ghost   & $\Delta_{\mu\nu,\alpha\beta}^{(SG)}$ & $-\frac{1}{3}$ & $+\frac{1}{6}$ &$-\frac{1}{12}$ \\                
	\hline
	IV: & "Scalar" & $\Delta_{\mu\nu,\alpha\beta}^{(S)}$ & $\frac{1}{3}$ & $-\frac{1}{6}$ &$+\frac{1}{12}$ \\                
\hline
\end{tabular}
\end{center}
\end{table}
In Table~\ref{tab:coef-exch} the coefficients A,B, and C
are listed for the exchanges calculated in this paper.
	The propagators are explicitly defined in Eqn's (\ref{eq:10.1a}-\ref{eq:10.1c}).
	Adding the contributions II and III gives I, as expected.

	\vspace*{2mm}

	\begin{center}
\fbox{\rule[-5mm]{0cm}{1.0cm} \hspace{5mm}
\begin{minipage}[]{12.9cm}
	\vspace{2mm}

{\red The results in Table~\ref{tab:coef-exch} show that in the $\lim_{\mu_G \rightarrow 0}$ the $h^{\mu\nu}$ 
propagator leads to the massless graviton contribution for the perihelion precession, due to the 
combination of the massive spin-2 and ghost contributions: from Table~\ref{tab:coef-exch} we obtain 
$AC=(4/3-1/3)*(-5/3-1/12)= -7/4$, which corresponds to the value for the massless propagator. 
Then, ${\cal V}^{(1)}(r) = 2AC[{\cal V}^{(0)}(r)]^2/m= -(7/2m) [{\cal V}^{(0)}(r)]^2$.
}\\
\end{minipage} \hspace{5mm} }\\
\end{center}

\noindent In this treatment, using the ${\cal V}^{(1)}$-interaction the change in the gravitational field
energy due to the presence of the sun and the planet is not included. In Appendix~\ref{app:PC3}
we review the derivation of this effect, which gives a contribution $\left[{\cal V}^{(0)}\right]^2/2m$.
Including this we get in total ${\cal V}^{(1)} = -(3/m) \left[{\cal V}^{(0)}\right]^2$, which
agrees with Einstein's result.\\

		\begin{center}
\fbox{ \begin{minipage}[b][6.5cm][l]{14cm}
\noindent {\bf Remark}: 
{\it We note that this $1/r^2$ correction to the Newton potential
does agree with the $-(3V^2/m+V^2/2m)$-correction in Ref.~\cite{VDam74} 
below formula (52). In \cite{VDam74,Schw70} the $-3V^2/m$ comes from
\begin{eqnarray*}
 E_{int}&=& -\frac{2GM}{r}\left(\sqrt{p^2+m^2}
 -\frac{1}{2}\frac{m^2}{\sqrt{p^2+m^2}}\right) \nonumber\\
  &\approx& -\frac{GMm}{r}\left[1+\left(1+\frac{1}{2}\right)\frac{p^2}{m}
 \right],
\label{app:C.51}\end{eqnarray*}
and $-V^2/2m$ comes from the relativistic correction to the kinetic energy
of the planet: $T \rightarrow \sqrt{p^2+m^2}-m$ replacing T by 
$T-T^2/2m \sim T-(E-V)^2/2m$.
}
\end{minipage} }\\
		\end{center}
\vspace*{2mm}

\noindent The planetary equation for the massless graviton, with the inclusion of the
gravitational field energy between the planet and the sun reads,
see \cite{Schw70} Eqns~(2-4.55)-(2-4.60), 
\begin{subequations} \label{SPP.6}
\begin{eqnarray}
 \frac{d^2u_0}{d\varphi^2} + u_0&=&-\frac{1}{L_1^2}\frac{d}{du}\frac{V^{(0)}_{eff}}{m}, \\
 V^{(0)}_{eff} &=& V-3V^2/m,\ V=-GMm/r, 
\end{eqnarray}\end{subequations}
where $u=1/R \equiv 1/d$ and $L_1$ is the angular momentum per unit planetary mass.
This gives (in units c=1) 
\begin{equation}
\frac{d^2u_0}{d\varphi^2} + \left(1-\frac{6G^2M^2}{L_1^2}\right)u_0 = 
\frac{GM}{L_1^2},
\label{SPP.7}\end{equation}
	leading to the non-Newtonian correction to the perihelion precession angle 
\begin{equation}
\Delta\varphi_E = 6\pi G^2M^2/L_1)^2 = 
	6\pi \left(1+\frac{m}{M}\right) G M/L,
\label{SPP.8}\end{equation}
	where $L^{-1}=(1/r_+ +1/r_-)/2= \mu G M m/J^2$. 
	Here, $r_\pm$ are the aphelion and perihelion distances.
The connection with Einstein's result \cite{Ein16} is given by the relation
$ GM/L_1 = 2\pi(a/T)(1-e^2)^{-1/2}$, where a is the semimajor axis,
T is the period, and $e$ is the eccentricity.

\noindent In Table~\ref{tab:solar-system} the results for the $\Delta\varphi_E$ 
	correction to the perihelion precession
per century are listed for Icarus, Mercury, Venus, Earth, and Mars. 
	Here, and in the following Table, the astronomical data are taken from Ref.~\cite{SW1992}.

\begin{table}[hbt]
\begin{center}
	\caption{ The Solar System. Mass planet $m_{pl}$ in earth masses, $r_{min}(AU)$,
	$\epsilon$ eccentricity, $J= 2\pi m_{pl} r_{min}^2/T$: orbital angular momentum (units
	$10^{40} kg\ m^2/s$), 
	$L_1$: angular momentum per unit mass ( units $10^{16} m^2\ s^{-1})$,
	n (orbits per century), $n\Delta\varphi$: (arc sec/century),
	and GRT/RFT results. Earth mass $M_{\oplus}=5.97\times 10^{24}$ kg 
	and the solar mass $M_{\odot}= 1.98892\times 10^{30}$ kg.
	}
\label{tab:solar-system}
\begin{tabular}{c|c|c|c|c|c} \hline
Planet& Icarus & Mercury & Venus & Earth & Mars \\ \hline
	Mass  & $0.245 10^{-12}$ & 0.055 &  0.820 &  1.000 &  0.110 \\        
$r_{min}$(AU) & 0.186 & 0.307 & 0.717 & 0.981 & 1.524 \\       
$\epsilon$ & 0.827 & 0.206 & 0.0068 & 0.0167 & 0.0915\\       
	Period & 409  & 87.97 &  224.70 &  365.26 & 686.98 \\      
	$J$  & $0.157\times 10^{-12}$ & 0.091 & 1.80 & 2.70 & 0.35 \\         
	$L_1$  & 0.48 & 0.111 & 0.154 & 0.182 & 0.235 \\        
 n     & 89.3 & 415.2 & 162.5 & 100.0 & 53.17  \\        
 $n\Delta\varphi$ & 9.8 $\pm$ 0.8 & 43.1 $\pm$ 0.5 & 8.4 $\pm$ 4.8 & 5.0 $\pm$ 1.2 & 1.52 \\
 GRT/RFT & 10.0 & 43.0 & 8.6 & 3.8 & 1.63 \\              
\hline
\end{tabular}
\end{center}
\end{table}
\begin{table}[hbt]
\begin{center}
	\caption{ The Solar System II. Mass planet $m_{pl}$ in earth masses, $r_{min}(AU)$,
	$\epsilon$ eccentricity, $J= 2\pi m_{pl} r_{min}^2/T$: orbital angular momentum (units
	$10^{40} kg\ m^2/s$), 
	$L_1$: angular momentum per unit mass ( units $10^{16} m^2\ s^{-1})$,
	n (orbits per century), $n\Delta\varphi$: (arc sec/century),
	GRT/FT results. Earth mass $M_{\oplus}=5.97\times 10^{24}$ kg 
	and the solar mass $M_{\odot}= 1.98892\times 10^{30}$ kg.
	Program calculation $^*)$, Literature $^{**})$. 
	}
\label{tab:solar-system2}
\begin{tabular}{c|c|c|c|c} \hline
	Planet& Jupiter & Saturn & Uranus & Neptune \\ \hline
	Mass  &  318 & 95.4  & 14.5   & 17.1 \\                   
$r_{min}$(AU) & 4.995 & 9.041 & 18.330& 29.820 \\       
$\epsilon$ & 0.0484& 0.0539& 0.0473 & 0.0095 \\       
	Period & 4333 & 10759 & 30687   & 60190  \\      
	$J$  & 791 & 321   & 69.2 & 102 \\         
	$L_1$  & 0.416& 0.562 & 0.799 & 1.000 \\        
 n     & 8.42 & 3.40  & 1.19  & 0.61  \\        
	$n\Delta\varphi ^{*}$ & 0.634$\times 10^{-1}$ &0.137$\times 10^{-1}$ & 0.240$\times 10^{-2}$ & 
	0.777$\times 10^{-3}$\\
	GRT/RFT$^{**}$   & 0.621$\times 10^{-1}$ & 0.136$\times 10^{-1}$ & 0.237$\times 10^{-2}$ & 
	0.773$\times 10^{-3}$\\
\hline
\end{tabular}
\end{center}
\end{table}

\subsection{Finite-mass and Ghost correction Perihelion-precession}                 
\label{sec:10.a}   
The finite-mass corrections are due to the differences
\begin{subequations} \label{eq:10.11} 
\begin{eqnarray}
 \delta^{(1)}\Delta_F^{\mu\nu,\alpha\beta}(x-y;\mu_G^2) &=& 
 \frac{1}{2}\left(\eta^{\mu\alpha}\eta^{\nu\beta}+\eta^{\mu\beta}\eta^{\nu\alpha}
 -\eta^{\mu\nu} \eta^{\alpha\beta}\right) \cdot\nonumber\\ && \times
 \left[\Delta_F(x-y;\mu_G^2) -\Delta_F(x-y;\mu_G^2=0)\right], \\
 \delta^{(2)}\Delta_F^{\mu\nu,\alpha\beta}(x-y;\mu_G^2) &=& 
 \frac{1}{6}\eta^{\mu\nu}\eta^{\nu\alpha}
 \left[\Delta_F^{(S)}(x-y;\mu_G^2) -\Delta_F^{(SG)}(x-y;\mu_G^2=0)\right].    
\end{eqnarray} 
\end{subequations}
From the inspection of Schwinger's computation \cite{Schw70} Eqs.~(2-4.36,37),
\begin{eqnarray*}
&&  E_{int}(y^0) = -GM \int d^3y\ \frac{1}{|{\bf x-y}|}
 \left[t^{00}-t_{kk}\right](y^0,{\bf y}),
\end{eqnarray*} 
and using (\ref{eq:10.11}) we get
\begin{subequations} \label{eq:10.12} 
\begin{eqnarray}
\delta^{(1)} E_{int}(y^0) &=& -GM \int d^3y\ \frac{1}{|{\bf x-y}|}
 \left[e^{-\mu_G |{\bf x-y}|}-1\right] t^{00}(y^0,{\bf y}), \\
\delta^{(2)} E_{int}(y^0) &=& -\frac{1}{3}GM \int d^3y\ \frac{1}{|{\bf x-y}|}
 \left[e^{-\mu_G |{\bf x-y}|}-\cos(\mu_G|{\bf x}-{\bf y}|)\right] t^{00}(y^0,{\bf y}).
\end{eqnarray} 
\end{subequations}
With the Sun in the center ${\bf x}=0$, using the limit $\mu_G < 10^{-57}m_e$
and for the distance Mercury-Sun 
$ d := |{\bf r}_{\bigodot} -{\bf r}_m| \approx 60\ 10^{60} m = 1.44\ 10^{23} \hbar/m_e c$, we 
find that $\mu_G|{\bf y}| < 1.44\ 10^{-34} <<<< 1$. 
Let $t^{00}(y^0,{\bf y}) \sim m/V_m$, i.e. a homogeous and static mass distribution inside
Mercury. Then,
\begin{eqnarray}
\delta E_{int}(y^0) &=& \delta^{(1)} E_{int}(y^0) + \delta^{(2)} E_{int}(y^0) 
= G M m\biggl[ \left(e^{-\mu_G d}-1\right)
 +\frac{1}{3}\left(e^{-\mu_Gd}-\cos(\mu_Gd)\right) \biggr]/d 
 \nonumber\\ 
 &\approx& [GMm] \left[-\frac{4}{3} +\frac{5}{6} (\mu_G d)\right] \mu_G,
\label{eq:10.13}\end{eqnarray} 
where we expanded the exponential and cosine in (\ref{eq:10.13}) keeping only terms up to the
quadratic ones in the graviton mass.
The first term in (\ref{eq:10.13}) adds a constant to the potential energy, 
which gives no contribution to the gravitational force and hence no contribution 
to the perihelion-precession
\footnote{
Mass of the Sun M$_{\bigodot}$ = M = 1.99\ 10$^{30}$ kg, 
mass of the Earth M$_{\bigoplus}$ = 5.97\ 10$^{24}$ kg, 
mass of Mercury m= 0.053 M$_{\bigoplus}$, 
electron mass m$_e$= 9.11\ 10$^{-31}$ kg. 
}.\\
The second term: the equation of an orbit becomes     
\begin{equation}
 \frac{d^2u}{d\varphi^2} + u = -\frac{1}{L_1^2}\frac{d}{du}\frac{V_{eff}}{m},\
 V_{eff}/m = V^{(0)}_{eff}/m + \frac{5}{3}GM\ \mu_G^2/u,
\label{eq:10.14}\end{equation} 
where $u=1/R \equiv 1/d$ and $L_1$ is the angular momentum per unit planetary mass.
From (\ref{eq:10.13}) we have
\begin{subequations} \label{eq:10.15} 
\begin{eqnarray} 
 V_{eff}/m) &=& -GM\ u -3G^2M^2 u^2 +\frac{5}{3}GM\ \mu_G^2/u, \\
 \frac{d}{du}\frac{V_{eff}}{m} &=& -GM -6G^2 M^2 u
 -\frac{5}{3}G M \left(\frac{\mu_G}{u}\right)^2,
\end{eqnarray} 
\end{subequations}
so that 
\begin{equation}
 \frac{d^2u}{d\varphi^2} + \left(1-\frac{6G^2M^2}{L_1^2}\right)\ u =
\frac{GM}{L_1^2} +\frac{5}{3L_1^2}\left(\frac{\mu_G}{u}\right)^2.          
\label{eq:10.17}\end{equation} 
If $u_0$ is the orbit for $\mu_G=0$ we have $1/u^2 \approx 1/u_0^2-2\Delta u/u_0^3$.
Writing (\ref{eq:10.17}) as 
\begin{eqnarray*}
&& u'' + A u = B+C/u^2 \approx B+C\mu_G^2\left(\frac{1}{u_0^2}-2\frac{u-u_0}{u_0^3}\right),\ {\rm or}\\
&& u'' + (A+2C\mu_G^2/u_0^3) u \approx B+3C\mu_G^2/u_0^2.
\end{eqnarray*} 
and (\ref{eq:10.17}) can be written approximately as
\begin{eqnarray}
 \frac{d^2u}{d\varphi^2} + 
 \left(1-\frac{6G^2M^2}{L_1^2}+\frac{10GM}{3L_1^2}\frac{\mu_G^2}{u_0^3}\right)\ u   
&=& \frac{GM}{L_1^2}\left( 1+ 5\left(\frac{\mu_G}{u_0}\right)^2\right). 
\label{eq:10.18}\end{eqnarray} 
The factor $(...)$ on the r.h.s. means a constant multiple of the Newtonian
potential $V= -G Mm /R$ and does not produce a perihelion precession, it
only changes slightly the scale of the orbit. 
The factor on the LHS multiplying the u-term brings a scaling factor for
the angle, and leads to a shift in the perihelion precession, quadratic in the graviton mass,
\begin{equation}
 \delta\varphi = -2\pi\frac{5GM}{3L_1^2}\frac{\mu_G^2}{u_0^3}. 
\label{eq:10.19a}\end{equation} 
The ratio with the $\Delta\varphi_E$ (\ref{SPP.8}) is           
\begin{equation}
 \frac{\delta\varphi}{\Delta\varphi_E} = -\frac{5}{9}(\mu_G R)^2
 \left(\frac{GM}{Rc^2}\right)^{-1}.
\label{eq:10.19c}\end{equation} 
Now, $G m_F^2/\hbar c$ is dimensionless, which implies that
\begin{eqnarray*}
 \frac{GM}{Rc^2} &=& \left[G \frac{m_F^2}{\hbar c}\right]\ 
 \frac{M}{m_F} \left[\frac{\hbar}{m_Fc}/R\right]
\label{eq:10.19d}\end{eqnarray*}
is also dimensionless. Here $m_F= \hbar c =197.32$ MeV is the Fermi mass.
So, in atomic units $\hbar=c=1$ we have
\begin{equation}
 \frac{\delta\varphi}{\Delta\varphi_E} = -\frac{5}{9}
 \left(\frac{\mu_G}{m_F}\right)^2 (m_FR)^3\frac{m_F}{M}
 \left[G m_F^2\right]^{-1}.
\label{eq:10.19e}\end{equation} 
{\it\bf The correction to Einstein's result is proportional
to the square of the graviton mass and vanishes for $\mu_G \rightarrow 0$}\\

\noindent {\bf Estimation}: Using 
$\sqrt{G}= 1.62\!\times\! 10^{-33}\ cm = 1.62\!\times\! 10^{-20}\ fm$, one has     
 $G m_F^2 \approx 2.7\!\times\!10^{-40}$. The ratio $m_F/M \approx 400 m_e/M =
 2\!\times\!\! 10^{-58}$. 
Assuming $R \approx 10^8 km$, gives $m_F R \approx 10^{26}$, 
insertion these numbers in (\ref{eq:10.19e}) gives
\begin{equation}
 \frac{\delta\varphi}{\Delta\varphi_E} \approx  -\frac{5}{9}
 \left(\frac{\mu_G}{m_F}\right)^2\cdot 10^{+78}\cdot 2\!\times\!\! 10^{-58}\cdot
 10^{+40}/2.7 \approx -\frac{10}{27}
 \left(\frac{\mu_G}{m_F}\right)^2\cdot 10^{60}.
\label{eq:10.19f}\end{equation} 
From more recent work \cite{Damour91,Finn02,Choud04,Gold10} 
the upper limit for the graviton mass seems to be 
$\mu_G \leq 7\!\times\!\! 10^{-32}\ {\rm eV}\ = 2\!\times\!\! 10^{-38}\ m_e 
\approx 0.5\!\times\!\! 10^{-40}\ m_F$. This gives
$ \left|\delta\varphi/\Delta\varphi_E\right| \leq 10^{-21}$, 
which is very tiny.\\
The Einstein correction $\Delta\varphi_E= 43^{\prime\prime}.03$/century, 
and experiment gives 
$\Delta\varphi_{exp} = 41^{\prime\prime}.4 \pm 0^{\prime\prime}.90$/century
 \cite{MTW73}.
For a deviation of the order of the error 
$\delta\varphi/\Delta\varphi_E \approx 0.01$ we find from
(\ref{eq:10.19e}) $\mu_G \approx 10^{-28} m_e$, {\it i.e.} ten orders of 
magnitude larger than the upper limit above.

\begin{flushleft}
\rule{16cm}{0.5mm}
\end{flushleft}

\subsection{Non-Newtonian Modified Gravity}                                             
\label{sec:10.c}   
The non-relativistic two-body gravitational potential is
\begin{eqnarray}
 V(r) &=& V_N(r)+V_{SC}(r)+V_{SG}(r) = -[GMm]\frac{e^{-\mu_Gr}}{r}-\frac{GMm}{3r}
 \left[ e^{-\mu_G r}- \cos(\mu_G r)\right].
\label{eq:10.31}\end{eqnarray}
with $\mu_G \approx 10^{-40} m_F$. The range of the Yukawa part
is $r_0= 10^{25} m \approx 10^9$ light year.
The correction to the Newton-potential in (\ref{eq:10.31}) is 
$\Delta V(r) = (2GMm/3c^2)\left[ 1+\mu_G r/4 + .... \right]\mu_Gc^2 $,
having a long-range repulsive part, which is very small.
For $\mu_G d \approx 1 \rightarrow d \approx 2\times 10^9$ ly, whereas the radius of the 
 universe $ r_U \approx 46.5\times 10^9$ ly.

\begin{flushleft}
\rule{16cm}{0.5mm}
\end{flushleft}

\section{Discussion and Conclusions}                                      
\label{sec:11}   
In studying the spin-2 theory with the axilary-field method, exploiting a spin-1 and a scalar spin-0, we
found, analyzing the possible theories, only in the limiting case 
$b \rightarrow \infty$, a model with a massive spin-2 field combined
with an imaginary-ghost spin-0 field has a smooth and proper massless limit. This
	gives a (linearized) relativistic gravitation theory (RGT-AF) in Minkowski space. 
In the framework of RGT-AF we performed the expansion in the (small) graviton mass, without
destroying the correct prediction for the perihelion precession of Mercury. 
Therefore, in the treatment of the massive spin-2 field with auxiliary 
fields and the Dirac quantization method, a
continuous and smooth change in the perihelion precession as a function of the
graviton mass can be realized, albeit necessitating to introduce an imaginary ghost-field.
This ghost-field $\epsilon(x)$ plays a crucial role in (i) the field commutators, (ii) the Feynman
propagator, and (iii) the correct massless limit. In this paper, we have
the quantization of the imaginary-ghost field performed, 
the Gupta subsidiary-condition discussed   
for the Hilbert-space structure, the ghost-propagator analyzed, and
the contribution to the perihelion precession calculated for the massive graviton.
In the massless limit, it converts the massive spin-projection operator into the
massless one, which is very important for the LRGT-AF model.\\
The role of the $\epsilon$-field is very similar to that of the B-field in 
the massive vector-field \cite{NO90}. In \cite{NO90} section (2.4.2), the field equations are
\begin{eqnarray*}
	&& (\Box + m_V^2) A^\mu -(1-\alpha) \partial^\mu B = -ej^\mu\ ,\  
	 \partial^\mu A_\mu + \alpha B =0,
\end{eqnarray*}
which are similar to equations (\ref{HMUNU.4}). Also, the vector-field commutator
has the vacuum expectation value
\begin{eqnarray*}
	&& \left(0|\left[A_\mu(x),A_\nu(y)\right]|0\right)= im_V^{-2}\partial_\mu\partial_\nu
	\Delta(x-y;\alpha m_V^2) \\ &&
	+i\int_0^\infty ds \rho(s)\left(-\eta_{\mu\nu}-s^{-1}\partial_\mu\partial_\nu\right)\
	\Delta(x-y;s),
\end{eqnarray*}
which is the parallel of (\ref{HMUNU.6}). Similarly, for the Feynman propagator. So, 
the function of the ghost-fields $B(x)$ and $\epsilon(x)$ is very much the same. 
A notable difference is the imaginary $\epsilon$-mass. 
They participate in the coupling to the electric current,
 respectively to the matter-energy-momentum tensor.
The introduction of asymptotic fields reveals the physical contents of LRGT-AF, as shown in the spectral
analysis of the theory. It shows that in the $\lim M \rightarrow 0$ also the 
"renormalized" mass vanishes with the ratio $M_r/M \rightarrow 1$.


The eventual connection with the cosmological constant gives strong conditions
on the possible mass $\mu_G$ of the graviton, leading to negligible corrections.
Also, the non-Newtonian corrections are tiny for the Solar system.\\
 In contrast to the RGT of Ref.~\cite{Log04}, which needs to have $\mu_G \ne 0$ in essence, 
 the introduction of the imaginary scalar ghost does not have dramatic consequences 
 for the black-holes.
 In the RTG calculation of the gravitational effects in the Solar system,
 the graviton mass $\mu_G$ is set to zero, see \cite{Log04} chapter 11. Hence, 
 the vDVZ-discontinuity is not addressed.

 Alternative solutions to the vDVZ-discontinuity problem in massive gravitation theory
 have been tried \cite{Por01,Kog01,Duff01}. 
 For more recent work, see {\it e.g.} 
 (i) massive conformal gravity (MCG) \cite{Faria2019}, 
 (ii) 4D gravity on a brane in 5D Minkowski space \cite{Por2000},
 and (iii) non-linearized gravity \cite{Dalmazi23}.

 The general-relativity theory (GRT) \cite{Ein16} has proven to be very successful in
describing an impressive number of phenomena, particularly in astrophysics and
cosmology. The experimental search for deviations from GRT has been 
extensive, but till now there is no sign of the presence of a Yukawa component in the 
graviton-exchange potential \cite{Gold10,FT98}, with the upper limit 
$\mu_G \le 2\times 10^{-38}m_e$ for the gravition mass.
Nevertheless, the vDVZ-discontinuity for very tiny graviton masses seems unreal and unsatisfactory.\\

 {\it In the study of massive gravity in this paper,
 with auxiliary scalar- and vector-ghost fields, 
 it appeared that it is not possible to
remove the vDVZ-discontinuity within the context of "standard" field theory models. 
In the "non-standard" spin-2 model in this paper, having the imaginary scalar-ghost,
the vDVZ-discontinuity disappears.
The quantization of the complex scalar-ghost field 
needs further study, see notes on this in Ref.~\cite{Nak72}.
}

\appendix

\section{Group Theoretical Intermezzo}
\label{app:B}   
In this appendix we describe the relation between the little group L(p) for the
four-vector $\bar{p}^{\mu}=(p^0,0,0,p)$ and          
the rotation-group $SO(3)$ and the Eucledian group in two-dimensions $E(2)$.
The latter are the little (or invariance) groups for the four-vectors
$\bar{p}^\mu=(p^0=M,0,0,0)$ respectively 
$\bar{p}^\mu = (p,0,0,p)$. The first denotes the 4-momentum of a 
particle of mass M in rest, and the second one the 4-momentum of a 
massless particle with $p^2=0$. To connect these two cases, we consider
the Lie algebra pertinent to the 4-vector $p^\mu=(p^0,0,0,p)$. A basis for this
Lie-algebra is given by
\begin{eqnarray}
 L_1 &=& J_1 + \tanh \chi_p\ K_2\ \ ,\ \ L_2 = J_2 - \tanh \chi_p\ K_1\ \ ,\ \
 J_3\ ,
\label{B.1}\end{eqnarray} 
where $\cosh\chi_p=p^0/M, \sinh\chi_p=p/M$. The elements $K_1,K_2$ are the
generators of the special Lorentz transformation along the x-, respectively
the y-axis, and $J_3$ for the rotations in the xy-plane. The Lie algebra for the
Lorentz group is  
\begin{subequations}
\label{B.2} 
\begin{eqnarray}
 \left[J_i,J_j\right] &=& i \epsilon_{ijk}\ J_k\ , \\
 \left[J_i,K_j\right] &=& i \epsilon_{ijk}\ K_k\ , \\
 \left[K_i,K_j\right] &=& i \epsilon_{ijk}\ J_k\ .        
\end{eqnarray} 
\end{subequations}
Using this algebra, we derive the Lie algebra for L(p): 
\begin{subequations}
\label{B.3} 
\begin{eqnarray}
\left[J_3,L_1\right] &=& \left[J_3,J_1\right] + \tanh\chi_p\left[J_3,K_2\right]
\nonumber\\ &=& i\left(J_2-\tanh\chi_p K_1\right) = i L_2\ , \\
\left[J_3,L_2\right] &=& \left[J_3,J_2\right] - \tanh\chi_p\left[J_3,K_1\right]
\nonumber\\ &=& -i\left(J_1+\tanh\chi_p K_2\right) = -i L_1\ , \\
\left[L_1,L_2\right] &=& \left[J_1,J_2\right] - \tanh^2\chi_p\left[K_2,K_1\right]
\nonumber\\ &=& i \cosh^{-2}\chi_p J_3 = i \frac{M^2}{p^2+M^2}\ J_3\ .
\end{eqnarray} 
\end{subequations}
We find from this algebra:\\

\noindent 1.\ \underline{Massive case}: For the particle at rest ${\bf p}=0$ and taking as 
a basis the elements
\begin{equation}
A_1 = \cosh\chi_p\ L_1\ \ ,\ \ A_2=\cosh\chi_p\ L_2\ \ ,\ \ A_3 = J_3\ ,
\label{B.5}\end{equation} 
which for $p=0$ satify the Lie-algebra isomorphic to $SO(3)$:
\begin{equation}
\left[A_i,A_j\right] = i \epsilon_{ijk}\ A_k\ .
\label{B.6}\end{equation} 

\noindent 2.\ \underline{Massless case}: For a massless particle $M=0$, and the algebra     
in (\ref{B.3}) reduces to a Lie-algebra isomorphic to the 
Euclidean group in two dimensions $E_2$
\begin{equation}
\left[L_1,J_3\right] = -i L_2\ \ ,\ \              
\left[L_2,J_3\right] = +i L_1\ \ ,\ \              
\left[L_1,L_2\right] = 0\ .                        
\label{B.7}\end{equation} 
The consequence of the abelian subalgebra, spanned by $(L_1, L_2)$, 
is that the helicities $\lambda$ for massless particles of spin j 
can only assume the values $\lambda = \pm j$.
Note, that since the representations of $E_2$ are 1-dimensional, orthochronous\
Lorentz transformation $L^\uparrow_+$ do not mix massless 
neutrino's and antineutrino's.

\noindent The representations of the $E_2$-group are given in Ref.~\cite{Tal68}.
For spin-2, the difference between the number of helicity states for the massive  
and massless case, 5 and 2, respectively, is the reason of the vDVZ-discontinuity
issue.

\section{Quantization imaginary-mass field}                             
\label{sec:9}   
We rescale the $\epsilon(x)$-field
\begin{equation}
 \epsilon(x) := \frac{1}{2}\sqrt{3}\sqrt{\frac{b(1-b)^2}{(3+b)^3}}
 \frac{M^2}{{\cal M}^2}\ \widetilde{\epsilon}(x)
  \Rightarrow \frac{1}{2}\sqrt{3}\frac{M^2}{{\cal M}^2}\ \widetilde{\epsilon}(x),
\label{eq:9.1}\end{equation} 
where $\Rightarrow$ indicates the limit $\lambda \rightarrow 1 ( b \rightarrow \pm\infty)$.
The basic commutation relation (\ref{40.8a}) for the scalar field now becomes normalized
\begin{equation}
 \left[\widetilde{\epsilon}(x),\widetilde{\epsilon}(y)\right] = 
 -i\Delta(x-y; M^2_\epsilon), \\
\label{eq:9.2}\end{equation} 
with $M_\epsilon = i \sqrt{2\lambda} M \Rightarrow i \sqrt{2} M$.                
The quantization of spinless complex-ghost fields has been discussed in 
e.g. \cite{Tan60,Schr71,Nak72}. In our discussion below, we will follow these
references. In particular we analyse the $\widetilde{\epsilon}(x)$ fields in terms
of the spin-less complex-ghost fields
$\phi(x)$ and $\phi^\dagger(x)$ having $M_\epsilon$ and $M_\epsilon^*$, respectively.
Compared to these references, the (-)-sign) in the commutator (\ref{eq:9.2}) is
different, and we will discuss the ensuing differences.
Apart from the imaginary masses, the situation is quite similar to that
of the B-field in the so-called B-field formalism for (massive) vector fields,
cfr.~\cite{NO90}, subsection 2.4.2. So, we introduce a Gupta subsidiary-condition
for the physical states $|f\rangle$ in the total Fock-space
\begin{equation}
 \phi^{(+)}(x)|f\rangle = \phi^{\dagger (+)}(x)|f\rangle = 0,
\label{eq:9.3}\end{equation} 
and the quantization procedure is quite analogous to that described in 
\cite{Tan60,Schr71,Nak72} for the complex-ghost fields.\\

\noindent {\it Notice that sofar we have not defined $\Delta(x-y;M^2_\epsilon)$.
This is the topic of the rest of this section.}


\subsection{Imaginary-ghost Quantization }                 
\label{app:NAK}    
To obtain real potentials, we follow the quantization method
given by Nakanishi \cite{Nak72} for the scalar field with an
imaginary mass.
We make the identification
\begin{equation}
 \widetilde{\epsilon}(x) =:            
 \frac{1}{\sqrt{2}}\left[ \phi(x) + \phi^\dagger(x)\right].
\label{app:NAK.1}\end{equation}
Here, $\phi$ and $\phi^\dagger$ are spinless free complex-ghost fields having
$\mu = +i\sqrt{2}\mu_G$ and $\mu^*=-i\sqrt{2}\mu_G$, respectively.
The Lagrangian \cite{Nak72} is given by
\begin{equation}
 {\cal L}_\phi = \frac{1}{2}\left( \vphantom{\frac{A}{A}}
 \partial^\alpha\phi \partial_\alpha\phi -\mu^2 \phi^2
 +\partial^\alpha\phi^\dagger \partial_\alpha\phi^\dagger 
 -\mu^{* 2} \phi^{\dagger 2} \right)
\label{app:NAK.2}\end{equation}
The expansion of the field operator $\phi(x)=\phi^{(+)}(x)+\phi^{(-)}(x)$
in terms of annihilation and creation operators is, see \cite{Nak72}, section 16,
\begin{subequations}
\label{app:NAK.3}
\begin{eqnarray}
 \phi^{(+)}(x) &=& \int\frac{d^3p}{2\omega_p(2\pi)^3}\ \alpha({\bf p})
 \exp\left(i{\bf p}\cdot{\bf x}-i\omega_p x_0\right), \\
 \phi^{(-)}(x) &=& \int\frac{d^3p}{2\omega_p(2\pi)^3}\ \beta^\dagger({\bf p})
 \exp\left(-i{\bf p}\cdot{\bf x}+i\omega_p x_0\right).    
\end{eqnarray}
\end{subequations}
The canonical commutation relations for a ghost with a negative metric imply
\footnote{
Note that, in contrast to the imaginary-mass case treated in 
\cite{Nak72}, section 16 and 17, 
we have here a ghost with a {\bf negative metric}. This is taken care of
by the (-)-sign on the r.h.s. in (\ref{app:NAK.4a}).
}
\begin{subequations} \label{app:NAK.4b}
\begin{eqnarray}
 \left[\alpha({\bf p}),\beta^\dagger({\bf q})\right] &=&
 \left[\beta({\bf p}),\alpha^\dagger({\bf q})\right] = 
 -(2\pi)^3\delta({\bf p}-{\bf q}), \label{app:NAK.4a}\\
 \left[\alpha({\bf p}),\alpha^\dagger({\bf q})\right] &=&
 \left[\beta({\bf p}),\beta^\dagger({\bf q})\right] = 0,\ {\rm etc}
\end{eqnarray}
\end{subequations}
from which follow the field-commutators
\begin{subequations}
\label{app:NAK.5}
\begin{eqnarray}
 \left[\phi(x),\phi(y)\right] &=& -i\Delta(x-y, i\mu ), \\
 \left[\phi(x),\phi^\dagger(y)\right] &=& 0, \\
 \left[\phi^\dagger(x),\phi^\dagger(y)\right] &=& -i\Delta(x-y, -i\mu), 
\end{eqnarray}
\end{subequations}
with
the two-point vacuum expectation values \cite{Nak72} eq.~(16.32),
\begin{subequations}
\label{app:NAK.6b} 
\begin{eqnarray}
 \langle 0| \phi(x) \phi(y)|0\rangle &=& -\int\frac{d^3p}{2\omega_p(2\pi)^3}
 \exp\left[i{\bf p}\cdot({\bf x}-{\bf y})-i\omega_p(x_0-y_0)\right], \label{eq:10.46a}\\
 \langle 0| \phi(x) \phi^\dagger(y)|0\rangle &=& 0, \\                                
 \langle 0| \phi^\dagger(x) \phi^\dagger(y)|0\rangle &=& -\int\frac{d^3p}{2\omega_p^*(2\pi)^3}
 \exp\left[i{\bf p}\cdot({\bf x}-{\bf y})-i\omega_p^*(x_0-y_0)\right].   
\end{eqnarray} 
\end{subequations}
where $\omega^*=-\omega$. This gives
\begin{subequations}
\begin{eqnarray}
 \langle 0|[\phi(x),\phi(y)|0\rangle &=& -\int\frac{d^3p}{(2\pi)^3\omega_p}
 \exp\left[i{\bf p}\cdot({\bf x}-{\bf y})\right] \sin\omega_p(x^0-y^0), 
 \label{app:NAK.7a}\\
 \langle 0|[\phi^\dagger(x),\phi^\dagger(y)|0\rangle &=& 
 -\int\frac{d^3p}{(2\pi)^3\omega_p^*}
 \exp\left[i{\bf p}\cdot({\bf x}-{\bf y})\right] \sin\omega_p^*(x^0-y^0), 
 \label{app:NAK.7b}.  
\label{app:NAK.7}\end{eqnarray} 
\end{subequations}

\noindent The field-commutator for the $\widetilde{\epsilon}(x)$-field becomes
\begin{equation}
 \left[\widetilde{\epsilon}(x),\widetilde{\epsilon}(y)\right] = 
 -\frac{i}{2}\left(\vphantom{\frac{A}{A}}
 \Delta(x-y;i\mu) + \Delta(x-y;-i\mu)\right).
\label{app:NAK.12}\end{equation}

\subsection{Imaginary-ghost Propagator }                 
Here, to emphasize the difference in mass, we used in the argument of the
invariant $\Delta$-function, $\mu$ instead of $\mu^2$.
This implies for the Feynman propagator of the $\widetilde{\epsilon}(x)$-field
\cite{note:difference}
\begin{eqnarray}
 i\Delta(x-y; M^2_\epsilon) &\equiv& 
 \langle 0|T\left[\widetilde{\epsilon}(x)\widetilde{\epsilon}(y)\right]|0\rangle =
 \left\{\vphantom{\frac{A}{A}}\langle 0|T\left[\phi(x)\phi(y)\right]|0\rangle + 
        \langle 0|T\left[\phi^\dagger(x)\phi^\dagger(y)\right]|0\rangle \right\} 
\nonumber\\ &=& -\frac{i}{2}\left[\vphantom{\frac{A}{A}} 
\Delta_F(x-y; M=i\mu) + \Delta_F(x-y;M=-i\mu)\right]. 
\label{app:NAK.13}\end{eqnarray}


The proper integration contour $\Gamma$, see Fig.~\ref{fig:Gamma.1}, 
 in the complex $p_0$-plane is given as \cite{Nak72} 
$\Gamma= R-\delta(\omega_p)+\delta(-\omega_p)$, where $\delta(\pm \omega_p)$
denotes a {\it counterclockwise circle} around the poles at $\pm\omega_p$
\cite{Leray}. (Note that for a real mass $M_\epsilon$, the $\Gamma$-contour 
becomes the usual contour for the Feynman propagator $C_F$.)\\
\noindent {\bf Conjecture}: 
The Feynman propagator function in (\ref{app:SG.12}) is
\begin{subequations}
\label{app:NAK.21}
\begin{eqnarray}
 \widetilde{\Delta}(p;M_\epsilon^2) &=& \frac{1}{2}\left[
 \widetilde{\Delta}^{(1)}(p;M_\epsilon^2) + 
 \widetilde{\Delta}^{(2)}(p;M_\epsilon^2)\right], \\
 \widetilde{\Delta}^{(1)}(p;M_\epsilon^2) &=& 
 \frac{1}{p^2-M_\epsilon^2+i0} + 2\pi i\ \delta(p^2-M_\epsilon^2), \\
 \widetilde{\Delta}^{(2)}(p;M_\epsilon^{* 2}) &=& 
 \frac{1}{p^2-M_\epsilon^{* 2}+i0}.     
\end{eqnarray}
\end{subequations}
To show this, we first evaluate for the integration contour 
$\Gamma= R -\delta(\omega_p)+\delta(-\omega_p)$ the contribution
from the Leray coboundaries to $\widetilde{\Delta}^{(1)}(p;M_\epsilon^2)$  
in the complex $p_0$-plane:
\begin{eqnarray*}
&& \delta(\omega_p) \sim 
 -i\pi\ e^{i{\bf p}\cdot{\bf x}-i\omega_p x^0}/\omega_p,\ \ 
 \delta(-\omega_p) \sim 
 +i\pi\ e^{i{\bf p}\cdot{\bf x}+i\omega_p x^0}/\omega_p, 
\end{eqnarray*}
and therefore 
\begin{equation}
 -\delta(\omega_p)+\delta(-\omega_p) = i\pi\ 
 \left[ e^{-i\omega_p x^0} +e^{+i\omega_p x^0}\right]
 \frac{e^{i{\bf p}\cdot{\bf x}}}{\omega_p}.
\label{app:NAK.22}\end{equation}
Next we evaluate
\begin{eqnarray}
 I_\delta &\equiv& \int\frac{d^4p}{(2\pi)^4}\ \delta(p^2-M^2) 
 = \int\frac{d^3p}{(2\pi)^4}\ e^{i{\bf p}\cdot{\bf x}}\cdot
 \nonumber\\ && \times
 \int \frac{dp_0}{|2p_0|}\ \left[\delta(p^0-\omega_p) + 
 \delta(p^0+\omega_p) \right]\ e^{-ip_0 x^0} \nonumber\\ 
 &=& \int\frac{d^3p}{(2\pi)^4}\ \frac{1}{2\omega_p}\
 \left[ e^{-i\omega_p x^0} +e^{+i\omega_p x^0}\right] 
 e^{i{\bf p}\cdot{\bf x}}.     
\label{app:NAK.23}\end{eqnarray}
Note: in the application here $M=i\mu$ and $\omega_p^2= {\bf p}^2-\mu^2$ is
real. This justifies the use of $|p_0|$ in (\ref{app:NAK.23}). 
From the last two equations, we conclude that
\begin{equation}
 -\delta(\omega_p)+\delta(-\omega_p) \sim 2\pi i\ \delta(p^2-M^2),    
\label{app:NAK.24}\end{equation}
which proves the conjecture (QED).\\
For $M_\epsilon = i\mu$ and $M^*_\epsilon= -i\mu$ we have 
\begin{eqnarray*}
 \left[p^2-M_\epsilon^2+i0\right] &\rightarrow& 
 {\cal P}(p^2+\mu^2)^{-1} -i\pi\delta(p^2+\mu^2), \\
 \left[p^2-M_\epsilon^{* 2}+i0\right] &\rightarrow& 
 {\cal P}(p^2+\mu^2)^{-1} -i\pi\delta(p^2+\mu^2). 
\end{eqnarray*}
\begin{center}
\fbox{\rule[-5mm]{0cm}{1.5cm} \hspace{5mm}
\begin{minipage}[]{12.9cm}
Therefore, using (\ref{app:NAK.21}), we find that
with $M_\epsilon=-M_\epsilon^*=i\mu$
\begin{eqnarray}
 \widetilde{\Delta}(p;M_\epsilon^2) &=& \frac{1}{2}\left[
 \widetilde{\Delta}^{(1)}(p;M_\epsilon^2) + 
 \widetilde{\Delta}^{(2)}(p;M_\epsilon^2)\right] 
 = {\cal P}\frac{1}{p^2+\mu^2}.
\label{app:NAK.25}\end{eqnarray}
\end{minipage} \hspace{5mm} }\\
\end{center}

\noindent We note that in the one-graviton-exchange diagram 
Fig.~\ref{fig.grav1} we have on-energy-shell external particles 
that $p^0=0$. Then, $p^2+\mu^2 \rightarrow -({\bf p}^2-\mu^2)$, 
which is used in (\ref{app:SG.14}).\\


\begin{flushleft}
\rule{16cm}{0.5mm}
\end{flushleft}
\section{Massless-limit Helicity-couplings}  
\label{app:helicity-coupling}
For massive gravitons the projection operator is \cite{Schw70}
\begin{eqnarray}
	\bar{\Pi}_{\mu\nu,\alpha\beta}(k) &=& \sum_{\lambda=1,5}
	\bar{\varepsilon}^{(\lambda)}_{\mu\nu}(k)\bar{\varepsilon}^{(\lambda)}_{\alpha\beta}(k)
	= \frac{1}{2}\left( \bar{\eta}_{\mu\alpha} \bar{\eta}_{\nu\beta} 
	+\bar{\eta}_{\mu\beta}  \bar{\eta}_{\nu\alpha}\right)
	-\frac{1}{3}\bar{\eta}_{\mu\nu}\bar{\eta}_{\alpha\beta},    
\label{app:helicity.1}\end{eqnarray}
with $ \bar{\eta}_{\mu\nu} = \eta_{\mu\nu}-k^\mu k^\nu/M^2$. 

\noindent  The coupling to the conserved energy-momentum $t_{M,\mu\nu}$ warrants the replacement 
\begin{eqnarray}
\bar{\Pi}_{\mu\nu,\alpha\beta}(k) &\rightarrow& 
P^{(2)}_{\mu\nu;\alpha\beta} =
\biggl\{\frac{1}{2}\left(\eta_{\mu\alpha}\eta_{\nu\beta}
	+\eta_{\mu\beta}\eta_{\nu\alpha}\right)
-\frac{1}{3}\eta_{\mu\nu}\eta_{\alpha\beta}\biggr\} 
\label{app:helicity.2}\end{eqnarray}
Next, we show that in the massless limit, owing to the properties of the polarization
vectors, the helicities $\lambda=-1,+1$ decouple smoothly from the $t_{M, \mu\nu}$-tensor
for $\lim_{M \rightarrow 0}$, and that the helicity $\lambda=0$ gives a contribution that is
cancelled in the massless limit by a contribution from the scalar imaginary ghost.

\subsection{Spin-2 Polarization Vectors}                           
\label{app:helicity.a}   
We work in a frame where $p^\mu= (p^0,0,0,p)$, for which we choose
the spin-1 polarization vectors in the standard form:
\begin{eqnarray}
 \epsilon^\mu(\pm 1) &=& \frac{1}{\sqrt{2}}\left(0,\mp1,-i,0\right)\ \ ,\ \ 
 \epsilon^\mu(0) = \frac{1}{M}\left(p,0,0,E_p\right)\ ,
\label{eq:2.1}\end{eqnarray} 
Then, the spin-2 polarization vectors are, up to a 'gauge' transformation,
\begin{subequations}
\begin{eqnarray}
 \varepsilon^{\mu\nu}(p,\lambda=+2) &=& \varepsilon^\mu(p,+1)\ \varepsilon^\nu(p,+1)\ ,
\\
 \varepsilon^{\mu\nu}(p,\lambda=+1) &=& \frac{1}{\sqrt{2}}\left(
\vphantom{\frac{A}{A}}\varepsilon^\mu(p,+1)\ \varepsilon^\nu(p, 0) +  
\varepsilon^\mu(p, 0)\ \varepsilon^\nu(p,+1)\right)\ , \\
 \varepsilon^{\mu\nu}(p,\lambda= 0) &=& \frac{1}{\sqrt{6}}\left(
\vphantom{\frac{A}{A}} 2 \varepsilon^\mu(p, 0)\ \varepsilon^\nu(p, 0) +  
\varepsilon^\mu(p,+1)\ \varepsilon^\nu(p,-1) +  
\varepsilon^\mu(p,-1)\ \varepsilon^\nu(p,+1)\right)\ , 
\label{eq:2.2}\end{eqnarray} 
\end{subequations}
and similarly for $\lambda=-1,-2$. 

\subsection{Polarization Vectors and Smooth Massless limit}
\label{app:helicity.b}
We note that for small M the leading terms of the $\lambda=0$ polarization are
\begin{eqnarray}
 \varepsilon^\mu(p,\lambda=0) &=& \frac{p^\mu}{M} + 
 \left(-\frac{M}{2p},0,0,\frac{M}{2p}\right) + O(M^3)
\label{eq:2.3}\end{eqnarray} 
which gives        
\begin{eqnarray}
 \varepsilon^{\mu\nu}(p,\lambda=+1) &\sim& \frac{1}{\sqrt{2}M}\left(\vphantom{\frac{A}{A}}
\varepsilon^\mu(p,+1)\ p^\nu + \varepsilon^\nu(p,+1)\ p^\mu\right)\ , \\
 \varepsilon^{\mu\nu}(p,\lambda= 0) &\sim& \sqrt{\frac{2}{3}}\frac{p^\mu p^\nu}{M^2}
+\frac{1}{\sqrt{6}}\left( \eta^{\mu\nu}-\frac{p^\mu\tilde{p}^\nu+\tilde{p}^\mu p^\nu}
{2p\cdot\tilde{p}}\right)\ , 
\label{eq:2.4}\end{eqnarray} 
where we introduced $\tilde{p}^\mu=(p^0,-{\bf p})$, and used the identity
\begin{equation}
\varepsilon^\mu(p,+1)\ \varepsilon^\nu(p,-1) +  
\varepsilon^\mu(p,-1)\ \varepsilon^\nu(p,+1) =          
 \eta^{\mu\nu}-\frac{p^\mu\tilde{p}^\nu+\tilde{p}^\mu p^\nu}
{2p\cdot\tilde{p}}\ . 
\label{eq:2.5}\end{equation} 
{\it Terms in the polarization tensors $\varepsilon^{\mu\nu}(p,\lambda) \propto p^\mu, p^\nu$ do not
contribute to the matrix elements because $p^\mu t_{M, \mu\nu}=t_{M,\mu\nu} p^\nu=0$. 
This is also in the
massless limit, since $\lim_{ M \downarrow 0} (p^\mu/M) t_{M,\mu\nu}=0$.} 
The only "false helicity", {\it i.e.} $\lambda \neq \pm 2$, 
that survives is $\varepsilon^{\mu\nu}(p,\lambda=0) \sim \eta^{\mu\nu}/\sqrt{6}$.
which contributes a term $-\eta_{\mu\nu}\eta_{\alpha\beta}/6$ in $P^{(2)}_{\mu\nu;\alpha\beta}$.
This term is canceled in the massless limit by a contribution from the $\epsilon$-ghost 
\begin{eqnarray*}
&& -\frac{1}{3}\eta_{\mu\nu}\eta_{\alpha\beta} \widetilde{\Delta}_F(p^2,M^2)
-\frac{1}{6}\eta_{\mu\nu}\eta_{\alpha\beta} \widetilde{\Delta}_F(p^2,M_\epsilon^2) = 
-\frac{1}{2}\eta_{\mu\nu}\eta_{\alpha\beta} \widetilde{\Delta}_F(p^2,M^2)
\nonumber\\ && 
+\frac{1}{6}\eta_{\mu\nu}\eta_{\alpha\beta} 
\left[\widetilde{\Delta}_F(p^2,M^2)-\widetilde{\Delta}_F(p^2,M_\epsilon^2)\right] 
\end{eqnarray*}
where on the r.h.s. the second ("mass correction") term 
$\propto M_\epsilon^2-M^2 \propto M^2$ ,
showing the smooth transition to the massless case.
\begin{flushleft}
\rule{16cm}{0.5mm}
\end{flushleft}
\section{BS-equation  and LS-equation}                              
\label{app:BS}    
In this appendix, we consider the Bethe-Salpeter equation (BSE) in the 
normalization used in \cite{BD65} for scalar external particles.
To start, we note that in \cite{BD65} the Feynman rules give 
the invariant amplitude $-i M$. For scalar-exchange the potential 
is readily seen using the Feynman rules \cite{BD65} to be given as
\begin{equation}
 V(P',p'; P,p) = \frac{g^2}{q^2-\mu^2+i\delta}
\label{app:BS.1}\end{equation}
Here, the external momenta are $(P',p')$ and $(P,p)$ for the final and
initial state, respectively. The exchange momentum is $q=p'-p=P-P'$.\\

\noindent The total and relative momenta for the initial, final, and
intermediate states are defined as
\begin{eqnarray}
 p_a &=& \mu_a P+p,\ \ p_b=\mu_b P-p,\ \ 
 p_a'= \mu_a P'+p',\ \ p_b'=  \mu_b P'-p', \nonumber\\
 k_a &=& \mu_a K_n + k,\ \ k_b= \mu_b K_n- k, K_n =k_a+k_b.
\label{app:BS.2a}\end{eqnarray}
In the following, we use for the weights $\mu_a=\mu_b=1/2$. 
From the conservation of the total momenta, {\it i.e.} $P_i=P_f=K_n \equiv W$,
the dependence of the amplitude and potential is given by
\begin{eqnarray}
 M(P',p';P,p) \equiv M(p_f,p_i;W),\ \ 
 V(P',p';P,p) \equiv V(p_f,p_i;W).    
\label{app:BS.2b}\end{eqnarray}

 \begin{figure}[hhhhtb]
 \resizebox{10.25cm}{!} 
 {\includegraphics[120,520][420,670]{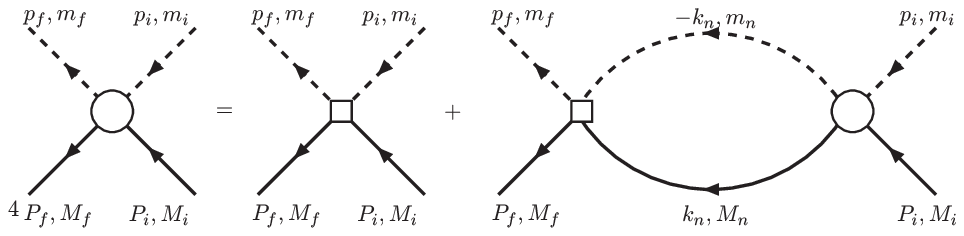}}
  \caption{\sl BS-Integral Equation}                         
  \label{fig:inteq3} 
  \end{figure}                     
 \begin{figure}[hhhhtb]
 \resizebox{11.25cm}{!}
 {\includegraphics[120,375][420,525]{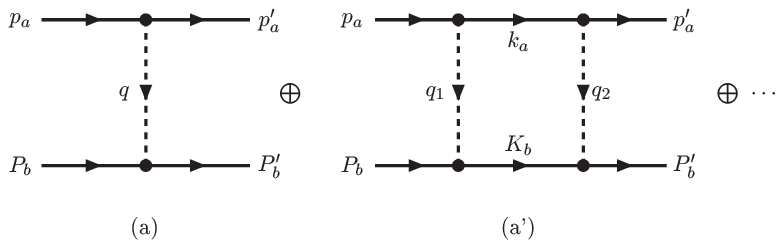}}
\caption{\sl One-meson and planar two-meson exchange, etc. Feynman graphs.                       
        The solid lines denote scalar heavy particles, e.g. the sun and the planet.
        The dashed lines refer to the scalar mesons.}
  \label{tmepar.kadysh.a2} 
  \end{figure}                     

\begin{figure}[htb]
 \resizebox{11.25cm}{!}
 {\includegraphics[110,385][390,565]{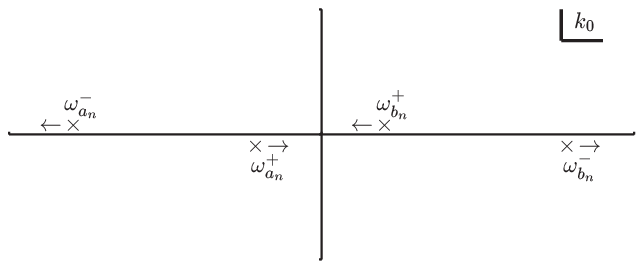}}
  \caption{\sl Poles of the two-particle Green function.  }    
  \label{fig:greenpoles}
  \end{figure}

\noindent From an analysis of the planar-box graph for scalar-exchange
we infer the BSE, see Fig.~\ref{fig:inteq3}, for scalar external particles as
\begin{eqnarray}
 M(P',p'; P,p) &=& V(p_f,k_n; W) + \int\!\! d^4k_n\ V(p_f,k_n;W)\   
 G_n(k_n;W)\ M(k_n,p_i;W) 
\label{app:BS.3}\end{eqnarray}
with 
\begin{eqnarray}
 G_n(k_n;W) &=& \frac{i}{\left[(\frac{1}{2}W+k_n)^2-M^2+i\delta\right]
 \left[(\frac{1}{2}W-k_n)^2-m^2+i\delta\right]}
\label{app:BS.4}\end{eqnarray}
Equation~(\ref{app:BS.4}) can easily be read off from the 
amplitude for the planar-box graph depicted in Fig.~\ref{tmepar.kadysh.a2}.
In the application to planetary motion in these notes, the particles,
planets and sun, off-energy-shell effects are non-existent. Therefore,
the amplitude and potential are $k_n^0$-independent. 
The poles of the Green function $G_n(k_n;W)$, see Fig.~\ref{fig:greenpoles}, are at
\begin{eqnarray}
 \omega^\pm_{a_n} &=& k^{0,\pm}_{n,a} = 
 -\frac{1}{2}\sqrt{s} \pm {\cal E}(k_n) \mp i\delta,
\nonumber\\
 \omega^\pm_{b_n} &=& k^{0,\pm}_{n,b} = 
 +\frac{1}{2}\sqrt{s} \pm E(k_n) \mp i\delta,
\label{app:BS.5}\end{eqnarray}
\noindent {\bf 1.\ Positive and negative energy contributions}:
Integrating over 
$k_n^0$, using the residue theorem, in the r.h.s. of (\ref{app:BS.4}) we get
\begin{eqnarray}
 && \int_{-\infty}^{+\infty} dk_n^0\ G_n(k_n;W) = 
 \pi \frac{{\cal E}(k_n)+E(k_n)}{{\cal E}(k_n) E(k_n)}\
 \frac{1}{s-({\cal E}(k_n)+E(k_n))^2}.
\label{app:BS.6}\end{eqnarray}
In the low-energy approximation, we have
\begin{eqnarray}
 && s-({\cal E}(k_n)+E(k_n))^2 \approx \frac{M+m}{m_{red}}
 \left({\bf p}_i^2 -{\bf k}_n^2\right),\ \          
 \frac{{\cal E}+E}{{\cal E}\ E}({\bf k}_n) \approx 
  \frac{M+m}{M m} \cdot\nonumber\\ && \times
 \left[1+\left(1-\frac{M^2+m^2}{M m}\right)
  \frac{{\bf k}_n^2}{2M m}\right] \sim \frac{1}{m}
 \left(1- \frac{{\bf k}_n^2}{2m^2}\right)\ \ (M \gg m).             
\label{app:BS.7}\end{eqnarray}
With this approximation the BSE (\ref{app:BS.3}) becomes
\begin{subequations}
\label{app:BS.8}
\begin{eqnarray}
 M(p_f,p_i; W) &=& V(p_f,k_n; W) + \int\!\! d^4k_n\ V(p_f,k_n;W)\   
 g_n(k_n;W)\ M(k_n,p_i;W), \\
 g_n(k_n;W) &=&                   
 \pi \frac{{\cal E}(k_n)+E(k_n)}{{\cal E}(k_n) E(k_n)}\
 \frac{1}{s-({\cal E}(k_n)+E(k_n))^2} \\
  &\approx & \frac{\pi}{Mm(1+{\bf k}^2_n/2m^2}\ 
 \frac{2m_{red}}{ {\bf p}_i^2-{\bf k}_n^2+i\delta}
\end{eqnarray}
\end{subequations}
The transition to the Lippmann-Schwinger equation (LSE) is made
by the transformation
\begin{eqnarray}
&& {\cal T}(p_f,p_i) = N(p_f)\ M(p_f,p_i;W)\ N(p_i),\ \
 {\cal V}(p_f,p_i) = N(p_f)\ V(p_f,p_i;W)\ N(p_i),   
\label{app:BS.9}\end{eqnarray}
where 
\begin{equation}
 N(p) = \sqrt{2\pi\frac{{\cal E}(p)+E(p)}{{\cal E}(p) E(p)}}
 \approx \sqrt{\frac{\pi}{2M m}}
  \left(1- \frac{{\bf p}^2}{4m^2}\right)\ \ (M \gg m).             
\label{app:BS.10}\end{equation}
So the potential for the LSE at low energy becomes
\begin{eqnarray}
 {\cal V}(p_f,p_i) &=& N(p_f)\ V(p_f,p_i;W)\ N(p_i)   
 \approx \frac{\pi}{2M m}\left(1-\frac{{\bf p}_f^2+{\bf p}_i^2}{4m^2}\right)
 V(p_f,p_i;W) \nonumber\\ &=&
 \frac{\pi}{2M m}\left(1-\frac{{\bf q}^2+{\bf k}^2/4}{4m^2}\right)
 V(p_f,p_i;W).                     
\label{app:BS.11}\end{eqnarray}

\noindent {\bf 2.\ No negative energy contributions}: We
split the intermediate state propagators in the "positive" and
"negative" part as follows:
\begin{eqnarray*}
&& \frac{1}{\left(\frac{1}{2}W+k_n\right)^2-M^2+i\delta} =
 \frac{1}{2{\cal E}(k_n)}\left[
\frac{1}{\left(\frac{1}{2}\sqrt{s}+k_n^0\right)-{\cal E}+i\delta}
-\frac{1}{\left(\frac{1}{2}\sqrt{s}+k_n^0\right)+{\cal E}-i\delta}
\right], \\
&& \frac{1}{\left(\frac{1}{2}W-k_n\right)^2-m^2+i\delta} =
 \frac{1}{2 E(k_n)}\left[
\frac{1}{\left(\frac{1}{2}\sqrt{s}-k_n^0\right)-E+i\delta}
-\frac{1}{\left(\frac{1}{2}\sqrt{s}-k_n^0\right)+E-i\delta}
\right].    
\label{app:BS.15}\end{eqnarray*}
Neglecting the contributions of the negative-energy states
the two-particle Green function is given by
\begin{eqnarray}
 G^{++}_n(k_n;W) &=& 
 \frac{1}{2{\cal E}(k_n)E(k_n)}\left[
\frac{1}{\frac{1}{2}\left(k_n^0+\sqrt{s}\right)-{\cal E}+i\delta}\cdot
\frac{1}{\frac{1}{2}\left(\sqrt{s}-k_n^0\right)-E+i\delta}
\right].     
\label{app:BS.16}\end{eqnarray}
As above, the $k_n^0$-integration gives
\begin{eqnarray}
 && \hspace{-1cm} \int_{-\infty}^{+\infty} dk_n^0\ G_n^{++}(k_n;W) = 
 \frac{2\pi}{{\cal E}(k_n) E(k_n)}\
 \frac{1}{\sqrt{s}-({\cal E}(k_n)+E(k_n))} \equiv g_n^{++}(k_n;W).
\label{app:BS.17}\end{eqnarray}
In the low-energy approximation, we obtain  
\begin{eqnarray}
 g_n^{++}(k_n;W) &=& \frac{\pi}{2{\cal E}(k_n) E(k_n)}\
 \frac{1}{\sqrt{s}-({\cal E}(k_n)+E(k_n))}\nonumber\\ 
  &\approx & \frac{\pi}{2M m}
 \left(1+\frac{M^2+m^2}{2M^2m^2} {\bf k}_n^2\right)
 \frac{2m_{red}}{ {\bf p}_i^2-{\bf k}_n^2+i\delta}
\label{app:BS.18}\end{eqnarray}
Again, the transition to the Lippmann-Schwinger equation (LSE) is made
by the transformation
\begin{eqnarray}
&& {\cal T}(p_f,p_i) = N(p_f)\ M(p_f,p_i;W)\ N(p_i),\ \
 {\cal V}(p_f,p_i) = N(p_f)\ V(p_f,p_i;W)\ N(p_i),   
\label{app:BS.19}\end{eqnarray}
where 
\begin{equation}
 N(p) = \sqrt{\pi/{2\cal E}(p) E(p)} \approx \sqrt{\frac{\pi}{2M m}}
  \left(1- \frac{M^2+m^2}{4M^2m^2}\ {\bf p}^2\right)\ \ (M \gg m),             
\label{app:BS.20}\end{equation}
and the potential for the LSE at low energy becomes
\begin{eqnarray}
 {\cal V}(p_f,p_i) &=& N(p_f)\ V(p_f,p_i;W)\ N(p_i)   
  \approx \frac{\pi}{2M m} \cdot\nonumber\\ && \times 
 \left(1-\frac{M^2+m^2}{4m^2 M^2} 
 ({\bf q}^2+{\bf k}^2/4)\right) V(p_f,p_i;W).                     
\label{app:BS.21}\end{eqnarray}
{\it Notice that for $ M \gg m$ the form with no negative energy \\
   contribution (\ref{app:BS.21}) is equivalent to (\ref{app:BS.11})}.

\noindent {\bf 3.\ Non-local Potential and Schr\"{o}dinger Equation:} 
For a non-local potential, i.e.
\begin{equation}
 {\cal V}({\bf k},{\bf q}) = v({\bf k}\left({\bf q}^2+\frac{1}{4}{\bf k}^2\right),
\label{app:BS.31}\end{equation}
the action on the wave function is \cite{NRS78}
\begin{equation}
 \langle {\bf r}|{\cal V}|\psi\rangle = 
-\frac{1}{2}\left(\bm{\nabla}^2 v({\bf r}) + v({\bf r}) \bm{\nabla}^2 
 \right)\ \psi({\bf r}).
\label{app:BS.32}\end{equation}
In \cite{NRS78} the $\phi$-function is introduced as $\phi(r)= m_{red}\ v(r)$,
and the radial Schr\"{o}dinger equation, orbital angular momentum integer l, 
after making the Green-transformation $u_l=(1+2\phi)^{-1/2} w_l$,reads
\begin{equation}
 w_l^{\prime\prime} + \left[k^2-2m_{red} W(r) -l(l+1)/r^2\right]\ w_l(r)=0.
\label{app:BS.33}\end{equation}
The "effective" potential $W$ is energy dependent and given by
\begin{eqnarray}
 W(r) &=& \frac{{\cal V}(r)}{1+2\phi}
 -\frac{1}{2m_{red}}\left(\frac{\phi'}{1+2\phi}\right)^2
 +\frac{2\phi}{1+2\phi}\frac{p_i^2}{2m_{red}} \nonumber\\
&\approx& {\cal V}(r) -2\phi(r)\left[{\cal V}(r)-\frac{p_i^2}{2m_{red}}\right]
\label{app:BS.34}\end{eqnarray}
Now, using a circular approximation the (classical) total energy is
\begin{equation}
 E = \frac{p_i^2}{2m_{red}} = +\frac{1}{2}{\cal V} < 0\ \ ({\bf b.s.})
\label{app:BS.35}\end{equation}


\noindent {\bf 4.\ Perihelion-precession Planets:}            
Here, we focus on the $1/r^2$-terms, which are responsible for
the perihelion-precession.
From the previous paragraph, we have that
\begin{equation}
 \phi(r) = \frac{7m_{red}}{4m^2}\ {\cal V}^{(0)}(r)
\label{app:BS.41}\end{equation}
There are now two possibilities:\\

\noindent {\bf a.}\ We treat E in (\ref{app:BS.34}) as a function of 
$r$ as given in (\ref{app:BS.35}), which gives
\begin{equation}
 W^{(a)}(r) = {\cal V}(r) -\phi(r)\ {\cal V}(r), 
\label{app:BS.42}\end{equation}
and consequently the $1/r^2$-correction is given by
\begin{equation}
 \Delta{\cal V} \approx -\phi(r)\ {\cal V}^{(0)}(r) = 
 -\frac{7}{4m}\left[{\cal V}^{(0)}\right]^2
\label{app:BS.43}\end{equation}
which is $7/12$ times Einstein's result.

\noindent {\bf b.}\ We treat E in (\ref{app:BS.34}) as a constant,      
like in \cite{Schw70,VDam74}. Then, the E-term in (\ref{app:BS.34})
is of $1/r$-type and only deforms the shape of the orbit a little
bit. The $1/r^2$-correction becomes 
\begin{equation}
 \Delta{\cal V} \approx -2\phi(r)\ {\cal V}^{(0)}(r) = 
 -\frac{7}{2m}\left[{\cal V}^{(0)}\right]^2,
\label{app:BS.44}\end{equation}
which agrees with Schwinger \cite{Schw70}, and 
leads to $7/6\times$ Einstein's result! \\

\noindent {\bf Note}: {\it In a circular, classical, motion
the kinetic and potential energy are connected by the
equilibrium equation:
\begin{eqnarray*}
&& |F_{grav.}| = |F_{centr.}| \rightarrow 
 |{\cal V}'| = \frac{|{\cal V}|}{r} = 
\frac{mv^2}{r}= \frac{p^2}{m r}
\end{eqnarray*}
 or $T = |{\cal V}|/2$, instead of $T=E-{\cal V}$!?
}
\begin{flushleft}
\rule{16cm}{0.5mm}
\end{flushleft}

\section{Scalar-interaction Imaginary-ghost Field }                 
\label{app:SIG}    
The Yukawa-interaction of a scalar fields $\psi(x), \chi(x)$ 
with the imaginary-ghost
field $\phi(x)$ we describe by the interaction Hamiltonian
\begin{equation}
 {\cal H}_I(x) = \frac{1}{2}
 \left[ g_\psi\psi^2(x) +g_\chi\chi^2(x)\right]\ 
 \left( \phi(x)+\phi^\dagger(x)\right),
\label{eq:SIG.1}\end{equation}
where the $\psi$ and $\chi$ masses are M and m respectively.
For the existence of the Dyson S-matrix, a Gaussian adiabatic factor is
necessary \cite{Nak72}
\begin{equation}
 {\cal H}_I^\varepsilon(x_0) = {\cal H}_I\ e^{-\varepsilon x_0^2},
\label{eq:SIG.2}\end{equation}
such that for the transition matrix $U_\varepsilon(x_0,y_0)$ the      
time limits $x_0 \rightarrow +\infty$ and $y_0 \rightarrow -\infty$ exist.
Then, the S-matrix is given by
\begin{eqnarray}
 S_\varepsilon &=& \sum_{n=0}^\infty \frac{(-i)^n}{n!}\ \int d^4x_1 \ldots
 \int d^4 x_n\ T\left[ {\cal H}_I^\varepsilon(x_1) \ldots 
 {\cal H}_I^\varepsilon(x_n)\right].
\label{eq:SIG.3}\end{eqnarray}
The 2nd-order S-matrix element is
\begin{eqnarray}
 \langle p', P'| S_\varepsilon^{(2)}| p,P\rangle &=& 
 -\frac{1}{2} \int d^4x \int d^4y\ 
 \langle p', P'|T\left[ {\cal H}_I^\varepsilon(x) 
 {\cal H}_I^\varepsilon(y)\right]| p,P\rangle.
\label{eq:SIG.4}\end{eqnarray}
The one-particle states of the scalar particles give the wave functions
\begin{subequations}
\label{eq:SIG.5}
\begin{eqnarray}
 \psi_P(x) = \langle 0|\psi(x)|P\rangle = 
 \left[(2\pi)^3 2\omega(P)\right]^{-1/2}\ e^{-iP\cdot x}, \\ 
 \chi_p(x) = \langle 0|\chi(x)|p\rangle = 
 \left[(2\pi)^3 2\omega(p)\right]^{-1/2}\ e^{-ip\cdot x}.    
\end{eqnarray}
\end{subequations}
The plane wave expansion of the imaginary-ghost field is \cite{Nak72}
\begin{subequations}
\begin{eqnarray}
 \phi(x) &=& \int\frac{d^3p}{\sqrt{(2\pi)^3 2\omega_p}}\left[
 \alpha({\bf p}) e^{-ip\cdot x} +\beta^\dagger({\bf p}) e^{+ip\cdot x}\right],
\label{eq:SIG.6a} \nonumber\\ 
 \phi^\dagger(x) &=& 
 \int\frac{d^3p}{\sqrt{(2\pi)^3 2\widetilde{\omega}_p}}\left[
 \alpha^\dagger({\bf p}) e^{-ip\cdot x} +\beta({\bf p}) e^{-ip\cdot x}\right].
\label{eq:SIG.6b}\end{eqnarray}
\end{subequations}
The quantization, such that $[\phi(x),\phi^\dagger(y)]=0$ and the
negative-metric, is given by the commutation relations
\begin{subequations}
\begin{eqnarray}
&& \left[\alpha({\bf p}), \beta^\dagger({\bf q})\right]=
   \left[\beta({\bf p}), \alpha^\dagger({\bf q})\right]=
 -(2\pi)^3\delta({\bf p}-{\bf q}), \label{eq:SIG.6c} \\
&& \left[\alpha({\bf p}), \alpha^\dagger({\bf q})\right]=
   \left[\beta({\bf p}), \beta^\dagger({\bf q})\right]= 0. 
\label{eq:SIG.6d}
\end{eqnarray}
\end{subequations}

For the imaginary-ghost exchange between the scalar particles, the relevant 
term in the Wick-expansion of $T\left[\ldots \right]$ is given by
\begin{equation}
 T\left[\psi^2(x)\phi(x)\ \chi^2(y) \phi(y)\right] \Rightarrow
 N \left[ \psi^2(x)\ \chi^2(y)\right]
 \langle 0|T\left[\phi(x)\ \phi(y)\right]|0\rangle, 
\label{eq:SIG.8}\end{equation}
where 
\begin{equation}
 \langle 0|T\left[\phi(x)\ \phi(y)\right]|0\rangle = 
 i \Delta_F(x-y; i\mu),
\label{eq:SIG.9}\end{equation}
and 
\begin{eqnarray}
 \langle p', P'| N \left[ \psi^2(x)\ \chi^2(y)\right] |p,P\rangle &=&
 (2\pi)^{-6}\left[16 \omega_{p'}\omega_{P'} \omega_p \omega_P\right]^{-1/2}
 \cdot\nonumber\\ && \times \left\{
 e^{i(p'-p)\cdot x} e^{i(P'-P)\cdot y} +
 e^{i(p'-p)\cdot y} e^{i(P'-P)\cdot x}\right\}.
\label{eq:SIG.10}\end{eqnarray}
The 2nd-order S-matrix element becomes 
\begin{eqnarray}
 \langle p', P'| S_\varepsilon^{(2)}|p,P\rangle &=& -g_\psi g_\chi\
 (2\pi)^{-3}\left[16 \omega_{p'}\omega_{P'} \omega_p \omega_P\right]^{-1/2}
 \int d^4x \int d^4y\ \cdot\nonumber\\ && \times  
 e^{i(p'-p)\cdot x} e^{i(P'-P)\cdot y}\ 
 i \Delta_F(x-y; i\mu)\ e^{-\varepsilon x_0^2} e^{-\varepsilon y_0^2}.
\label{eq:SIG.11}\end{eqnarray}
Using the variables
\begin{equation}
 Z = \frac{1}{2}(x+y),\ \ z=x-y, 
\label{eq:SIG.12}\end{equation}
and defining the M-matrix by
\begin{equation}
 S_\varepsilon(f,i) = \delta_{f,i} -(2\pi)^4 i\ 
 \delta(P_f-P_i)\ M_\varepsilon(f,i),
\label{eq:SIG.13}\end{equation}
we obtain 
\begin{eqnarray}
 \langle p', P'| M_\varepsilon^{(2)}|p,P\rangle &=& g_\psi g_\chi\
 (2\pi)^{-3}\left[16 \omega_{p'}\omega_{P'} \omega_p \omega_P\right]^{-1/2}
 \cdot\nonumber\\ && \times    
 \int d^4z\ e^{i(p'-p)\cdot z}\ \Delta_F(z;i\mu)\ e^{-\varepsilon z_0^2/2} 
\label{eq:SIG.14}\end{eqnarray}
A similar contribution to $M_\varepsilon^{(2)}$ comes from the exchange of
a $\phi^\dagger$ imaginary-ghost particle. 
In that case, the Feynman propagator is 
\begin{equation}
 i \Delta_F(z;-i\mu) = 
\theta(z_0)\ \langle 0|]\phi^\dagger(z) \phi^\dagger(0)|0\rangle +
 \theta(-z_0)\ \langle 0|]\phi^\dagger(0) \phi^\dagger(z)|0\rangle.   
\label{eq:SIG.15}\end{equation}
Working out the $z_0$-integrals in (\ref{eq:SIG.14}) for $\Delta_F(z;-i\mu)$
explicitly, we have
\begin{eqnarray}
&& \lim_{\varepsilon \rightarrow 0} \int^\infty_{-\infty} dz_0\
 e^{-\varepsilon z_0^2/2} \theta(z_0)\ \langle 0|]\phi^\dagger(z) \phi^\dagger(0)|0\rangle   
 \Rightarrow \nonumber\\ && 
 -\lim_{\varepsilon \rightarrow 0} \int^\infty_{-\infty} d\tau\
 e^{-\varepsilon \tau^2/2} \theta(\tau)\ e^{-i\widetilde{\omega}_p \tau}      
 = i\left[ {\cal P}\frac{1}{\widetilde{\omega}_p}+i\pi 
 \delta(\widetilde{\omega}_p)\right], 
\label{eq:SIG.16}\end{eqnarray}
and
\begin{eqnarray}
&& \lim_{\varepsilon \rightarrow 0} \int^\infty_{-\infty} dz_0\
 e^{-\varepsilon z_0^2/2} \theta(-z_0)\ \langle 0|]\phi^\dagger(0) \phi^\dagger(z)|0\rangle   
 \Rightarrow \nonumber\\ && 
 -\lim_{\varepsilon \rightarrow 0} \int^\infty_{-\infty} d\tau\
 e^{-\varepsilon \tau^2/2} \theta(-\tau)\ e^{+i\widetilde{\omega}_p \tau}     
 = i\left[ {\cal P}\frac{1}{\widetilde{\omega}_p}+i\pi 
 \delta(\widetilde{\omega}_p)\right], 
\label{eq:SIG.17}\end{eqnarray}
and we get 
\begin{eqnarray}
&& \int^\infty_{-\infty} dz_0\ \Delta_F(z; -i\mu) = 2\pi 
\int\frac{d^3p}{2\widetilde{\omega}_p(2\pi)^3}\ e^{i{\bf p}\cdot{\bf z}}\
 \left[ \frac{1}{\pi}{\cal P}\frac{1}{\widetilde{\omega}_p}
 +i \delta(\widetilde{\omega}_p)\right]. 
\label{eq:SIG.18a}\end{eqnarray}
Similarly, for $\Delta_F(z;+i\mu)$ 
\begin{eqnarray}
&& \int^\infty_{-\infty} dz_0\ \Delta_F(z; i\mu) = 2\pi 
\int\frac{d^3p}{2\omega_p(2\pi)^3}\ e^{i{\bf p}\cdot{\bf z}}\
 \left[ \frac{1}{\pi}{\cal P}\frac{1}{\omega_p}+i \delta(\omega_p)\right]. 
\label{eq:SIG.18b}\end{eqnarray}
 
 \begin{figure}[hhhhtb]
 \resizebox{!}{!}
 {\includegraphics[100,475][300,725]{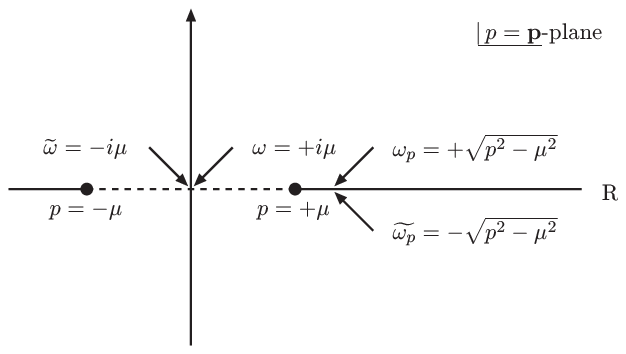}}
 \caption{Branchpoints $\omega(p)$ at $p = \pm \mu$, and the relation between 
$\omega(p)$ and $\widetilde{\omega}(p)$. The dashed line indicates the branch line.
On the upper rim of the branchline $\omega = i\sqrt{\mu^2-p^2}$, and 
$\widetilde{\omega} = -i\sqrt{\mu^2-p^2}$. 
	 For $\mu < p < \infty$ one has $\omega=\sqrt{p^2-\mu^2}$,  
whereas $\widetilde{\omega} = -\sqrt{p^2-\mu^2}$. With these choices of the branches of the
square-root $\widetilde{\omega}=-\omega$.}
  \label{fig.grav3}
  \end{figure}                     
Here, for the branches of $\omega_p$ and $\widetilde{\omega}_p$ 
see Fig.~\ref{fig.grav3}, 
\begin{subequations}
\label{eq:SIG.21}
\begin{eqnarray}
 \omega_p &=& \left\{\begin{array}{l} 
 +\sqrt{{\bf p}^2-\mu^2}\ \ ({\bf p}^2 > \mu^2), \\
 +i\sqrt{{\mu^2-\bf p}^2}\ \ ({\bf p}^2 < \mu^2) \end{array}\right.
 \\
 \widetilde{\omega}_p &=& \left\{\begin{array}{l}
 -\sqrt{\mu^2-{\bf p}^2}\ \ ({\bf p}^2 > \mu^2), \\ 
 -i\sqrt{{\mu^2-\bf p}^2}\ \ ({\bf p}^2 < \mu^2) \end{array}\right.
\end{eqnarray}
\end{subequations}
For the further evaluation of (\ref{eq:SIG.18b}) and (\ref{eq:SIG.18a}) we consider 
\begin{eqnarray}
 J &\equiv& \frac{1}{2\pi r}\int_0^\infty \frac{pdp}{\omega_p}\ \sin(pr)\ 
 \left[i \delta(\omega_p) + \frac{1}{\pi} {\cal P}\frac{1}{\omega_p}\right] \nonumber\\
 &=& \frac{1}{2\pi r}\left[-i \int_0^\mu\frac{pdp}{\sqrt{\mu^2-p^2}} + 
 \int_\mu^\infty \frac{pdp}{\sqrt{p^2-\mu^2}}\right]\ \sin(pr)\cdot\nonumber\\
 && \times 
 \left[i \delta(\omega_p) + \frac{1}{\pi} {\cal P}\frac{1}{\omega_p}\right] 
 \equiv \frac{1}{2\pi r}\left(K_1+K_2\right).
\label{eq:SIG.22}\end{eqnarray}
Using the identity
\footnote{
According to Eq.~(16.44) of \cite{Nak72} 
the Dirac $\delta$-function and Cauchy's Principal value
for a complex argument, being useful in extracting the finite
parts of the Dyson S-matrix, can be defined by
\begin{eqnarray*}
 \delta(\omega) &=& \lim_{\epsilon \rightarrow 0} 
 \delta_{\varepsilon}(\omega)\ \ ,\ \ 
 \delta_\varepsilon(\omega) = \int_{-\infty}^\infty \frac{d\tau}{2\pi}
 e^{-i\omega\tau} e^{-\frac{1}{2}\varepsilon\tau^2}, \\
 P\left(\frac{1}{\omega}\right) &=& \lim_{\epsilon \rightarrow 0} 
 P_{\varepsilon}\left(\frac{1}{\omega}\right)\ \ ,\ \ 
 P_\varepsilon\left(\frac{1}{\omega}\right) = \frac{1}{2}\int_{-\infty}^\infty d\tau\
 \sigma(\tau)\ e^{-i\omega\tau} e^{-\frac{1}{2}\varepsilon\tau^2}, \\
\end{eqnarray*}
From this definition it can be verified that $\delta(ix)=\delta(x)$.\\
}

\begin{equation}
 \delta(\omega_p) = \delta\left(\sqrt{\mu^2-p^2}\right) = \frac{1}{p}\sqrt{\mu^2-p^2}
 \left[\delta(p-\mu)+\delta(p+\mu)\right],
\label{eq:SIG.23}\end{equation}
we obtain for $K_1$ and $K_2$ 
\begin{eqnarray}
 K_1^\varepsilon &=& +\frac{1}{2}\sin(\mu r) 
 +\frac{1}{\pi} \int_0^{\mu-\varepsilon}\frac{pdp}{p^2-\mu^2} \sin(pr),
\nonumber\\ 
 K_2^\varepsilon &=& +\frac{i}{2}\sin(\mu r) 
 +\frac{1}{\pi} \int_{\mu+\varepsilon}^\infty \frac{pdp}{p^2-\mu^2} \sin(pr),
\label{eq:SIG.24}\end{eqnarray}
Here, a factor $1/2$ in included in the $\delta$-term because of the endpoint situation.

\noindent Next, we evaluate the similar integrals for the $\phi^\dagger$-propagation.
\begin{eqnarray}
 \widetilde{J} &\equiv& \frac{1}{2\pi r}\int_0^\infty \frac{pdp}{\widetilde{\omega}_p}\ \sin(pr)\ 
 \left[i \delta(\widetilde{\omega}_p) + 
 \frac{1}{\pi} {\cal P}\frac{1}{\widetilde{\omega}_p}\right] \nonumber\\
 &=& \frac{1}{2\pi r}\left[+i \int_0^\mu\frac{pdp}{\sqrt{\mu^2-p^2}} - 
 \int_\mu^\infty \frac{pdp}{\sqrt{p^2-\mu^2}}\right]\ \sin(pr)\cdot\nonumber\\
 && \times 
 \left[i \delta(\widetilde{\omega}_p) + 
 \frac{1}{\pi} {\cal P}\frac{1}{\widetilde{\omega}_p}\right] 
 \equiv \frac{1}{2\pi r}\left(\widetilde{K}_1+\widetilde{K}_2\right).
\label{eq:SIG.25}\end{eqnarray}
For $\widetilde{K}_1$ and $\widetilde{K}_2$ we obtain
\begin{eqnarray}
 \widetilde{K}_1^\varepsilon &=& -\frac{1}{2}\sin(\mu r) 
 +\frac{1}{\pi} \int_0^{\mu-\varepsilon}\frac{pdp}{p^2-\mu^2} \sin(pr),
\nonumber\\ 
 \widetilde{K}_2^\varepsilon &=& -\frac{i}{2}\sin(\mu r) 
 +\frac{1}{\pi} \int_{\mu+\varepsilon}^\infty \frac{pdp}{p^2-\mu^2} \sin(pr),
\label{eq:SIG.26}\end{eqnarray}
From these results, we find 
\begin{eqnarray}
 \frac{1}{2}\left(J+\widetilde{J}\right) &=& 
 +\frac{1}{\pi} {\cal P} \int_0^\infty \frac{pdp}{p^2-\mu^2} \sin(pr).
\label{eq:SIG.27}\end{eqnarray}
The time-integral over the propagator for the field 
$\widetilde{\epsilon} = (\phi+\phi^\dagger)/\sqrt{2}$ becomes
\begin{eqnarray}
&& \int^\infty_{-\infty} dz_0\ \frac{1}{2}\left[\Delta_F(z; i\mu)+\Delta_F(z;-i\mu)\right] = 
 \nonumber\\ &&
 \frac{1}{2\pi^2 r} {\cal P} \int_0^\infty \frac{qdq}{q^2-\mu^2} \sin(qr) = 
  {\cal P} \int\frac{d^3q}{(2\pi)^3}\ 
 \frac{\exp(i{\bf q}\cdot{\bf r})}{{\bf q}^2-\mu^2}.
\label{eq:SIG.28}\end{eqnarray}
and we obtain 
\begin{eqnarray}
 \langle p', P'| M_\varepsilon^{(2)}|p,P\rangle &=& g_\psi g_\chi\
 (2\pi)^{-3}\left[16 \omega_{p'}\omega_{P'} \omega_p \omega_P\right]^{-1/2}
 \cdot\nonumber\\ && \times    
 \int d^3r\ e^{-i({\bf p}'-{\bf p})\cdot{\bf r}}\cdot
 {\cal P} \int\frac{d^3q}{(2\pi)^3}\ \frac{\exp(i{\bf q}\cdot{\bf r})}{{\bf q}^2-\mu^2} 
 \nonumber\\ &=& g_\psi g_\chi\
 (2\pi)^{-3}\left[16 \omega_{p'}\omega_{P'} \omega_p \omega_P\right]^{-1/2}
 {\cal P} \frac{1}{({\bf p}'-{\bf p})^2-\mu^2},  
\label{eq:SIG.29}\end{eqnarray}
which justifies the use of the principal-value integral in Eq.~(\ref{app:SG.14}) etc.

\begin{flushleft}
\rule{16cm}{0.5mm}
\end{flushleft}
\section{Perihelion precession: -1/6-correction }
\label{app:PC3}    
The order $G^2$ contributions to the perihelion precession 
evaluated in sections \ref{sec:10}, \ref{app:S} and \ref{app:SPP}, 
differ from the Einstein result by a factor of 7/6.
Here, we evaluate in detail an additional effect of order $G^2$ 
in the interaction between m and M. 
This is associated with the gravitational energy between the planet (mas m) 
and the Sun (mass M), which is not localized on either mass \cite{Schw70}. It is distributed
in space and can be calculated from the Newtonian field strength:
\begin{equation}
 {\bf g}({\bf x};{\bf r}) = G\ \bm{\nabla}\left[\frac{M}{|{\bf x}|} + 
 \frac{m}{|{\bf x}-{\bf r}|}\right].
\label{app:PC3.1} \end{equation}
\begin{table}[hbt]
\begin{center}
\caption{ Newtonian Gravity and Electrostatics.
The positions and charges of the masses M and m are 
${\bf x}_M$ and ${\bf x}_m$, respectively Q and q. The relative distance is 
 ${\bf r}= {\bf x}_m-{\bf x}_M$.
}
\label{tab:grav-coul}
\begin{tabular}{c|l|l} \hline
 & Newtonian Gravity & Electrostatics \\ \hline 
 Force between & & \\[-3mm]
 & ${\bf F}_{N   } = -\frac{GmM}{r^2} \hat{\bf r}$ & 
   ${\bf F}_{C   } = +\frac{q Q}{r^2} \hat{\bf r}$ \\[-3mm]
 two sources & & \\
 Force derived & & \\[-3mm]
  & ${\bf F}_{N   } = -m \bm{\nabla} \Phi_{N   }({\bf x}_M)$ & 
   ${\bf F}_{C   } = -q \bm{\nabla} \Phi_{C   }({\bf x}_M)$ \\[-3mm]
 from potential & & \\
 Potential outside & & \\[-3mm]
 & $\Phi_{N   } = -\frac{GM}{r}$,\ \ ${\bf g} = -\bm{\nabla} \Phi_N({\bf x}_M$          
& $\Phi_{C   } = \frac{Q}{r}$,\ \ ${\bf E} = -\bm{\nabla} \Phi_C({\bf x}_M$ \\[-3mm]
 spherical source & & \\
 Field equation & & \\[-3mm]
 & $\bm{\nabla}^2 \Phi_{N   } = 4\pi G \mu({\bf x}_M)$         
 & $\bm{\nabla}^2 \Phi_{C   } =  -\rho_{elec}({\bf x}_M) $ \\[-3mm]
 for potential & & \\
\hline
\end{tabular}
\end{center}
\end{table}
The energy density in a Newtonian gravitational field can be derived as follows:
Consider the assembling of a system of N particles of mass $M_A$ at the positions
${\bf x}_A$. The Newtonian potential energy W of the system is the needed energy 
by bringing them one by one from infinity in the gravitational field of all particles 
already assembled, which is
\begin{equation}
  W({\bf r}) = -\frac{1}{2}\sum_{A \neq B} \frac{G M_A M_B}{|{\bf x}_A-{\bf x}_B|}.
\label{app:PC3.2} \end{equation}
For a continuous distribution of mass with density $\mu({\bf x})$ this is
\begin{equation}
  W({\bf r}) = -\frac{1}{2}\int\!\! d^3x\! \int\!\! d^3x'\ 
 \frac{G \mu({\bf x})\mu({\bf x}')}{|{\bf x}-{\bf x}'|} =
 \frac{1}{2} \int\!\! d^3x\ \mu({\bf x})\ \Phi_N({\bf x}).
\label{app:PC3.3} \end{equation}
Via the Newtonian field equation
$\bm{\nabla}^2 \Phi_N({\bf x}) = 4\pi G\ \mu({\bf x})$ 
one can eliminate the source $\mu({\bf x})$ in the last expression of 
Eq.~(\ref{app:PC3.3}) and applying the divergence theorem leads to 
\begin{equation}
  W({\bf r}) = -\frac{1}{8\pi G}\int\!\! d^3x\ 
 \bm{\nabla}\Phi_N({\bf x})\cdot\bm{\nabla}\Phi_N({\bf x}) =
  -\frac{1}{8\pi G}\int\!\! d^3x\ [{\bf g}({\bf x})]^2 \equiv 
 \int\!\! d^3x\ \rho_{grav}({\bf x}).
\label{app:PC3.4} \end{equation}
Here $\rho_{grav}({\bf x})$ is the energy density associated with the Newtonian gravitational field.

\noindent The gravitational energy density in the 
gravitational field  of the masses M and m is 
\begin{equation}
 \rho_{grav}({\bf x};{\bf r}) = 
-\frac{1}{8\pi} G\ \bm{\nabla}\left(\frac{M}{|{\bf x}|}+\frac{m}{|{\bf x}-{\bf r}|}\right)
  \cdot\bm{\nabla}\left(\frac{M}{|{\bf x}|}+\frac{m}{|{\bf x}-{\bf r}|}\right).
\label{app:PC3.7} \end{equation}
Here, the gravitational self-energy terms, proportional to $M^2$ and $m^2$, are in principle
incorporated into M and m and independent of each others presence. (Moreover, these self-energy
terms are independent of the positions and hence can not contribute to the perihelion
precession.)
The cross term, which contains the correlation of the two bodies, is ${\bf r}$-dependent 
and given by
\begin{equation}
 \rho_{cross}({\bf x};{\bf r}) = -\frac{1}{4\pi} G\ \bm{\nabla}\left(\frac{M}{|{\bf x}|}\right)
  \cdot\bm{\nabla}\left(\frac{m}{|{\bf x}-{\bf r}|}\right).
\label{app:PC3.8} \end{equation}
{\it Because of the mass-energy equivalence, the energy density $\rho_{cross}({\bf x};{\bf r})$ 
implies also a mass distribution $\mu_{cross}= \rho_{cross}/c^2$, and hence gives a gravitational pull 
to the planet and sun.}
The interaction energy of this energy density with M is of order $G^2$ and given by (units c=1)
\begin{eqnarray}
 {\cal V}_{cross,M}({\bf r}) &\equiv& 
-\int\!\! d^3x\ \frac{MG}{|{\bf x}|}\ \mu_{cross}({\bf x};{\bf r}) =
  \frac{1}{4\pi} G^2 M^2 m\ \int\!\! d^3x\ 
\left(\frac{1}{|{\bf x}|}\bm{\nabla}\frac{1}{|{\bf x}|}\right)\cdot
 \bm{\nabla}\frac{1}{|{\bf x}-{\bf r}|} \nonumber\\ &=& \frac{1}{8\pi} G^2 M^2 m\ 
\int\!\! d^3x\ \bm{\nabla}\frac{1}{|{\bf x}|^2}\cdot       
 \bm{\nabla}\frac{1}{|{\bf x}-{\bf r}|} = 
  -\frac{1}{8\pi} G^2 M^2 m\cdot\nonumber\\ && \times 
  \int\!\! d^3x\ \frac{1}{|{\bf x}|^2}\cdot       
 \bm{\nabla}^2\frac{1}{|{\bf x}-{\bf r}|}  = 
  +\frac{1}{2} G^2 M^2 m\ \frac{1}{|{\bf r}|^2}  = + \frac{V^2}{2m}.
\label{app:PC3.9} \end{eqnarray}
Here, we used 
$|{\bf x}|^{-1} \bm{\nabla} |{\bf x}|^{-1} = \bm{\nabla}|{\bf x}|^{-2}/2$,  
 applied partial integration, $\bm{\nabla}^2(1/r)= -4\pi \delta({\bf r})$, and
the Newtonian potential $V \equiv -GMm/r$. In total, one has for the perihelion precession
$-7V^2/2m+ V^2/2m = -6V^2/2m$, which is Einstein's result.
\begin{flushleft}
\rule{16cm}{0.5mm}
\end{flushleft}

\begin{center}
\fbox{\rule[-5mm]{0cm}{1.5cm} \hspace{5mm}  
\begin{minipage}[]{12.9cm}
\noindent {\bf Note}: 
In this note, we give a detailed derivation that is symmetric 
between the planet and the sun. 
Similar to the mass M, the interaction energy of the energy density 
$\rho_{cross}$ with the planet, mass m, is given by
\begin{eqnarray}
 \hspace{-2.0cm} {\cal V}_{cross,m}({\bf r}) &\equiv& 
-\int\!\! d^3x\ \frac{mG}{|{\bf x}|}\ \mu_{cross}({\bf x};{\bf r}) =
  \frac{1}{4\pi} G^2 M m^2\cdot \int\!\! d^3x\ 
\left(\frac{1}{|{\bf x}-{\bf r}|}\bm{\nabla}\frac{1}{|{\bf x}|}\right)\cdot
\nonumber\\ &&\hspace{-1.0cm} \cdot
 \bm{\nabla}\frac{1}{|{\bf x}-{\bf r}|}= 
  +\frac{1}{2} G^2 M m^2\ \frac{1}{|{\bf r}|^2}  = + \frac{V^2}{2M} \ll {\cal V}_{cross,M}({\bf r})\ \
 {\rm for}\ \ m \ll M.
\label{app:PC3.11} \end{eqnarray}
The total potential energy due to $\mu_{cross}({\bf x},{\bf r})$ is
\begin{eqnarray}
 {\cal V}_{cross}({\bf r}) &=& {\cal V}_{cross,M}({\bf r}) +        
 {\cal V}_{cross,m}({\bf r}) \nonumber\\ &=& \frac{1}{2} G^2 M m\ (M+m)
 \frac{1}{|{\bf x}_A-{\bf x}_B|^2} = \frac{M+m}{2M m}\ V^2, 
\label{app:PC3.12} \end{eqnarray}
where ${\bf x}_A$ and ${\bf x}_B$ are the position of the sun and the
planet respectively, ${\bf r}={\bf x}_B-{\bf x}_A$, and $V=-GMm/r$. 
{\it The potential energies ${\cal V}_{cross,M}$ and ${\cal V}_{cross,m}$ are both sensitive to the 
position of the sun and the planet, and therefore leads to a force between the
planet and the sun}. Notice the symmetry with respect to $ M \leftrightarrow m$ which
ensures the {\it $<$ action = -reaction $>$} rule.

\noindent In the center-of-mass, the separation of the relative motion, 
taking $M_A=M$, $M_B=m$, is as follows: 
\begin{subequations}
\label{app:PC3.13} 
\begin{eqnarray}
 M \frac{d^2 {\bf x}_A}{dt^2} &=& +\bm{\nabla}_A {\cal V}_{cross} =
 +\frac{1}{2} \frac{M+m}{Mm}\ \bm{\nabla}_r V^2, \\
 m \frac{d^2 {\bf x}_B}{dt^2} &=& +\bm{\nabla}_B {\cal V}_{cross} =
 -\frac{1}{2} \frac{M+m}{Mm}\ \bm{\nabla}_r V^2, 
 \end{eqnarray}
\end{subequations}
This gives for the center of mass $d^2 {\bf R}_{c.m.}/dt^2=0$, 
and for the relative motion
\begin{subequations}
\label{app:PC3.14}
\begin{eqnarray}
\frac{d^2}{dt^2}({\bf x}_B-{\bf x}_A) &=&
 -\frac{1}{2}\left(\frac{1}{M}+\frac{1}{m}\right) \frac{M+m}{Mm}\
 \bm{\nabla} V^2 = -\frac{1}{2}\left(\frac{M+m}{Mm}\right)^2 \bm{\nabla} V^2
 \Rightarrow \\ 
 \mu_{red} \frac{d^2 {\bf r}}{dt^2} &=& -
 \bm{\nabla}\left(\frac{V^2}{2\mu_{red}}\right) 
 \approx -\bm{\nabla}\frac{V^2}{2m}.
 \end{eqnarray}
\end{subequations}
Here, $\mu_{red} = M m/(M+m) \approx m$ for $m \ll M$.\\

\noindent {\it Application of (\ref{app:PC3.14}) to the planet-sun system demonstrates that 
the potential in Eq.~(\ref{app:PC3.9}) indeed represents to a very good approximation
the proper extra potential from the $\rho_{cross}$ energy distribution.}

\end{minipage} \hspace{5mm} }\\
\end{center}
\begin{flushleft}
\rule{16cm}{0.5mm}
\end{flushleft}
\section{Cosmological constant and Graviton mass}
\label{app:COSMO}  

\noindent The gravitational action including the cosmological term 
reads \cite{Weinberg72,LL.V2}   
\begin{eqnarray}
 S_g &=& -\frac{c^3}{16\pi G} \int d^4x\ \sqrt{-g}\ R  
         -\frac{1}{c} \int d^4x\ \sqrt{-g}\ \Lambda. 
\label{cosmo.1}\end{eqnarray}
The Einstein equation follows from 
\begin{eqnarray}
 \frac{\delta S_g}{\delta g_{\mu\nu}} &=& -\frac{c^3}{16\pi G} \left(
 R^{\mu\nu}-\frac{1}{2} g^{\mu\nu} R \right)\sqrt{-g}
 +\frac{\Lambda}{2c}\ g^{\mu\nu}\sqrt{-g} = 0, 
\label{cosmo.2}\end{eqnarray}
or, with the inclusion of the matter term $\delta S_M/\delta g_{\mu\nu}$, 
\begin{equation}
 \left( R^{\mu\nu}-\frac{1}{2} g^{\mu\nu} R \right)
 +\lambda\ g^{\mu\nu} = -\kappa T^{\mu\nu}/c^2, 
\label{cosmo.3}\end{equation}
with Einstein's constant $\kappa= 8\pi G/c^2$ and 
$\lambda = 8\pi G \Lambda/c^4= \kappa \Lambda/c^2$.
Here is used, see \cite{diffdet},
$\delta\sqrt{-g}/\delta g_{\mu\nu} = -g^{\mu\nu} \sqrt{-g}/2$.
Eqn.~(\ref{cosmo.3}) is Einstein's equation Ref.~\cite{Einstein19}, see also 
Ref.~\cite{Rindler} equation (8.1.39).\\
\noindent Incorporation of the cosmological term in the spin-2 formalism of
this paper, in the weak-field approximation, is achieved by 
the change 
${\cal L}^{(2)} \rightarrow {\cal L}^{(2)} + c_0\ \sqrt{-g}$ 
in the spin-2 Lagrangian, with $c_0=\Lambda/c$.
In the weak field approximation, this becomes
, see Appendix~\ref{app:Miscel},
\begin{eqnarray}
 {\cal L}^{(2)} &\rightarrow& {\cal L}^{(2)} + c_0\ \left[
\vphantom{\frac{A}{A}} 1+\frac{1}{2} \kappa h^\mu_\mu -\kappa^2\left(
 \frac{1}{4} h^\mu_\nu h^\nu_\mu -\frac{1}{8}(h^\mu_\mu)^2\right)\right]
\label{cosmo.5}\end{eqnarray}
With the constraint $h^\mu_\mu=0$, coming from 
$\partial{\cal L}_{\eta\epsilon}/\partial\epsilon(x)=0$, only the 
$c_0\ h^{\mu\nu}h_{\mu\nu}/4$-term is relevant. This implies that in the
Klein-Gordon equation (\ref{eq:20.10}) for the $h^{\mu\nu}$-field 
$M_2^2 \rightarrow M_2^2+c_0 \kappa^2$.
Assuming that the origin of the gravitational mass is entirely   
due to the cosmological constant we have $\mu_G=M_2 = \sqrt{\Lambda/c}\ \kappa$.


\noindent The Friedmann equation reads \cite{Borner88}, see also 
\cite{Rindler} eqn.~(9.73) with $\Lambda=\lambda$, 
\begin{equation}
 H^2 \equiv \left(\frac{\dot{R}}{R}\right)^2 = \frac{8\pi G}{3}
\rho -\frac{kc^2}{R^2}+\frac{1}{3} \lambda c^2.
\label{cosmo.7}\end{equation}
Note that for $R \gg 1$ the density becomes
\begin{equation}
 \rho = \frac{3H^2}{8\pi G}-\frac{\lambda c^2}{8\pi G}
	= \rho_c-\frac{1}{16\pi G} \left(\frac{\mu_Gc^2}{\hbar}\right)^2. 
\label{cosmo.8}\end{equation}
The sign of the $\lambda$-term is in agreement with \cite{Borner88} Eqn.~(9.1), but is
opposite to that in 
Ref.~\cite{Log04} Eqn.~(10.27) which has $\Lambda \Rightarrow -\mu_G^2c^2/\hbar^2$,
and implies the presence of "dark matter". $\Lambda < 0$ leads           
to an Anti-deSitter space for an empty universe, which seems unphysical.   
At the present epoch, the Hubble constant is
\begin{equation}
 H_0^2 = \frac{8\pi G}{3} \rho_0 -\frac{kc^2}{R_0^2}+\frac{1}{3}\lambda c^2.
\label{cosmo.9}\end{equation}
The deceleration parameter $q_0=-\ddot{R}/RH^2$ 
satisfies \cite{Islam92}, with $\Omega_0= (8/3) \pi GH_0^{-2} \rho_0$, 
\begin{equation}
q_0 \equiv -\ddot{R}/RH^2  = \frac{1}{2} \Omega_0 -c^2 \Lambda/3H_0^2.
\label{cosmo.10}\end{equation}
From observations, the deceleration parameter  
 $|q_0| < 5$ \cite{Haw83}, which gives 
\begin{equation}
 |\Lambda| \leq 21 H_0^2/c^2 \approx 10^{-54}\ cm^{-2},
\label{cosmo.11}\end{equation}
for $H_0 \leq 100$ km\ s$^{-1}$\ Mpc$^{-1}$.
With $M_{Pl} = 10^{-33}$ cm$^{-1}$, one has $|\Lambda|/M_{Pl}^2 < 10^{-120}$.\\

\noindent In the GUT picture, before the breakdown of the GUT gauge-symmetry via a
first-order phase transition at the critical temperature 
$T_c \approx 10^{14}$ GeV, the (GUT) cosmological constant is much
larger than the present one, and is given by
\begin{equation}
 \Lambda \equiv \frac{8\pi G}{3} T_c^4 \approx (10^{15} GeV)^4 M_{Pl}^{-2}.
\label{cosmo.12}\end{equation}
Interpretation of the cosmological constant term as a mass term in
the equation of the $h^{\mu\nu}$-field we have \cite{Log04}
\begin{equation}
 \mu_G = \sqrt{2\Lambda} = \sqrt{\frac{16\pi G}{3}}\ T_c^2=
 \sqrt{\frac{16\pi}{3}} \left(\frac{T_c}{M_{Pl}}\right)^2\ M_{Pl} 
 = 4.1\ 10^{-10} M_{Pl},
\label{cosmo.13}\end{equation}

\noindent The perihelion precession of Mercury imposes a limit on the
present cosmological constant, which follows from the modification of the
Schwarzschild metric, namely
\begin{equation}
 ds^2 = c^2\left(1-2M/r-\frac{1}{3}\Lambda r^2\right)\ dt^2 -
 \left(1-2M/r-\frac{1}{3}\Lambda r^2\right)^{-1}\ dr^2 -
 r^2\left(d\theta^2 + \sin^2\theta\ d\phi^2\right), 
\label{cosmo.14}\end{equation}
where $M=M_\odot G/c^2$. From the accuracy of the value of the 
perihelion precession of Mercury, one derives that, see \cite{Islam92},
\begin{equation}
 |\Lambda| < 10^{-42} cm^{-2} =  10^{-108}\ M_{Pl}^2 \rightarrow 
 \mu_G=\sqrt{2\Lambda}= 1.4\ 10^{-54} M_{Pl} \approx 2.8\ 10^{-32}\ m_e, 
\label{cosmo.15}\end{equation}
where is used $M_{Pl}^{-1} = 10^{-33}$ cm, in units $\hbar=c=1$.              

\noindent The upper limit from the LIGO-Virgo Collaboration is
$\mu_G < 1.76\times 10^{-23} eV/c^2 = 3.44\times 10^{-29} m_e$ \cite{Abbott21}.

\noindent The solar system upper limit of the graviton mass is 
$m_g^{SS} < 4.4\times 10^{-22} eV/c^2$, giving 
$\lambda^{SS}_g > 2.8\times 10^{12}$ km. LISA measurements with an "ideal source",
could give an improvement by a factor $\sim 50$ \cite{CHL2018}.

\noindent {\it The transition between the large cosmological constant $\Lambda$ in
(\ref{cosmo.12}) and the tiny one in (\ref{cosmo.15}) can be understood within  
the inflational phase transition scenario \cite{Guth81}. 
For this, the "latent heat" $\Lambda M_{pl}^2$ is during this 
phase transition transformed into radiation, diminishing enormously 
the cosmological constant. }

\section{Miscellaneous formulas }             
\label{app:Miscel}   
For a diagonalizable matrix A
\begin{eqnarray}
	\det(I+\varepsilon A) &=& 1 + \varepsilon\ f_1(A) + \varepsilon^2\ f_2(A) + \ldots
	\label{Mis.1}
\end{eqnarray}
The first order term is $f_1(A)= Tr(A)$. To calculate the second order,
we use
\begin{eqnarray}
	I+ \varepsilon A &=& \exp(\varepsilon B),\ \det(\varepsilon B)= \exp[\varepsilon Tr(B)]
	\label{Mis.2}
\end{eqnarray}
This leads to
\begin{eqnarray}
\det(I+\varepsilon A) &=& \det[\exp(\varepsilon B)] = \det\left(I + \varepsilon B
+\frac{\varepsilon^2}{2} B^2 + \dots\right) \nonumber\\ &=&
1 + \varepsilon\ Tr(B) +\frac{\varepsilon^2}{2} \left(Tr\ B\right)^2 + \ldots.
	\label{Mis.3}
\end{eqnarray}
Also, we can rewrite
\begin{eqnarray}
	\det(I+\varepsilon A) &=& 
	\det[\exp(\varepsilon B)] = \det\left[I + \varepsilon \biggl( B
	+\frac{\varepsilon}{2} B^2 + \dots\biggr)\right] \nonumber\\ &=&
1 + \varepsilon\ Tr \left(B +\frac{\varepsilon}{2} B^2 + \ldots\right) 
	+\varepsilon^2 f_2\bigl(B+ \ldots\bigr) + O(\varepsilon^3).
	\label{Mis.4}
\end{eqnarray}
Taking $\lim\varepsilon \rightarrow 0$ leads to
\begin{eqnarray}
 f_2(B) &=& \frac{1}{2}\bigl[\left(Tr\ B\right)^2- Tr\ \left(B^2\right)\bigr].
	\label{Mis.5}
\end{eqnarray}
Since in this limit A=B we have $f_2(A)=f_2(B)$.

\noindent For $g_{\mu\nu}= \eta_{\mu\nu}+\kappa\ h_{\mu\nu}$ or in term of
matrices $g = \eta +\kappa\ h$ we introduce
\begin{eqnarray}
	I &=& D\ \eta\ D\ ,\ A= D\ h\ D\ ,\ \varepsilon = \kappa,
	\label{Mis.6}
\end{eqnarray}
where the diagonal matrix D has $D_{00}=1, D_{mm}=i\ (m=1,2,3)$. Then,
\begin{eqnarray}
	\det(I + \varepsilon\ A) &=& \det\bigl(D\left(\eta + \kappa\ h\right)D\bigr)=
	 -\det\bigl(\eta+\kappa\ h\bigr),
	\label{Mis.7}
\end{eqnarray}
since $\det\ D=-i$. 
Using the result above, we obtain
\begin{eqnarray}
	\det\bigl(\eta + \kappa\ h\bigr) &=& -\det(I+\varepsilon\ A) =
	-\bigl(1+\varepsilon\ f_1(A) +\varepsilon^2 f_2(A) + \dots\bigr)
	\label{Mis.8}
\end{eqnarray}
This gives 
\begin{eqnarray}
	\sqrt{-\det(g)} &=& \left[-\det\bigl(\eta + \kappa\ h\bigr)\right] 
	\nonumber\\ &=&
	\left(1+\varepsilon\ f_1(A) + \varepsilon^2 f_2(A) + \ldots\right)^{1/2}
	\nonumber\\ &=&
	1+\frac{1}{2}\varepsilon\ f_1(A)+\frac{1}{2}\varepsilon^2 f_2(A)
	-\frac{1}{8}\varepsilon^2 f_1^2(A) + O(\varepsilon^3)
	\label{Mis.9}
\end{eqnarray}
Now, $Tr\ A= Tr(DhD)=Tr(D^2h) = Tr(\eta h)$ which gives
\begin{eqnarray}
	\sqrt{-\det(g)} &=& 1+\frac{1}{2}\varepsilon\ Tr(\eta h) +\frac{1}{2}\epsilon^2
	\left[ \frac{1}{2}(Tr(\eta h))^2-\frac{1}{2} Tr((\eta h)^2)
	-\frac{1}{4} (Tr(\eta h))^2\right]  +\ldots
	\label{Mis.10}
\end{eqnarray}
Up to the second order in the gravitation constant $\kappa$ we obtained
\begin{eqnarray}
\sqrt{-\det(g)} &=& 1+\frac{1}{2}\kappa\ h^\mu_\mu -\kappa^2 \left[
	\frac{1}{4} h^{\mu\nu} h_{\nu\mu}-\frac{1}{8} h^\mu_\nu  h^\nu_\nu \right].
	\label{Mis.11}
\end{eqnarray}



\begin{thebibliography}{99}
\bibitem{neglect.1} 
Consider a plane wave with $k \approx 2\pi/R_m$, where $R_m$ is the radius of the 
planet m in its orbit around the sun. Then, 
$\left| \bm{\nabla}V^{(0)}\cdot\bm{\nabla}\psi\right| \sim (G M m/R_m^3)$, which gives the estimate
\begin{eqnarray*}
&& \left|\frac{ \bm{\nabla} V^{(0)}\cdot\bm{\nabla}\psi}{M m\ V^{(0)} 
\psi}\right| \sim [M m\ R_m^2]^{-1} \ll 1,   
\end{eqnarray*}
justifying the neglect of the $\bm{\nabla} \psi$ term in the planetary motion.
\bibitem{note:difference}  
In contrast to the treatment in \cite{Nak72} 
the commutation relations etc., have an extra (-)-sign  
in order to have a negative-metric. 
This implies that for the Feynman propagator, there occurs a $(-)$-sign:
\begin{eqnarray*}
 {\rm positive-metric} &:& \langle 0|T\left[\phi(x)\phi(y)\right]|0\rangle =
 +i\Delta_F(x-y), \\
 {\rm negative-metric} &:& \langle 0|T\left[\phi(x)\phi(y)\right]|0\rangle =
 -i\Delta_F(x-y).    
\end{eqnarray*}
This is obvious from the derivation of the vacuum expectation values 
of $\langle 0| \phi(x)\phi(y)|0\rangle$ given in \cite{BD65}, section (12.4).
\bibitem{diffdet}  
Consider the matrix with elements $a_{ij}, i,j=1 \ldots n$. Development 
of the determinant with respect to the i-th row is
\begin{equation}
 det A = \sum_{j=1}^n a_{ij} A_{ij},\ A_{ij}= (-)^{i+j} \Delta_{ij}, 
\label{diffdet.1}\end{equation}
where $\Delta_{ij}$ is the minor with respect to the ij-element of A, i.e. the 
determinant of the matrix obtained from A by deleting row i and column j.
The minor is related to the inverse of A 
\begin{equation}
 (A^{-1})_{ij} = (det A)^{-1} A_{ji}
\label{diffdet.2}\end{equation}
which implies that for each row i
\begin{equation}
 det A = \sum_{j=1}^n a_{ij}\cdot (A^{-1})_{ji}(det A),
\label{diffdet.3}\end{equation}
which implies
\begin{equation}
 \frac{\partial det A}{\partial a_{ij}} = (A^{-1})_{ji} (det A).
\label{diffdet.4}\end{equation}
Because
\begin{equation}
  g_{\mu\nu} g^{\nu\lambda} = \delta^\lambda_\mu,\ g^{\mu\nu}=g^{\nu\mu},
\label{diffdet.5}\end{equation}
application of (\ref{diffdet.4}) leads to 
\begin{equation}
 \delta\sqrt{-g}/\delta g_{\mu\nu} = -\frac{1}{2} g^{\mu\nu} \sqrt{-g}.
\label{diffdet.6}\end{equation}
\end{thebibliography}
\end{document}